\documentclass[journal]{IEEEtran}
\usepackage[colorlinks,bookmarksopen,bookmarksnumbered,citecolor=red,urlcolor=red]{hyperref}
\usepackage{tabularx,booktabs,caption}
\usepackage{cite}
\usepackage{amsmath}
\usepackage{amssymb}
\usepackage{latexsym}
\usepackage{float}
\usepackage{epsfig}
\usepackage{subfigure,epstopdf,graphicx,array,algorithm}
\usepackage[]{graphicx}
\usepackage{algpseudocode}

\algnewcommand{\Inputs}[1]{%
	\State \textbf{Inputs:}
	\Statex \hspace*{\algorithmicindent}\parbox[t]{.7\linewidth}{\raggedright #1}
}
\algnewcommand{\Initialize}[1]{%
	\State \textbf{Initialize:}
	\Statex \hspace*{\algorithmicindent}\parbox[t]{.7\linewidth}{\raggedright #1}
}

\ifCLASSINFOpdf
\else
\fi

\begin{document}
	\title{{A Survey on Design and Performance of Higher-Order QAM Constellations}}
		\author{
		\IEEEauthorblockN{Praveen Kumar Singya$^{1}$, Parvez Shaik$^{2}$, Nagendra Kumar$^{3}$, Vimal Bhatia$^{2}$, \IEEEmembership{Senior~Member,~IEEE},\\ and Mohamed-Slim Alouini$^{1}$, \IEEEmembership{Fellow, IEEE}}
		
		\thanks{$^{1}$P. K. Singya and M.-S. Alouini are with the Computer, Electrical, and Mathematical Science and Engineering (CEMSE) Division, King Abdullah University	of Science and Technology (KAUST), Thuwal 23955-6900, Saudi Arabia (e-mail: praveen.singya@kaust.edu.sa, slim.alouini@kaust.edu.sa)}
		\thanks{$^{2}$P. Shaik and V. Bhatia are with the Discipline of Electrical Engineering, Indian Institute of Technology Indore, Indore-453552, India (e-mail: phd1601202003@iiti.ac.in, vbhatia@iiti.ac.in)}
		\thanks{$^{3}$N. Kumar is with the Department of Electronics \& Communication Engineering, National Institute of Technology Jamshedpur-831014, India (e-mail: kumar.nagendra86@gmail.com)}
	}
	
\maketitle

\begin{abstract} As the research on beyond 5G heats up, we survey and explore power and bandwidth efficient modulation schemes in details.   In the existing publications and in various communication standards, initially square quadrature amplitude modulation (SQAM) constellations (even power of 2) were considered.  However, only the square constellations are not efficient for varying channel conditions and rate requirements, and hence, odd power of 2 constellations were introduced. For odd power of 2 constellations, rectangular QAM (RQAM) is  commonly used.  However, RQAM is not a good choice due to its lower power efficiency, and a modified cross QAM (XQAM) constellation is preferred as it provides improved power efficiency over RQAM due to its energy efficient two dimensional (2D) structure.
The increasing demand for high data-rates has further encouraged the research towards  more compact 2D constellations  which lead to hexagonal lattice structure  based constellations, referred to as hexagonal QAM (HQAM).  In this work, various QAM constellations are discussed and detailed study of star QAM, XQAM, and HQAM constellations is presented. Generation, peak and average energies, peak-to-average-power ratio, symbol-error-rate, decision boundaries,  bit mapping,  Gray code penalty, and bit-error-rate of star QAM, XQAM, and HQAM constellations with different constellation orders are presented. Finally, a comparative study of various QAM constellations is presented which justifies the supremacy of HQAM over other QAM constellations.
  With this, it can be claimed that the use of the HQAM in various wireless communication systems and standards can further improve the performance targeted for beyond 5G wireless communication systems.  
\end{abstract} 
%
\section {Introduction}
%
\normalsize
	 With 5G deployments beginning to take place, the basic need of high data-rate multimedia applications can be met with optimum bandwidth and power efficient modulation schemes.  Over a wider prospective, digital modulation schemes play a key role in attaining high data-rates with bandwidth and power efficiency. The basis of 5G and future wireless communication systems depends on the robustness of the employed modulation schemes. A digital modulation scheme for information transmission  depends on various factors such as data-rates, robustness to channel impairments, bandwidth, power, and cost efficiency \cite{goldsmith2005wireless}. Data-rate describes the maximum information transmission through a channel whereas the bandwidth efficiency describes maximum utilization of the limited spectrum by accommodating more information. Power efficiency describes transmission of reliable information with optimum power. However, in practice optimization of all these factors may not be possible at the same time. For example, if power efficiency is targeted then lower order modulation is preferred which leads to lower bandwidth efficiency and lower data-rates. Hence, there is a trade-off between various expectations from modulation schemes. The optimization/trade-off of these parameters is application oriented in digital radio frequency (RF) system design. Considering a terrestrial microwave radio
	link design, bandwidth efficiency with low bit-error-rate (BER) is given high priority, since, the RF stations are connected to power source. Hence, power efficiency is not a prime concern  and also receiver's cost/complexity is not considered because  few receivers are required. On the other hand, in cellular communication, power efficiency is mainly focused since the mobile phones operate on a limited battery power. In mobile communication, both power and cost efficiency are stronger constraints than bandwidth efficiency.

	Digital modulation schemes are highly impacted by noise however, this provides  flexibility of multiplexing various information formats at high data-rates with good quality-of-service (QoS). Digital modulation is classified based on the variation in transmitted signal in terms of amplitude, phase or frequency with respect to the digital message signal. If the amplitude or phase of the transmitted signal is varied with respect to the message signal, the resultant signal is termed as amplitude shift keying (ASK) or phase shift keying (PSK), respectively. ASK and PSK are known as the linear modulation techniques since, they follow the principle of superposition and scaling. If the transmitted signal frequency varies with respect to the message signal, this results in frequency shift keying (FSK). Since, FSK is a non-linear modulation technique, it is not spectrally efficient as compared to the linear modulation schemes. Thus, linear modulation techniques are widely employed in wireless communications \cite{rappaport1996wireless}. There is yet another advance modulation technique where both the amplitude and phase of the transmitted signal varies with respect to the digital message signal, known as quadrature amplitude modulation (QAM).  Among the available digital modulation schemes, QAM and quadrature phase shift keying (QPSK) are widely used in communication standards because of the bandwidth and power efficiency. Further, $M$-ary QAM is more power efficient than $M$-ary PSK modulation \cite{rappaport1996wireless} and thus,  is widely preferred in the modern wireless communication standards. \par{}

	In modern and future wireless communication systems, power efficient high data-rate transmission with efficient utilization of the limited bandwidth is a major concern. Power efficiency can be achieved by reducing the average transmit power of the constellation for a fixed BER. However, high data-rates for a limited bandwidth require high transmit power to maintain the same performance. Adaptive modulation is one such solution for spectrally efficient high data-rate communication with optimum power efficiency for the present and future wireless communication systems. In wireless communication, received signal-to-noise ratio (SNR) varies due to the multipath fading. In such scenario, adaptive modulation plays an important role by associating high data-rates to the channels with good channel conditions (high SNR) and low data-rates to the channels with bad channel conditions (low SNR) to improve overall system capacity for a limited bandwidth. Hence, an appropriate modulation scheme with variable constellation order is selected based on the channel condition. In adaptive modulation, constant SNR is maintained by varying various parameters such as transmitted power, data-rates,  and modulation orders \cite{goldsmith2005wireless, hayes1968adaptive, cavers1972variable, otsuki1995square, webb1995variable, kamio1995performance,kallel1994adaptive, ue1995symbol, goldsmith1997variable}.  Adaptive modulation is embodied in the present day wireless standards since the $3^{rd}$ generation (3G) \cite{UMTSHSPA} mobile networks.  Adaptive modulation finds its applications in high-speed modems \cite{bingham1990multicarrier, chow1995practical}, satellite links \cite{filip1990optimum, monk1995open, rose}, and in the applications where high QoS is to be maintained \cite{cox1991subband}.
 
	Performance of the communication systems can be improved further by combining the spatial diversity techniques\cite{alouini1999capacity} with the adaptive transmission. Spatial diversity can be achieved by employing multiple antennas at the transmitter or  receiver, or at both. By using multiple antennas at  both the transmitter and the receiver, multiple copies of the information signal can be transmitted and hence, high data-rates can be achieved within the limited bandwidth at  the cost of increased hardware and signal processing complexity. However, it is not always possible to deploy multiple antennas at each node due to the device size, cost, and implementation complexity. In such cases, spatial diversity can be achieved through cooperative relaying which is considered as  virtual multiple-input and multiple-output (MIMO) system. In cooperative relaying, the end-to-end information transmission takes place through indirect links (relay nodes) by enabling the cooperation with the relays employed. \par{}
	
	Motivated with this, in this work, a detailed study of various higher order QAM constellations is presented. This study explores star QAM, cross QAM (XQAM), and hexagonal QAM (HQAM) constellations in details. Their generation, peak and average energies, peak-to-average-power ratio (PAPR), symbol-error-rate (SER), decision boundaries,  bit mapping, Gray code penalty, and BER  for different constellation orders are presented. Further, applications of different QAM constellations in various communication fields are explored. In the following subsections, a detailed list of abbreviations used in this work and paper organization are presented.
	\subsection{List of Abbreviations}
	Various abbreviations used in this work are shown in Table \ref {Abbre}. 
	\begin{table}[!h]
		\centering
		\caption {Various abbreviations used in this work.}
		\label{Abbre}
		\begin{tabular}{|p{1.2cm}|p{6.7cm}|}
			\hline
			{1D}     & {1 dimensional} \\
			\hline
			{2D}     & {2 dimensional} \\
			\hline
			{3G}     & {$3^{rd}$ generation} \\
			\hline
			{4G}     & {$4^{th}$ generation} \\
			\hline
			{5G}     & {$5^{th}$ generation} \\
			\hline
			{ADSL}   & {Asymmetric digital subscribers line} \\
			\hline
			{AF}     & {Amplify-and-forward} \\
			\hline
			{APSK}   & {Amplitude-phase shift keying} \\
			\hline
			{ASK}    & {Amplitude shift keying}  \\
			\hline	
			{ASEP}   & {Average symbol-error-probability} \\
			\hline
			{ASER}   & {Average symbol-error-rate} \\
			\hline
			{AWGN}   & {Additive white Gaussian noise} \\
			\hline
			{BER}    & {Bit-error-rate} \\
			\hline
			{BEP}    & {Bit-error-probability} \\
			\hline
			{CDF}    & {Cumulative distribution function} \\
			\hline	
			{CDMA}   & {Code division multiple access} \\
			\hline 
			{DF}     & {Decode-and-forward} \\
			\hline
			{DVB}    & {Digital video broadcasting} \\
			\hline  	
			{DVB-S2} & {Digital video broadcasting-satellite second generation} \\
			\hline
			{DVB-SH} & {Digital video broadcasting-satellite services to handheld} \\
			\hline
			{FEC}    & {Forward error correction} \\
			\hline  	
			{FSK}    & {Frequency shift keying} \\
			\hline 
			{FSO}    & {Free space optics} \\
			\hline 
			{HQAM}   & {Hexagonal QAM} \\
			\hline
			{i.n.i.d.}  & {Independent and non-identically	distributed} \\
			\hline
			{LED}    & {Light emitting diode} \\
			\hline
			{LoS}    & {Line-of-sight} \\
			\hline
			{MIMO}   & {Multiple-input and  multiple-output} \\
			\hline
			{ML}     & {Maximum likelihood} \\
			\hline
			{MMSE}   & {Minimum mean square error} \\
			\hline
			{MRC}    & {Maximum ratio combining} \\
			\hline
			{NLA}    & {Non-linear amplifier} \\
			\hline
			{NLoS}   & {Non-line-of-sight} \\
			\hline
			{OFDM}   & {Orthogonal frequency division multiplexing} \\
			\hline
			{OWC}    & {Optical wireless communication} \\
			\hline    	
			{PAPR}   & {Peak-to-average power ratio} \\
			\hline
			{PDF}    & {Probability density function} \\
			\hline
			{PSK}    & {Phase shift keying} \\
			\hline  
			{QAM}    & {Quadrature amplitude modulation} \\	 
			\hline
			{QoS}    & {Quality-of-service} \\
			\hline
			{QPSK}   & {Quadrature phase shift keying} \\
			\hline	
			{RF}     & {Radio frequency} \\
			\hline
			{RQAM}   & {Rectangular QAM} \\
			\hline
			{SQAM}   & {Square QAM} \\
			\hline
			{SDR}    & {Semi-definite relaxation} \\
			\hline 
			{SDPR}   & {Semi-definite programming relaxation} \\
			\hline 
			{SER}    & {Symbol-error-rate} \\
			\hline	
			{SEP}    & {Symbol-error-probability} \\
			\hline
			{SNR}    & {Signal-to-noise ratio} \\
			\hline
			{SSK}    & {Space shift keying} \\
			\hline
			{STBC}   & {Space time block codes} \\
			\hline
			{TDMA}   & {Time division multiple access} \\
			\hline
			{TQAM}   & {Triangular QAM} \\
			\hline
			{TAS}    & {Transmit antenna selection} \\
			\hline
			{UVC}   & {Ultraviolet communication} \\
			\hline
			{VDSL}   & {Very high speed digital	subscribers line} \\
			\hline
			{VLC}    & {Visible light communication} \\
			\hline
			{Wi-Fi}  & {Wireless fidelity} \\
			\hline
			{WiMAX}  & {Worldwide interoperability for microwave access} \\
			\hline
			{XQAM}   & {Cross QAM} \\
			\hline   
		\end{tabular}
	\end{table} 
\vspace{-1em}
	\subsection{Paper Organization}
	In Section II, family of QAMs is discussed is details. Section III, Section IV, and Section V discuss detailed study of some of the newer QAM constellations. In Section VI, probabilistic shaping is discussed. In Section VII, applications of various QAM constellations in wireless communication systems are discussed.  Finally, conclusion and future challenges are presented in Section VIII.
\section{Quadrature Amplitude Modulation (QAM)}
	\subsection{Evolution of QAM}
\begin{figure*}[!h]
		\centering
		\includegraphics[width=6in,height=2in]{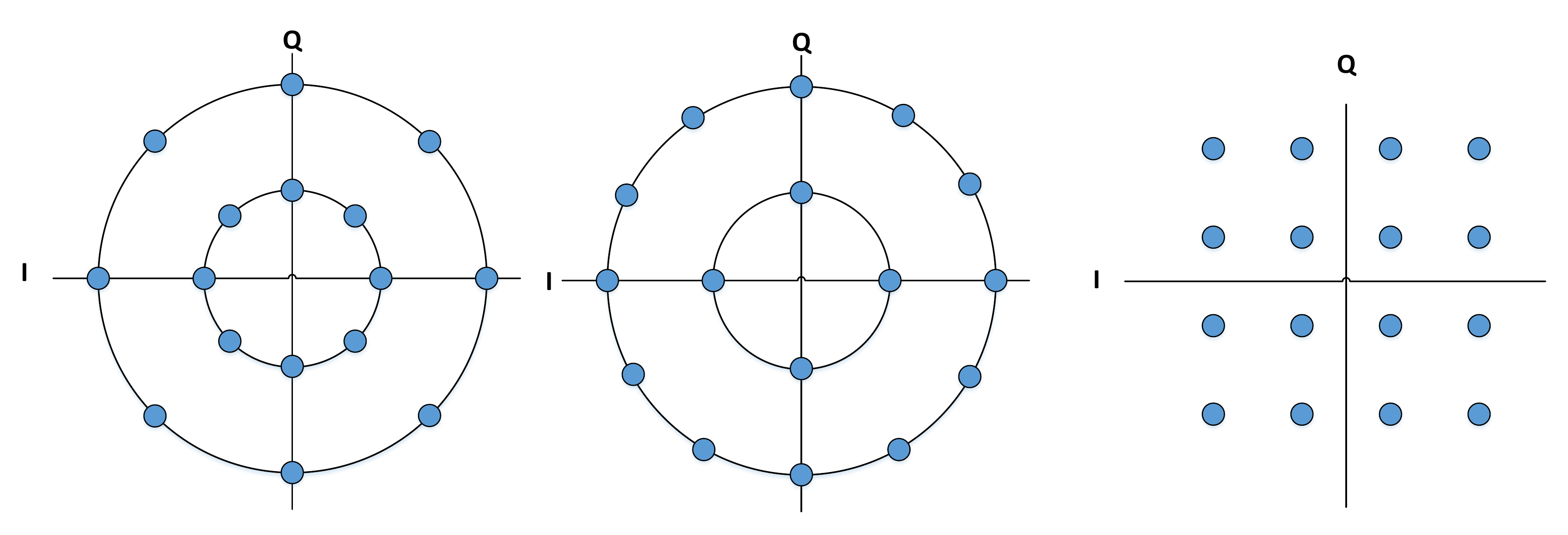}
		\caption{Type I, Type II, and Type III QAM constellations.}
		\label{Type}
	\end{figure*}

	In the late 1950s, digital phase modulation schemes gained considerable research attention along with the digital amplitude modulation for transmission. It was an extension to the amplitude modulation by considering both the amplitude and phase modulations for transmission.
	The QAM was first proposed by C. R. Cahn \cite{cahn1960combined}  in 1960. Cahn's work was extended by Hancock and Lucy in \cite{hancock1960performance} where it is realized that the performance of  a circular constellation can be improved by placing the constellation points on the concentric circles with lesser points in the inner circle and more points in the outer circles.  They named Type I to the Cahn's constellation and Type II to their newly proposed constellation. In 1962, Campopiano and Glazer  proposed  properly organized square QAM constellations and denoted it as Type III constellation \cite{campopiano1962coherent}. From the results it was concluded that the performance of Type II constellation is little better as compared to Type III constellation, but implementation and detection complexity of Type III constellation is significantly low as compared to Type I and II constellations.  All three constellations are shown in Fig. \ref{Type}. 
	Later in 1971, Salz et. al. of Bell Labs proposed some actual QAM constellations \cite{salz1971data}. They used both the coherent and non-coherent demodulations and implemented  circular constellations with the help of 
	2 and 4 amplitude positions along with 4 and 8 phase positions.
	In \cite{simon1973hexagonal}, Simon and Smith proposed some tightly packed 2D spherical shaped honeycomb type structure (hexagonal QAM constellations), and simple generation and detection techniques for such constellations were discussed.
	In \cite{foschini1973selection}, Foschini et al.  proposed optimum signal constellations with circular constellation envelop for large constellation size ($M$) which had optimum BER performance but their detection complexity was high. Foschini et al. extended their work in \cite{foschini1974optimization}, and error optimization technique for optimum 2D signal constellations was proposed. In the same year, Thomas et al. \cite{thomas1974digital} proposed empirically generated 29 amplitude-phase shift keying (APSK) signal sets, ranging from $M=4$ to $M=128$ to determine  an optimum signal constellation through symbol-error-probability (SEP) bound for both the average and peak SNRs.  
	In \cite{smith1975odd}, Smith proposed various odd bits QAM constellations and a comparative study of rectangular and symmetric constellations was performed. In \cite{weber1978differential}, Weber proposed a differential encoding technique for reducing the performance penalty by minimizing the phase ambiguity for multiple APSK systems. 
	In \cite{forney1984efficient}, Forney et al. presented a detailed survey on various power efficient rectangular and hexagonal QAM constellations with coding fundamentals for band limited channels like telephone channels.   Since then, applications of all these proposed constellations have widely been seen in various communication systems and standards which are discussed in the upcoming Section.
	%
	 \subsection {Various QAM Constellations}  
 	In wireless communication systems, a power efficient signal constellation which targets the best QoS with high spectral efficiency has always been an exciting research area since the genesis of early modern mobile communication systems. A breakthrough in the research on modulation schemes happened with the invention of QAM in the early 1960's. Interestingly, various QAM constellations proposed years ago are still actively being used in commercial communication systems. With the recent developments in advanced signal processing algorithms, family of QAM has gained increased attention in present mobile communication systems and is widely adopted in various wireless communication standards and commercial applications. In QAM, information is encoded in both the amplitude and phase of the transmitted signal. Hence, for a given average energy, more bits per symbol can be encoded in QAM which makes it more spectrally efficient. An
	overview of some of these prominent QAM constellations is presented next. 
  	\subsubsection{Square QAM (SQAM)}
\begin{figure}[!h]
		\centering
	 	\includegraphics[width=2in,height=1.7in]{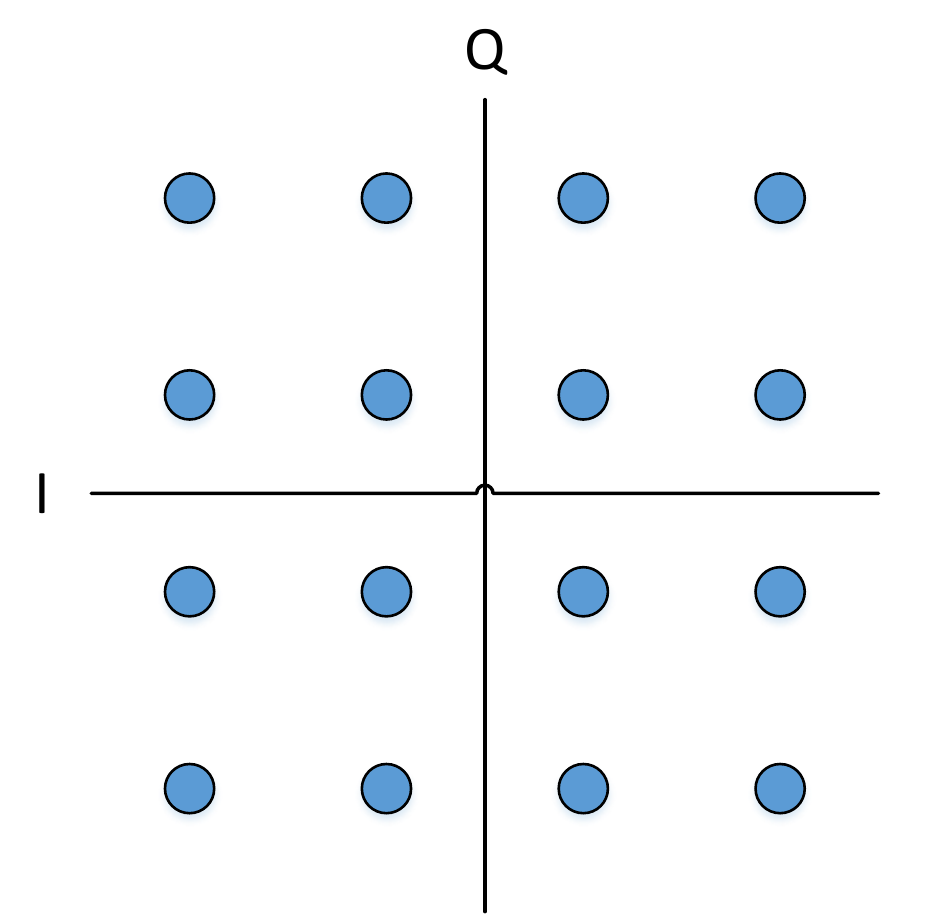}
		\caption{16-SQAM constellation.}
		\label{16SQAM}
	\end{figure}

   	%
	In 1962,  Campopiano and Glazer \cite{campopiano1962coherent} extended the works done by Cahn, Hancock and Lucy, by proposing  properly organized square or rectangle shaped QAM constellations. Campopiano and Glazer denoted it as Type III constellation which is later known as QAM constellation.  Type III QAM provides better error performance than the predecessors Type-I and Type-II QAMs.  For QAM constellations, simple maximum likelihood (ML) detection with rectangular or squared boundaries is preferred.
	SQAM usually takes a perfect square shape for the even power of 2 signaling points 4, 16, 64, 256, 1024, 4096, and so on.  SQAM has the maximum possible minimum Euclidean distance between constellation points for a given average symbol power and requires simple ML detection technique. $M$-ary QAM requires less carrier-to-noise power ratio than the $M$-ary PSK \cite{du2010wireless}. Thus, QAM is widely considered in various wireless communication systems and IEEE standards. Lower order QAM constellations have lower spectral efficiency, provide better cell overlap control and tolerance to distortion or SER/BER performance. However, high data-rates can be achieved with higher order QAM constellations at the cost of strict SER/BER requirements, severe cell-to-cell interference, smaller coverage radii, and hardware complexity. SQAM is widely deployed in various wireless standards such as in 3G/4G/5G digital video broadcast-cable/satellite/terrestrial communications, satellite communications, wireless fidelity (Wi-Fi), worldwide interoperability for microwave access (WiMAX), asymmetric digital subscribers line (ADSL), very high speed digital subscribers line (VDSL), power line ethernets, microwave backhaul systems and others. Various QAM applications with different constellation orders are presented in Table \ref{standards}. As an example, 16-SQAM constellation is shown in Fig. \ref{16SQAM}.
  	%
	 \subsubsection{Rectangular QAM (RQAM)}
\begin{figure}[h]
	 	\centering
	 	\includegraphics[width=3.5in,height=2.5in]{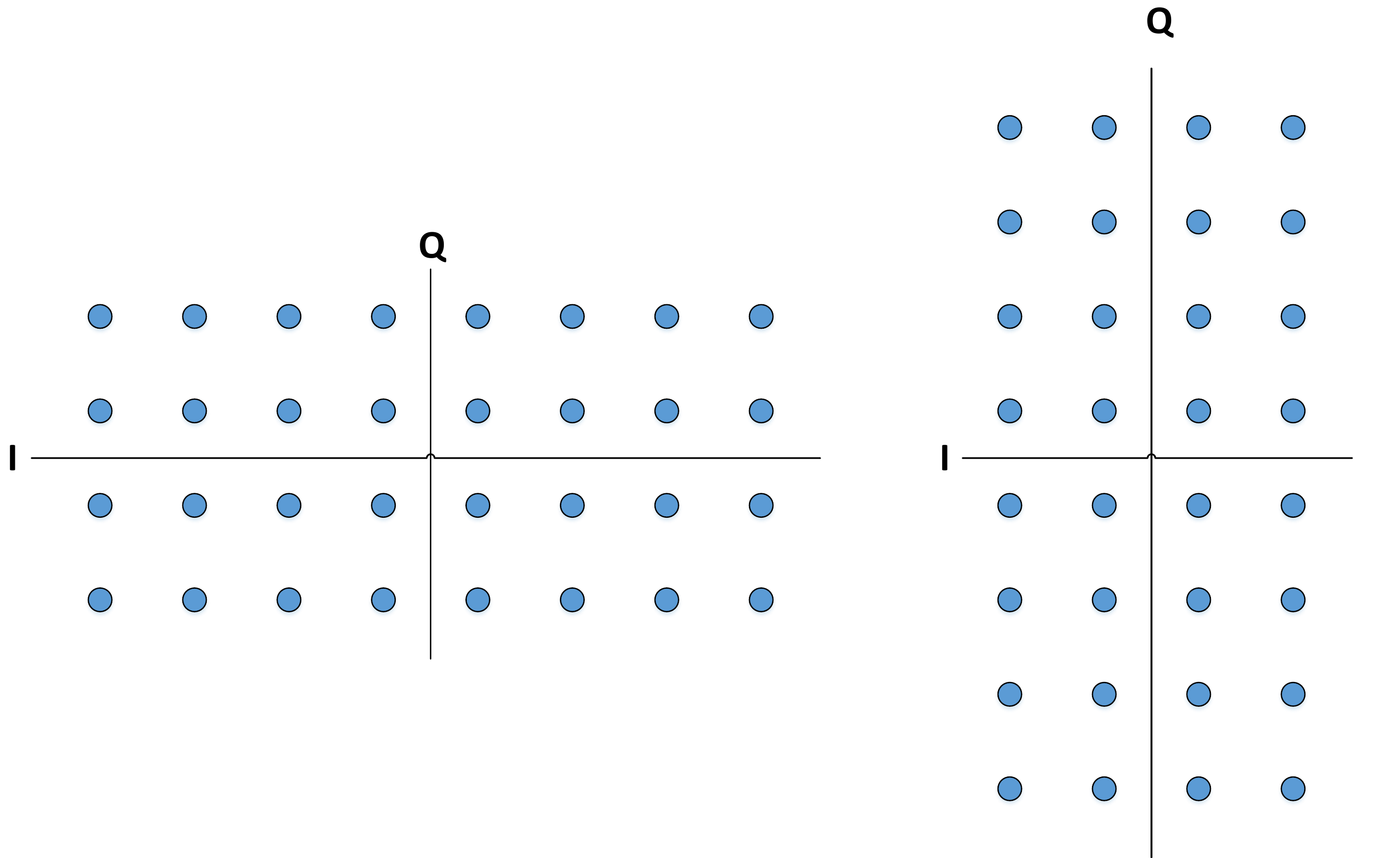}
	 	\caption{32-RQAM in-phase (left) and quadrature phase (right) constellations.}
	 	\label{RQAMcons}
	 \end{figure}

	 %
	 RQAM is a Type-III  QAM with non-square constellation which is commonly preferred for transmitting odd number of bits per symbol. As spectral efficiency depends on the  modulation order $M$, higher spectral efficiency is achieved with higher value of $M$. Spectral efficiency also depends on the channels conditions. To improve the spectral efficiency, adaptive modulation is used in practice to maximize spectral efficiency for a given channel conditions according to $M$ \cite{goldsmith2005wireless,webb1995variable}. Further, both the even as well as odd power of 2 constellations are required for granular adaptation based on channel conditions. For this, rectangular QAM (RQAM) is commonly  preferred due to its generalized behavior as various modulation schemes such as multilevel ASK, binary PSK (BPSK), QPSK, orthogonal binary FSK, and SQAM as its special cases \cite{dixit2014performance}. RQAM is a suboptimal QAM for  ${M}\geq16$, since the average transmitted power required to achieve minimum distance is slightly greater than the average power required for the best $M$-ary QAM signal constellation. Thus, general-order RQAM is preferred in practical telecommunication systems \cite{karagiannidis2006symbol,proakis2008masoud}. RQAM constellations can be obtained by considering the constellation points in a rectangular shape. The obtained rectangular constellation can be parallel to in-phase axis or quadrature-phase axis with identical average energy. The 32-RQAM constellations are shown in the Fig. \ref{RQAMcons}.
 
 \subsubsection{Star QAM} 
 %
 For  mobile radio applications, SQAM is widely preferred over the last few decades. However, SQAM is optimum if the communication channel is Gaussian. However, in practice, the Gaussian channel assumption does not apply and hence, severe shortcomings were observed for mobile radio channels over fading environment.  Carrier recovery and automatic gain control (AGC) are required when SQAM is opted even with differential coding which are complex to achieve and sustain in a real time communication system. To avoid the use of AGC and to mitigate the false lock problem,  star QAM
 constellation was proposed. Star QAM is a special case of circular APSK, which outperforms the SQAM in peak power limited systems. It consists of multiple concentric PSK circles with equal constellation points in each circle and identical phase angle between them. Amplitude and phase of the constellation points are mutually independent \cite {ishibashi2013peak, webb1991bandwidth}. Hence, differential detection can be applied successfully rather than the coherent detection, which omits the need for accurate phase tracking and channel estimation at the receiver \cite{may1998performance,ishibashi2005low}. Due to the above favorable features, star QAM is adopted in various satellite communication standards such as in  digital video broadcasting satellite standard-second generation  (DVB-S2), digital video broadcasting-satellite services to handheld (DVB-SH), advanced broadcasting system via satellite, and Internet protocol over satellite  \cite{ishibashi2013peak,yang2014star}. A detailed study of star QAM constellations is presented later.
 %
	 \begin{table*}[!h]
	 	\centering
	 	\caption{Applications of different QAM constellations in existing wireless communication systems and IEEE standards.}
	 	\label{standards}
	 	\begin{tabular}{@{} ccccccc @{}}
	 		\toprule
	 		\textbf{Standards} & \textbf{QAM Constellation Points} &	\multicolumn{1}{c}{\textbf{References }} \\ 
	 		\midrule
	 		Digital video broadcasting- cable (DVB-C)  & 16, 32, 64, 128, 256 & \cite{ANSI/SCTE072013,DVB-C,zhang2010exact,reimers2013dvb,Dirk} \\
	 		DVB-C2 & \begin{tabular}[c]{@{}l@{}}16 to 4096\end{tabular} & \cite{DVB-C2, Dirk} \\
	 		DVB-C2-future extensions & 16384 \& 65536 & \cite{Dirk}  \\
	 		Digital video broadcasting- terrestrial (DVB-T) &16 \& 64 &\cite{DVB-T}\\
	 		DVB-T2 &16, 64, \& 256 &\cite{DVB-T2} \\
	 		Digital video broadcasting– satellite (DVB-S) &16 &\cite{DVB-S}\\
	 		Asymmetric digital subscribers loop for copper twisted cables & upto 32768&\cite{adsl}\\
	 		Power line ethernets & upto 4096 &\cite{powerline}\\
	 		Ultra-high capacity microwave backhaul systems & 2048 &\cite{sasakiultra}\\
	 		IEEE 802.11n  &16 \& 64 &\cite{charfi2013phy} 	\\
	 		IEEE 802.11g  &16 \& 64 &\cite{sauter2010gsm} 	\\
	 		IEEE 802.11ad &16 \& 64 &\cite{banerji2013ieee} 	\\
	 		IEEE 802.11ac &16, 64 \& 256 &\cite{verma2013wifi} \\
	 		IEEE 802.11ad &16 \& 64 &\cite{verma2013wifi} \\
	 		IEEE 802.11ay &16, 64 &\cite{ghasempour2017ieee} \\
	 		IEEE 802.11af &16, 64 \& 256 &\cite{ieee2009ieee} \\
	 		IEEE 802.11ah &16, 64 \& 256 &\cite{sun2013ieee, oliveira2019mac}\\
	 		IEEE 802.11ax &16, 64, 256 \& 1024  &\cite{ieee2010ieee} \\
	 		IEEE 802.22   &16 \& 64 &\cite{lekomtcev2012comparison} \\
	 		IEEE 802.22b  &16, 64 \& 256 &\cite{ieee2007ieee} \\
	 		CDMA 2000 1x EV-DO  &16&\cite{tse2005fundamentals} \\
	 		Release 7-UMTS/HSPA- TR 25 999   &16 \& 64 &\cite{UMTSHSPA, sauter2010gsm}\\
	 		Release 8-LTE- TS 36.211  &16 \& 64 &\cite{LTE}\\
	 		Release 10-LTE Advanced- TS 36.300  &16 \& 64 &\cite{LTE-A}\\
	 		Release 14-LTE Advanced Pro- TS 36.306  &16, 64, \& 256 &\cite{LTE-Ap}\\
	 		Release 15-5G support- TS 36.331 &16, 64, 256 \& 1024 &\cite{5G}\\
	 		IEEE 802.16e &16 \& 64 & \cite{hanzo2011mimo}\\
	 		WiMAX 1.5/IEEE Std 802.16-2009 &16 \& 64 & \cite{pareit2011history}\\
	 		WiMAX 2/IEEE 802.16m &16 \& 64 & \cite{ieee209ieee}\\
	 		IEEE 802.16-2017 &16 \& 64 & \cite{802.16-2017}\\
	 		Telephone circuit modem V.29/34 &16 \& 64 &\cite{hanzo2001wireless}\\
	 		Optical modem &16 &\cite{olsson2011112}\\ 
	 		Digital multi-programme television distribution by cable networks &16, 32, 64, 128 \& 256 &\cite{ITU-J}	\\
	 		H.261 Reconfigurable Wireless Video Phone System & 4, 16, \& 64 &\cite{ITU-T}\\
	 		DCT Videophone features & 16-pilot assisted &\cite{streit1996quadtree}\\
	 		\bottomrule
	 	\end{tabular}
	 \end{table*}
	 %
   	\subsubsection{Cross QAM (XQAM)}
   	%
	For odd bits per symbol, RQAM is not a good choice as it has higher peak and average powers. Smith proposed \cite{smith1975odd} an improved cross shaped constellation to resolve this issue by removing the outer corner points and arranging them in a cross shape such that the average energy of the constellations is reduced. This type of constellation is named as XQAM.  XQAM has lower peak and average powers than RQAM, and provides at least 1 dB gain over the RQAM constellations \cite{kumar2018aser,singya2018impact,singya2019performance}. XQAM  has been adopted in various communication system such as in ADSL and VDSL with 5-15 bits \cite{ADSL1999,ADSL2004}, 32-XQAM and 128-XQAM in digital video broadcasting-cable (DVB-C) \cite{etsi1998}. XQAM has commonly been applied in applications requiring blind equalization \cite{colonnese2008high,cartwright2001blind,abrar2006blind}. A detailed study of XQAM constellations is presented later.
	%
	
   \subsubsection{Hexagonal QAM (HQAM)}
   %
	Requirement for high data-rates at low energy further directed the research towards more power efficient 2D  hexagonal shaped constellation referred to as HQAM. HQAM has the densest 2D packing which 
	reduces the peak and average power of the constellation, and makes the constellation more power efficient than the other existing constellations. In most of the existing communication systems, SQAM is widely preferred due to its simpler ML detection than the HQAM constellation, however, with the advancement in technology, implementation complexity of HQAM detection is reduced considerably. Due to superiority of HQAM over other constellations, it can be considered in various applications such as in multicarrier systems \cite{han2008use}, multiple-antenna systems \cite{kapetanovic2013optimal,srinath2013fast,leung2016design}, physical-layer network coding \cite{hekrdla2014hexagonal}, small cell \cite{hosur2013hexagonal}, optical communications \cite{doerr201128}, and  advanced channel coding \cite{tanahashi2009multilevel,hashimoto2013non,morita2015nearest}. This work focuses on the HQAM constellation and a detailed study of HQAM constellations is presented later.

	\subsection {Application of Different QAM Constellations in Existing Communication Systems and Standards} 
	%
	QAM and QPSK are the most widely preferred constellations in various communication standards. Further,  $M$-ary QAM  has better bandwidth and power efficiency than $M$-ary PSK constellation \cite{rappaport1996wireless} and thus, is widely employed in various modern communication standards. A list of various communication standards where different QAM constellations with various constellation orders are employed is shown in Table \ref{standards}.
 
	\section{Star QAM Constellations}
	\begin{figure*}[h!]
	\centering
	\includegraphics[width=7.3in,height=2.7in]{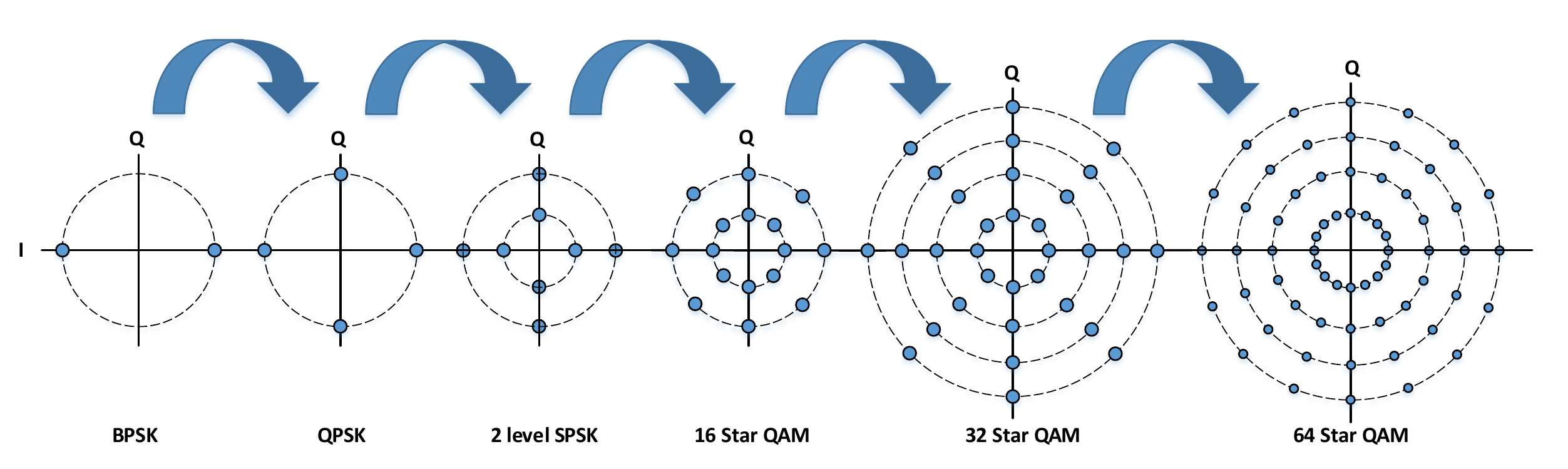}
	\caption{Modeling of various star QAM constellations.}
	\label{StarQAM}
\end{figure*}
	In the initial studies, SQAM was widely preferred in mobile radio applications. However, severe limitations of SQAM were observed over mobile radio channels (with multipath fading) \cite{webb1992qam}.
Usually, SQAM required both the carrier recovery and AGC at the receiver. In the literature, mainly two AGC methods; averaging method and forced update method were adopted for SQAM. Both performed similarly poor, however, averaging method was preferred over the other, as it has no bandwidth overhead. There were some serious issues with carrier recovery as false locks were observed not only at multiple of $90^0$, but also at $26+(l \times 90)$ and $52+(l \times 90)$ degrees with $l=0,1,2,3$ as highlighted in \cite{webb1992qam}. To improve the performance and to avoid the need of AGC and false lock problem, star QAM came into existence. Star QAM consists of multiple concentric PSK circles with equal constellation points in each circle and identical phase angle between them. Star QAM consists of 8 possible lock positions for lower order constellations (16 and 32-star QAM constellations) which increases for higher order constellations (64-star QAM constellation). All the constellation points are rotated with a fixed angle for these positions. After differential encoding, all the false lock problems can be eliminated. Modeling of various star QAM constellations is shown in Fig. \ref{StarQAM}. 

\begin{figure*}[h!]
	\centering
	\includegraphics[width=5in,height=2.8in]{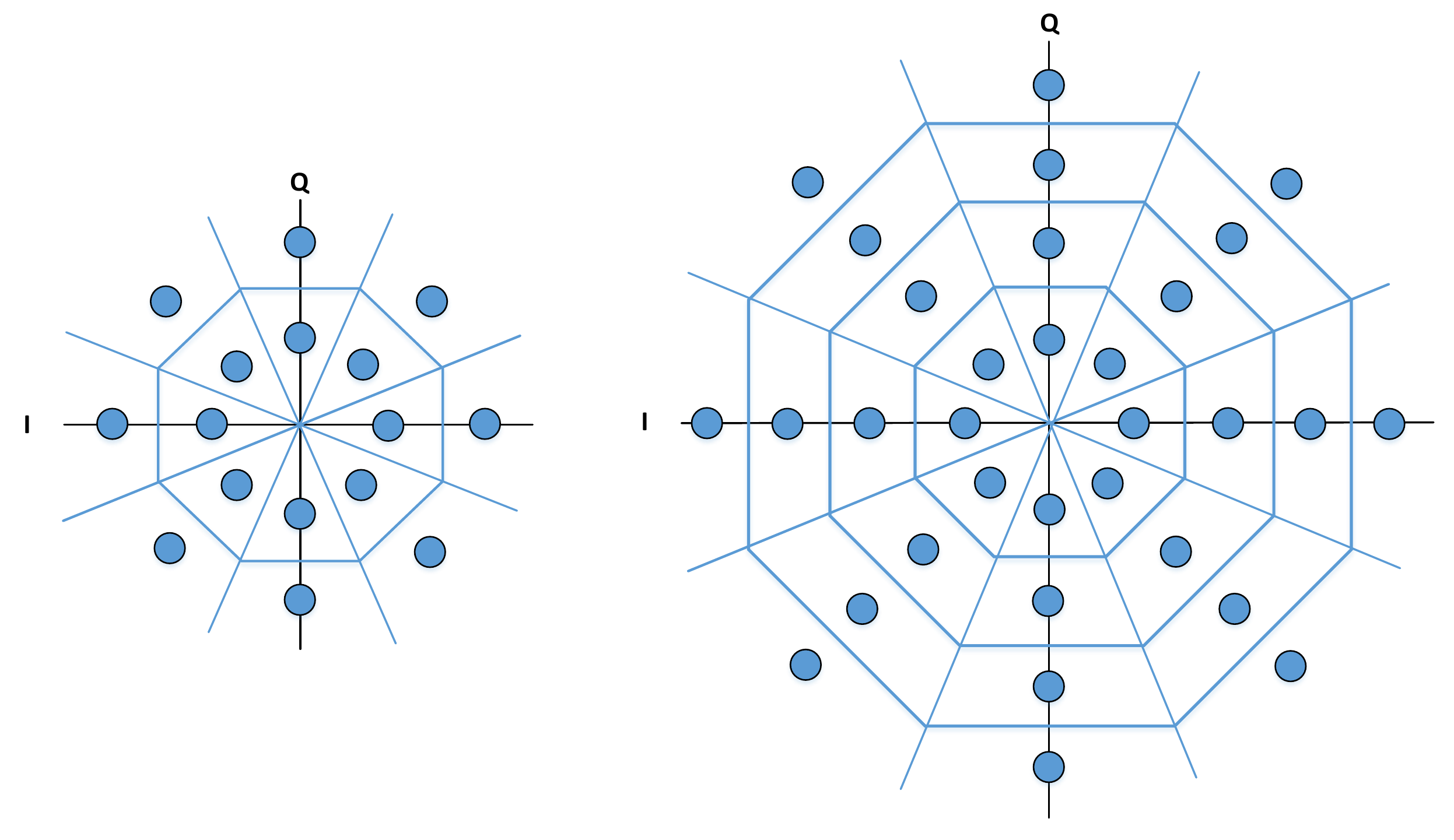}
	\caption{Decision regions for 16-star QAM (left) and 32-star QAM (right) constellations.}
	\label{stardec}
\end{figure*}

\subsection{Decision Regions}
Decision regions for star-QAM are not straight-forward as compared with SQAM/RQAM regions. It does not consist of  horizontal and vertical boundaries only. For decision making, inclined regions with some phase angles are also required.
Let us consider 16 and 32-star QAM constellations. For decision making, initially, the entire constellation is divided into 8 sub-regions with phase angle $(2l+1)\times \pi/8$, for the constellation points situated at an phase angle of $l\times \pi/4$, where $l=0,1,...,7$. Then in each sub-region, the constellation points are separated by a boundary placed at the mid-point of the two nearby points. 
Hence, the decision boundaries for inner circle points are triangular regions and for mid circles points are trapezoidal regions. Decision boundaries are left open for the outer most circle points. 
For 16-star QAM constellation (Fig. \ref{stardec} (left)),
4 bits are required to map 16 constellation points. Here, last 3 bits are used to show the change in phase angle, and remaining starting bits are used to represent the change in ring amplitudes. In this case, symbol decoding is converted into simple comparison between the current and previously decoded symbol. Let, the amplitude of inner and outer rings be $r_i$ and $r_o$, respectively. Let, the received symbol amplitude are $r_t$ and $r_{t+1}$ at time instance $t$ and $t+1$, respectively. Then the first bit is set to 1 only if \cite{webb199016}
\begin{align}
\frac{r_{t+1}}{r_{t}}>\frac{r_i+r_o}{2}, \nonumber\\
\frac{r_{t+1}}{r_{t}}<\frac{2}{r_i+r_o},
\end{align}
otherwise set to 0. If $\theta_t$ and $\theta_{t+1}$ are the phase angles of the received symbol at time instance $t$ and $t+1$, respectively, then the demodulated angle will be \cite{webb1992qam}
\begin{align}
\theta_{dem}=(\theta_{t+1}-\theta_{t})\text{mod}2\pi.
\end{align}
Finally, the decoded angle is quantized to the closest multiple of $\pi/4$, and then a lookup table is prepared to decode the remaining bits.
For 16 and 32 points star QAM constellations, decision regions are shown in Fig. \ref{stardec}. Similar procedure can be applied for the higher-order star QAM constellations.

\subsection{SER Analysis}
\begin{figure}[!h]
	\centering
	\includegraphics[width=3.4in,height=2.7in]{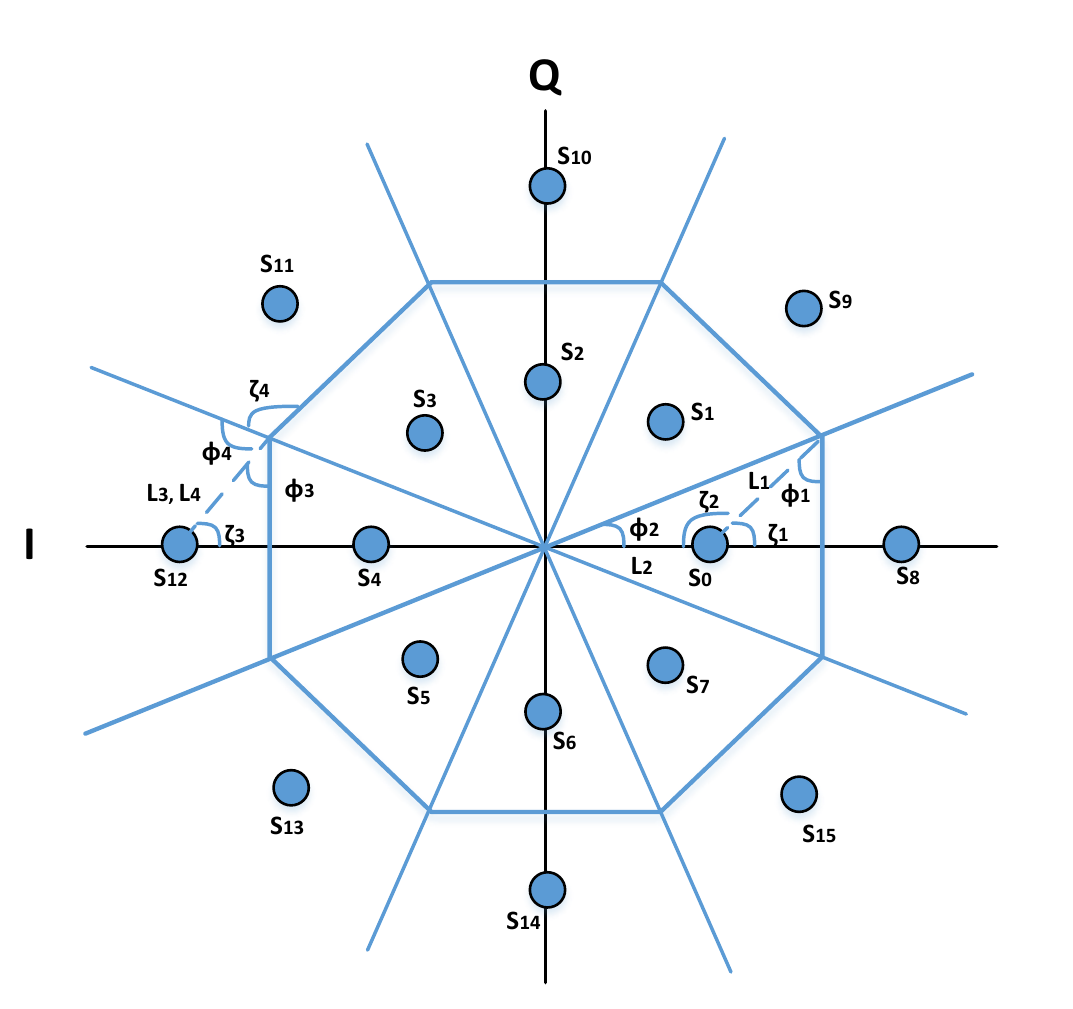}
	\caption{16-star QAM.}
	\label{16star}
\end{figure}

For an $M$-ary equiprobable digital modulation signal with arbitrary placement of constellation points in 2D plane, SEP expression over the additive white Gaussian noise (AWGN) channel can be given by the weighted sum of
probabilities of each sub-regions for all the constellation points as \cite{dong1999signaling}
\begin{align}\label{SEPS}
\mathcal{P}_e=\sum_{u=1}^{N_S}\frac{\mu_u}{2\pi}\int_{0}^{\zeta_u}\exp\Big[{\frac{-a_uL^2 \text{sin}^2(\phi_u)}{N\text{sin}^2(\theta+\phi_u)}}\Big]d\theta,
\end{align} 
where $N_S$ denotes the number of sub-regions, $\mu_u$ represents a priori probability that a symbol belongs to the $u^{th}$ sub-region, and $a_u$, $\zeta_u$, and $\phi_u$ are some important parameters related to the $u^{th}$ sub-region for analysis which are shown in Fig. \ref{16star}.
Considering a 16-star QAM constellation for average SEP (ASEP) calculation as shown in Fig. \ref{16star}. For analysis, perfect phase tracking is assumed at the receiver. Inner and outer ring's radius are denoted by $r_i$ and $r_o$, respectively, which corresponds to a ring ratio $\mathbb{R}=r_i/r_o$. For symmetry, the ASEP is written in terms of error probabilities when symbol $S_0$ and $S_{12}$ are transmitted as
\begin{align}
\mathcal{P}_{S_0}=& \frac{1}{\pi}\sum_{u=1}^{2}\int_{0}^{\zeta_u}\exp\Big[{\frac{-L_u^2 \text{sin}^2(\phi_u)}{2N\text{sin}^2(\theta+\phi_u)}}\Big]d\theta,  \nonumber\\
\mathcal{P}_{S_{12}}=& \frac{1}{\pi}\sum_{u=3}^{4}\int_{0}^{\zeta_u}
\exp\Big[{\frac{-L_u^2 \text{sin}^2(\phi_u)}{2N\text{sin}^2(\theta+\phi_u)}}\Big]d\theta, 
\end{align}
respectively, where $L_u$ denotes the lengths (u = 1,...,4) as shown in Fig. \ref{16star}. Thus, the ASEP expression of 16-star QAM can be given as \cite{dong1999error}
\begin{align}
\mathcal{P}_{e}=&\frac{1}{2\pi}\sum_{u=1}^{4}\int_{0}^{\zeta_u}\exp\Big[{\frac{-L_u^2 \text{sin}^2(\phi_u)}{2N\text{sin}^2(\theta+\phi_u)}}\Big]d\theta, \nonumber\\
=&\frac{1}{2\pi}\sum_{u=1}^{4}\int_{0}^{\zeta_u}\exp\Big[{\frac{-a_uL^2 \text{sin}^2(\phi_u)}{2N\text{sin}^2(\theta+\phi_u)}}\Big]d\theta,
\end{align}
where $\zeta_1=\zeta_3=\text{tan}^{-1}\Big[(\sqrt{2}-1)\frac{\mathbb{R}+1}{\mathbb{R}-1}\Big]$, $\zeta_2=(\pi-\zeta_1)$, $\zeta_4=(7\pi/8-\zeta_1$), $\phi_1=\phi_3=(\pi/2-\zeta_1$), $\phi_2=\pi/8$, $\phi_4=(\pi/8+\zeta_1$),
$a_1=a_3=a_4=\frac{1}{4}\Big[(\mathbb{R}-1)^2+(\sqrt{2}-1)^2(\mathbb{R}+1)^2\Big]$, $a_2=1$, and $\frac{L^2}{2N}=\frac{r_0^2}{2N}=\frac{8\gamma}{1+\mathbb{R}^2}$.
Similar procedure can be followed for the higher order star QAM constellations.

\begin{figure*}[h!]
	\centering
	\includegraphics[width=4.7in,height=3in]{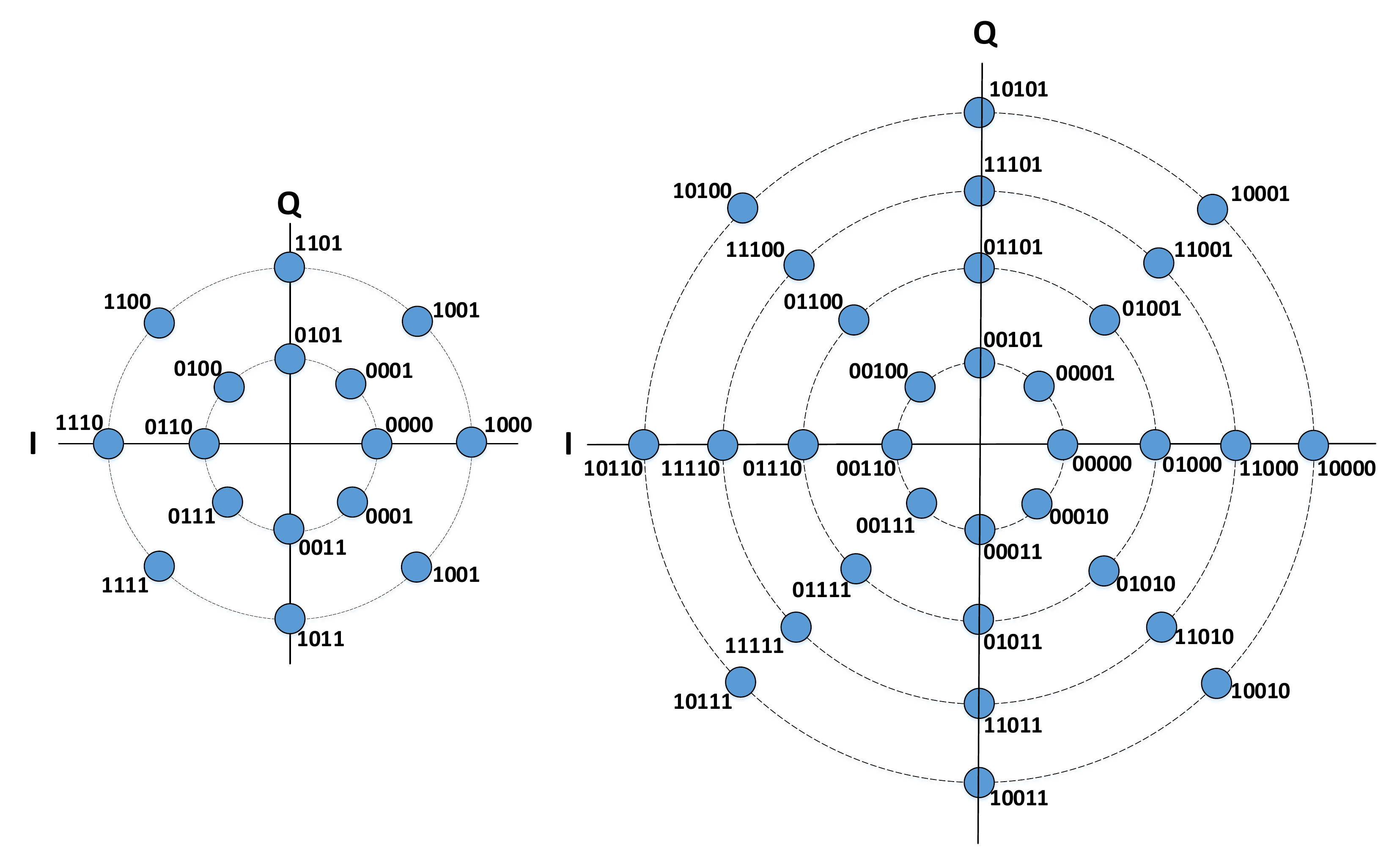}
	\caption{Bit mapping for 16-star QAM (left) and 32-star QAM (right) constellations.}
	\label{bitstar}
\end{figure*}
\subsection{Bit Mapping, Gray Code Penalty, and BER}
Star QAM is a special case of circular APSK, where both the amplitude and phase of constellation points vary. It consists of multiple concentric PSK circles with equal constellation points in each circle. Further, there are total 8 possible phase angles for lower order constellations which increase with the constellation size. Constellation points in all the circles have identical phase angle between them. 
Let us consider 16 and 32-star QAM constellations whose bit mapping is shown in Fig. \ref{bitstar}. As there are total 8 possible phase angles, last 3 bits of the bit stream are used to represent the phase angles. Remaining starting bits are used to represent the change in amplitude of the constellation points in the concentric circles.  For 16-star QAM (Fig. \ref{bitstar}, left), there are 2 concentric circles, each have 8 constellation points. There are total 8 phase angles $l\times \pi/4$ with $l=0,1,...,7$ \footnote{The star QAM constellation can be rotated with any arbitrary angle, however, a fixed phase difference should be maintained between the adjacent constellation points in a circle.}. Hence, the last 3 bits of the bit stream represent the change in phase angle by $\pi/4$ with the adjacent constellation point in the circle. If 000 is considered it means the symbol is transmitted with no phase change, if 001 is considered that means the signal is transmitted with $\pi/4$ phase shift than the previous symbol and so on.
 Remaining 1 bit (starting bit) represents the change in amplitude of the concentric circles. If first bit is 0, inner circles is selected, else outer circle is considered. For 32-star QAM (Fig. \ref{bitstar}, right), there are 4 concentric circles, each have 8 constellation points. 
Hence, the starting 2 bits are used to represent the change in amplitude and the last 3 bits are used to show the change in phase.
Similar procedure can be used for all the star QAM constellations. For 64-star QAM constellation, there are 16 possible lock positions. Hence, the last 4 bits of the 6 bits stream will be used to represent the change in phase, and the starting 2 bits denotes the change in amplitude ring.

In star QAM, each symbol shares its boundaries with maximum of 4 adjacent symbols (as can be seen from 32-star QAM). The inner-most and outer-most ring symbols share their boundaries with 3 adjacent symbols. Hence, bit mapping is performed in such a manner that each symbol differ with maximum of 1 bit from its nearby symbol. Hence, star QAM constellations have perfect Gray coding, and their Gray code penalty ($G_P$) is 1.
The $G_p$ is introduced by Smith which  is defined as the average bit difference between the two neighboring symbols in a constellation as
\begin{align}\label{GP}
G_P=\frac{1}{M}\sum_{k=1}^{M}{G_P^{s_k}},
\end{align}
where $M$ represents the constellation order, $s_k$ denotes the $k^{th}$ data symbol, $G_P^{s_k}$ is the Gray code penalty of the $k^{th}$ data symbol.  For a given SEP, approximated BEP expression can be given as \cite{smith1975odd}
\begin{align}\label{BER}
\mathcal{P}_b=G_P\frac{\mathcal{P}_s}{\text{log}_2{M}},
\end{align}
where $\mathcal{P}_s$ represents the SEP which is given in (\ref{SEPS}) for the star QAM. Finally, from (\ref{BER}), the BEP of star QAM can be achieved.

However, if perfect coherent detection is assumed, the BER performance of SQAM is better than the star QAM as in \cite{adachi1992performance}, where BER performance of different 16 point constellations is compared. For the locally generated reference carrier, ambiguity arises in exact phase orientation due to the symmetric constellation which can be resolved through differential encoding. However, error propagation occurs in differentially encoded bits which degrades the BER performance. Still, the  differential encoded SQAM (DE-SQAM) outperforms the star QAM which is explained mathematically in \cite {adachi1992performance} for 16 point constellations.   

 In this work, comparative study of various QAM constellations with a fixed separation of $2d$ between the constellation points is presented. However,  fixed separation of $2d$ between constellation points is not maintained in star QAM, and hence, its comparison with the other QAM constellations is not presented due to the lack of common platform.

\section{XQAM Constellations}
 	%
 		 \begin{figure*}[!t]
 		\centering
 		\includegraphics[width=6.8in,height=2.7in]{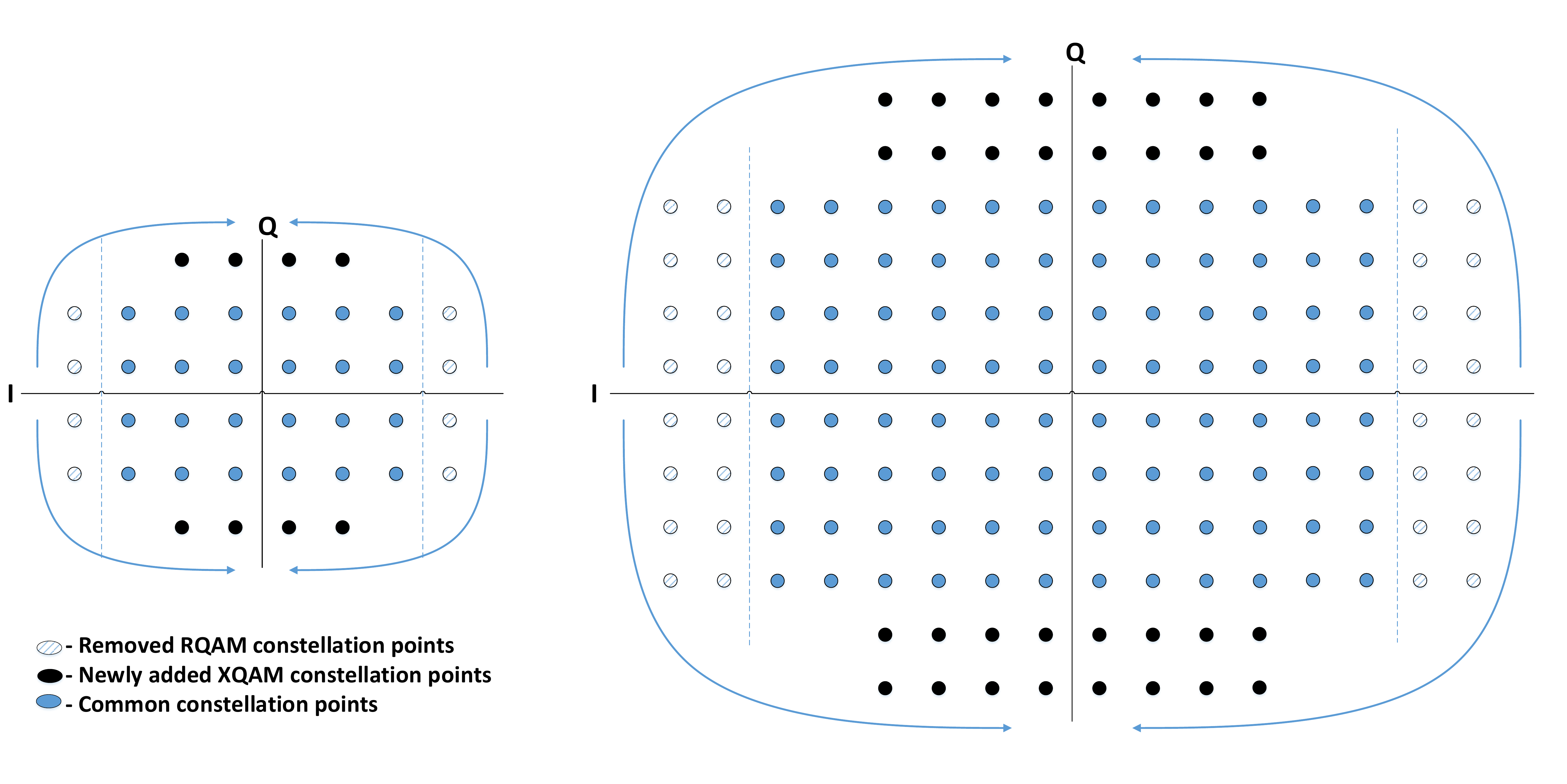}
 		\caption{32-XQAM and 128-XQAM constellations.}
 		\label{XQAMNew}
 	\end{figure*} 
	 For even bits per symbol, the transmitted signal is modulated with SQAM, whereas for odd bits per symbol, RQAM is commonly preferred. However, as discussed above, RQAM is not a good choice since RQAM constellations have high peak and average energies and hence, are power inefficient. To overcome this, Smith  proposed a XQAM constellation which  provides at least 1 dB gain over the RQAM constellations due to its reduced peak and average powers than RQAM  \cite{kumar2018aser,singya2018impact,singya2019performance}. If channel conditions are good, then instead of going for a constellation size of $M$ from $2^{2n}$ to $2^{2n+2}$ as in SQAM, a more gradual increase from $2^{2n}$ to $2^{2n+1}$ is possible with XQAM which can adapt the channel conditions more efficiently than RQAM \cite{vitthaladevuni2005exact,zhang2010exact}. Here $n$ denotes the number of bits per symbol which can take any value greater than or equal to 2. For XQAM constellation, $M=2^{2n+1}$, hence,  $M=32, 128, 512, 2048, ...$ are possible constellations with XQAM. Construction of 32-XQAM and 128-XQAM from 32-RQAM and 128-RQAM constellations are shown in Fig. \ref{XQAMNew}. Here, corner points are shifted to top-and-bottom near the y-axis such that the peak and average energies of the constellation are reduced.
	  In the following subsections, detailed study of the decision regions, SER, bit mapping, Gray code penalty, and BER of various XQAM constellations is presented.
 	 %
	 \subsection{Decision Regions}
\begin{figure}[!h]
	\centering
	\includegraphics[width=2.5in,height=2.2in]{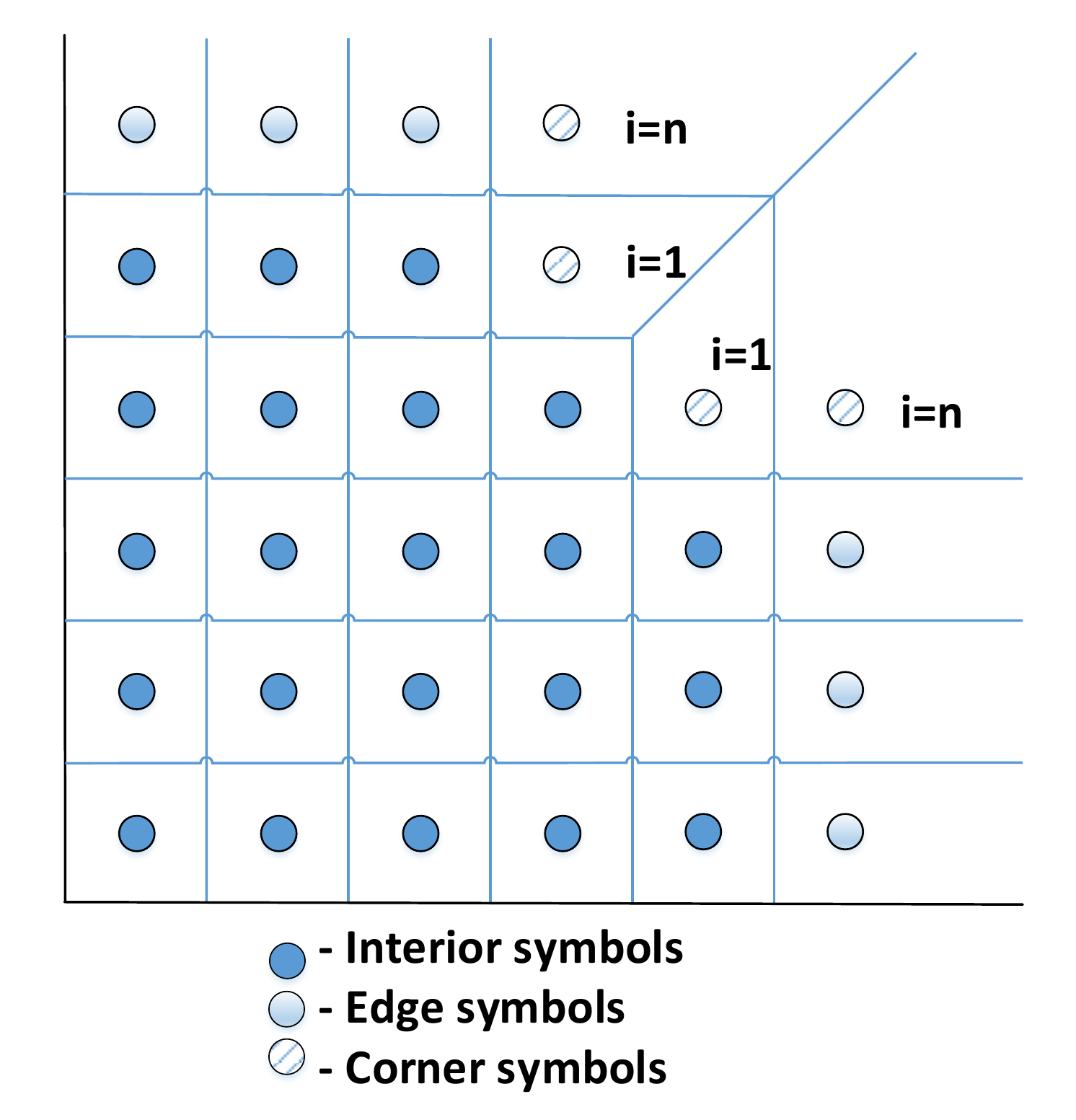}
	\caption{Single quadrant of 128-XQAM constellation.}
	\label{XQAMquad}
\end{figure}

	 %
	 Decision regions for both SQAM and RQAM are quite straight-forward, since, in-phase and quadrature-phase bits have vertical and horizontal decision regions, respectively. However, only the horizontal and vertical decision regions are not sufficient for XQAM constellations. 
 In XQAM, three types of symbols exist; edge symbols, corner symbols, and interior symbols. As the XQAM constellations are symmetric around the origin, we consider only one quadrant for analysis which is shown in Fig. \ref{XQAMquad}.
Number of interior, edge, and corner symbols/points in each quadrant are given as
\begin{align}
{N_{IS}}&=(2n)^2+2(n-1)(2n-1),\nonumber\\
{N_{ES}}&=2(2n-1),\nonumber\\
{N_{CS}}&=2n,
\end{align}
respectively.
	 For XQAM constellations, the end columns symbols are moved to new cross type positions in the constellation. For interior symbols, decision regions are closed squares, however, for edge symbols, decision regions are semi-infinite rectangles. Decision regions for corner symbols are not straight-forward and consist of $45^{0}$ lines along with the horizontal and vertical regions.
	 \begin{figure}[!h]
	 	\centering
	 	\includegraphics[width=3.6in,height=3in]{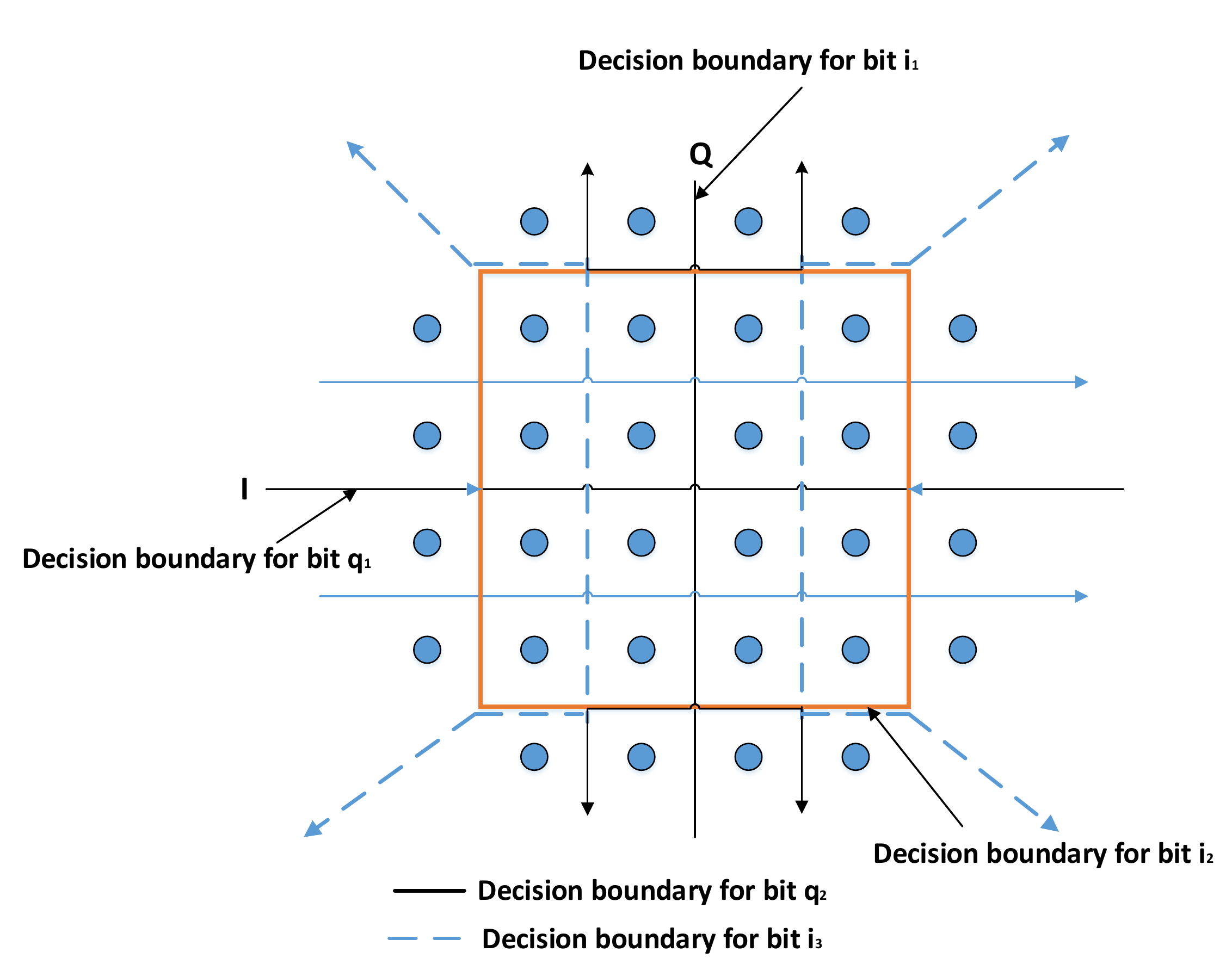}
	 	\caption{Decision regions for 32-XQAM constellation.}
	 	\label{32XQAMdecision}
	 \end{figure}
	 As an example, we consider a 5 bits ( $i_1i_2i_3q_1q_2$) 32-XQAM constellation. Decision regions for $i_1$ and $q_1$ bits are vertical and horizontal regions. However, for $i_2$, $i_3$, and $q_2$ bits, decisions regions are formed with horizontal, vertical, and ${45}^{0}$ inclined regions, which is shown in Fig. \ref{32XQAMdecision}. Similar procedure can be applied to obtain the decision regions for the other XQAM constellations.
	 \begin{figure*}[!h]
	 	\centering
	 	\includegraphics[width=5in,height=3.5in]{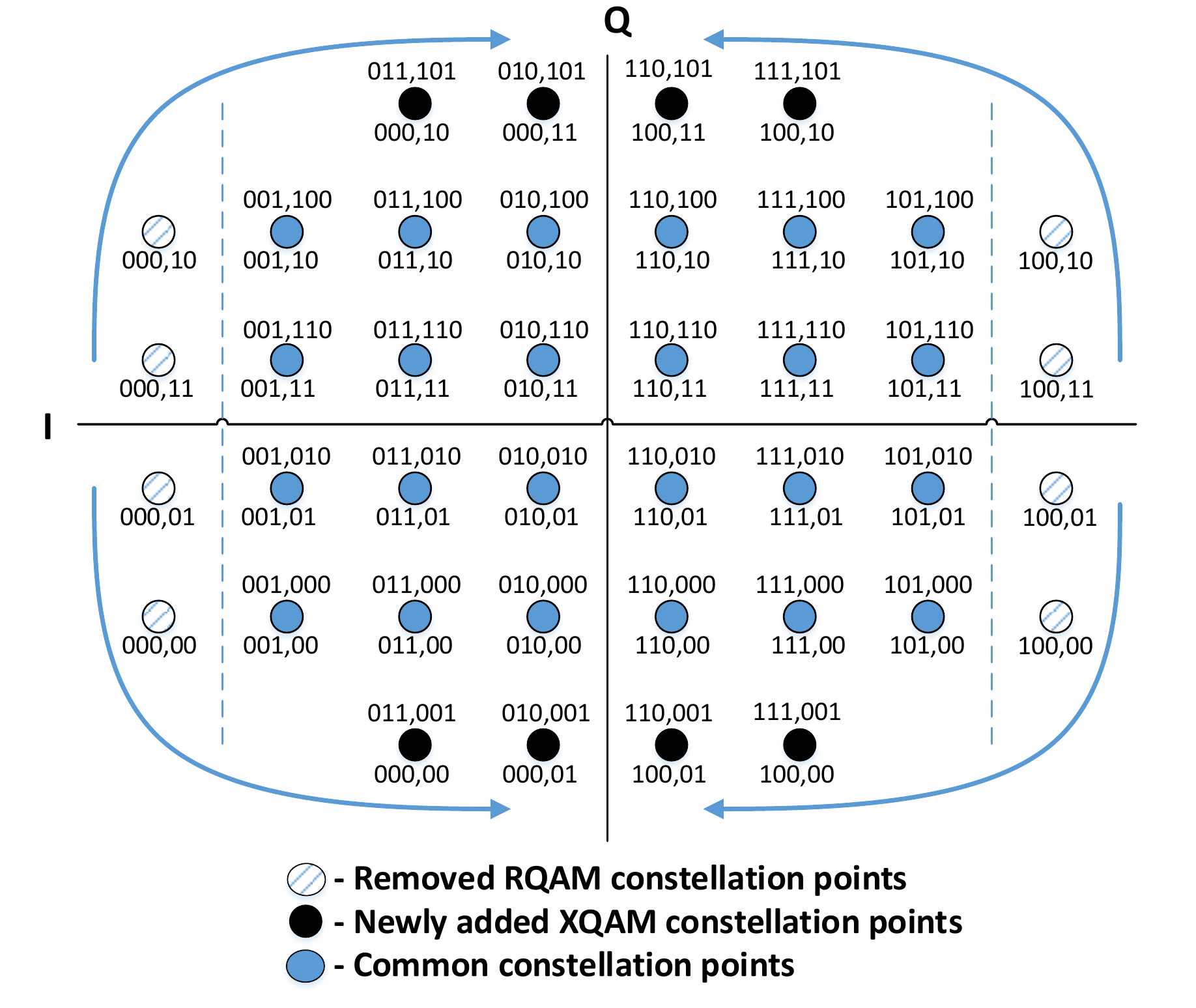}
	 	\caption{Bit mapping of 32-XQAM constellation.}
	 	\label{bitmap32XQAM}
	 \end{figure*}
	 %
	\subsection{SER Analysis}
	%
	The 32-XQAM and 128-XQAM constellations are shown in Fig. \ref{XQAMNew}. Here, corner points are shifted to top-and-bottom near the y-axis such that the peak and average energies of the constellation are reduced.
	For equiprobable constellation points, average symbol energy of XQAM is given as
	\begin{align}
	E_{avg}&=\frac{4}{M}\Big[2\times3n\sum_{i=1}^{3n}(2i-1)^2d^2-2n\sum_{i=2n+1}^{3n}(2i-1)^2d^2\Big],\nonumber\\&
	=\frac{d^2}{\lambda},
	\end{align}
	where $\lambda=\frac{48}{31M-32}$. Thus, the SNR can be given as
	\begin{align}\label{eta}
	\gamma=\frac{E_{avg}}{N_0}=\frac{d^2}{\lambda N_0}=\frac{\eta^2}{2\lambda},
	\end{align} 
	where $\eta=\frac{d}{\sqrt{N_0/2}}$ represents the minimum distance between  a constellation point and decision boundary, ${N_0}/{2}$ denotes the two sided power spectral density of noise, and $\eta=\sqrt{2\lambda\gamma}$ from (\ref{eta}). For the interior and edge symbols, respective SEP expressions in terms of $\eta$ can be given as \cite{proakis2008masoud}
	\begin{align}
	\mathcal{P}_{IS}&= 1-(1-2Q(\eta))^2=4Q(\eta)-4Q^2(\eta),\nonumber\\
	\mathcal{P}_{ES}&= 1-(1-Q(\eta))(1-2Q(\eta))=3Q(\eta)-2Q^2(\eta), 
	\end{align}
	where $Q(\cdot)$ represents the Gaussian-Q function. SEP expressions for the $i^{th}$ corner point from the bottom-to-top, or from left-to-right can be given as \cite{zhang2010exact}
	\begin{align}
	\mathcal{P}_{CP}^i&=3Q(\eta)-2Q^2(\eta)+\mathbb{F}(i,\eta;\eta),~~~~ i=1,2,...,n-1\nonumber\\
	\mathcal{P}_{CP}^n&= 2Q(\eta)-Q^2(\eta)+\mathbb{F}(n,\eta;\infty),\nonumber\\
	\end{align}
	where $\mathbb{F}(i,\eta,c_1)=\int_{\eta}^{c_1}Q(2i\eta+x)f(x)dx$, wherein $f(x)=1/\sqrt{2\pi}e^{-x^2/2}$ represents the probability density function (PDF) of the standard normal distribution. Finally, considering all the points, exact average SEP for the generalized XQAM is given as 
	\begin{align}\label{XSEP}
	 \mathcal{P}^{XQAM}(\eta)&=\frac{4}{M}\Big(N_{IP}\mathcal{P}_{IP}+N_{EP}\mathcal{P}_{EP}+2\sum_{i=1}^{n}\mathcal{P}_{CP}^i\Big)\nonumber\\&=
	 \frac{4}{M}\Big((M-6n)Q(\eta)-(M-12n+2)Q^2(\eta)\nonumber\\&
	 +2\;\mathbb{F}(n,\eta;\infty)+2\sum_{i=1}^{n-1}\mathbb{F}(i,\eta;\eta)\Big).
	\end{align}
	Finally, solving (\ref{XSEP}), the conditional SEP of general order XQAM constellation in AWGN channel can be given as \cite{zhang2010exact}
	\begin{align}\label{CEP}
	\mathcal{P}_e^{X}(e|\gamma)&=\Big[g_1Q(\sqrt{2\lambda\gamma})+\frac{4}{M}Q(2\sqrt{\lambda\gamma})-g_2Q^2(\sqrt{2\lambda\gamma})\nonumber\\&
	-\frac{16}{M}\sum_{i=1}^{n-1}Q(\sqrt{2\lambda\gamma},\alpha_i^+)
	-\frac{8}{M}\sum_{i=1}^{n-1}Q(2i\sqrt{\lambda\gamma},\beta_i^+)\nonumber\\&
	+\frac{8}{M}\sum_{i=2}^{n}Q(2i\sqrt{\lambda\gamma},\beta_i^-)\Big],
	\end{align}
	where  $g_1=4-\frac{6}{\sqrt{2M}}$,~ $g_2=4-\frac{12}{\sqrt{2M}}+\frac{12}{M}$, ~$\alpha_i^+=\text{arctan}(\frac{1}{2i+1})$, for $i=1,...,n-1$ \\	and $\beta_i^{\pm}=\text{arctan}(\frac{i}{i\pm1})$, for $i=1,...,n$.\\
	Further, $Q(x, \phi)$ is related to the 1D and 2D Gaussian Q-functions as
	\begin{align}
	Q(x,\phi)=\frac{1}{\pi}\int_{0}^{\phi}\text{exp}\Big(-{\frac{x^2}{2\;{\text{sin}}^2\theta}}\Big)d\theta~~~~~~\text{for}~~ x\geq0.
	\end{align}
	%
	\subsection {Bit Mapping, Gray Code Penalty, and BER}
	A detailed study about XQAM labeling is discussed in \cite{smith1975odd} and given for completeness.
	Binary and Gray codes for both the dimensions of an RQAM constellation are defined as
	\begin{align}
	g_k^1&=b_k^1\nonumber\\
	g_k^l&=b_k^l\oplus b_k^{l-1},~ 1\leq k\leq (2^{n+1}~\text{or}~ 2^n),\nonumber\\&
	 ~~~~~~~~~~~~~~~~~ 1<l\leq (n ~\text{or}~ n+1),
	\end{align}
	where $g_k^l$ and $b_k^l$ represent the $l^{th}$ bit of the $k^{th}$ symbol of Gray and binary codes, respectively. However, these relationships are not valid in XQAM where the far left and far right columns symbols are shifted to a new position to form a cross structure. Hence,  perfect Gray coding is not possible in XQAM. The problem of modulation and labeling was solved by Smith \cite{smith1975odd} by adding one extra bit to create a  $2n+2$ bits impure Gray coding. Here, this extra bit is not transmitted, however, it is only used for modulation and demodulation, and all the BER analysis is performed with $2n+1$ bits pseudo Gray codes.
	From Fig. \ref{bitmap32XQAM}, it is observed that with the use of the $2n+1$ bits pseudo Gray codes (5 bits Gray coding for 32-XQAM), topmost or bottom most blue colored symbols differ by 2 bits from its neighboring newly generated black symbols, however the other inner blue symbols only have one bit difference with its neighboring symbol. From Fig. \ref{bitmap32XQAM}, it is also observed that the $2n+1$ bits pseudo Gray codes are imperfect, and hence, $G_p$ arise which is defined in (\ref{GP}). For example, $G_p$ of 32-XQAM is calculated as
	 \begin{align}
	 G_p&=\frac{1}{32}\Big[\frac{3}{2}+\frac{4}{3}+\frac{4}{3}+\frac{3}{2}+1+\frac{5}{4}+\frac{5}{4}+\frac{5}{4}+\frac{5}{4}+1+1 \nonumber\\&
	 +1+1+1+1+1+1+1+1+1+1+1+1  \nonumber\\&
	 +\frac{5}{4}+\frac{5}{4}+\frac{5}{4}+\frac{5}{4}+1+\frac{3}{2}+\frac{4}{3}+\frac{4}{3}+\frac{3}{2}\Big]\nonumber\\&=
	 1.166.
	 \end{align}
	 As we increase the constellation order $M$, $G_p$ increases accordingly. Hence, a generalized $G_p$ for $M$-ary XQAM constellation can be given as $(1+\frac{1}{\sqrt{2M}}+\frac{1}{3M})$ \cite{vitthaladevuni2005exact}. Please note that, for a constellation with perfect Gray coding, $G_p$ is always 1. In any constellation, bit mapping should be performed very precisely and carefully to minimize the BEP for a given SEP for the constellation. Finally, substituting the SEP given in (\ref{CEP}) for XQAM, approximated BEP expression can be obtained.
	 From (\ref{BER}), it is clear that the BEP is directly dependent on $G_P$ since $\text{log}_2{M}$ is constant. Hence, optimum  bit mapping is required to minimize the $G_P$ for minimum BEP. Therefore, any other bit mapping results in higher $G_p$ which increases the BEP accordingly.

\section{HQAM Constellations}
\begin{figure}[!h]
		\centering
		\includegraphics[width=3in,height=2.5in]{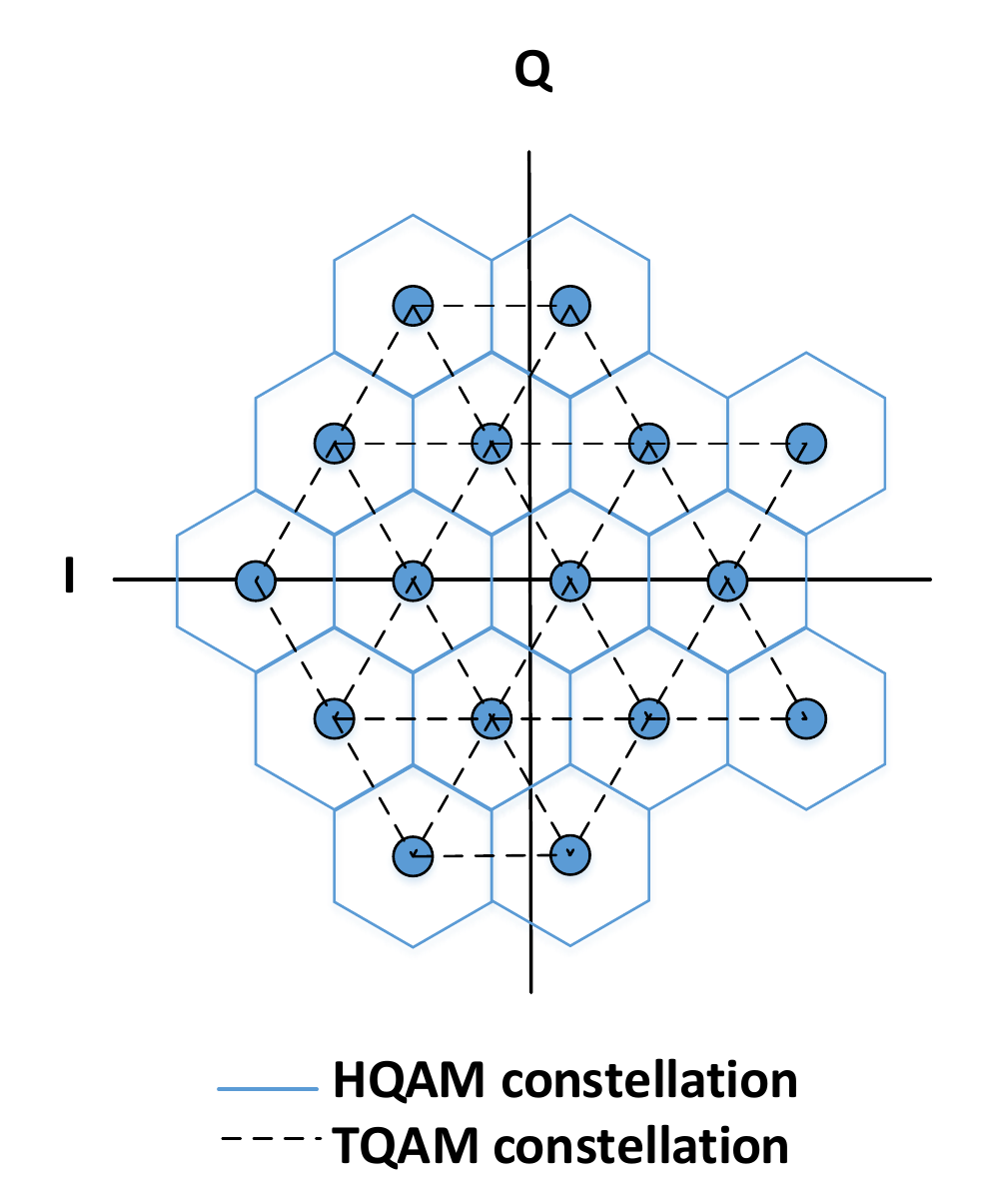}
		\caption{16-HQAM constellation.}
		\label{16TQAMHQAM}
	\end{figure}

\begin{figure*}
		\hfill
		\subfigure[]{\includegraphics[width=18 cm,height=9.5cm]{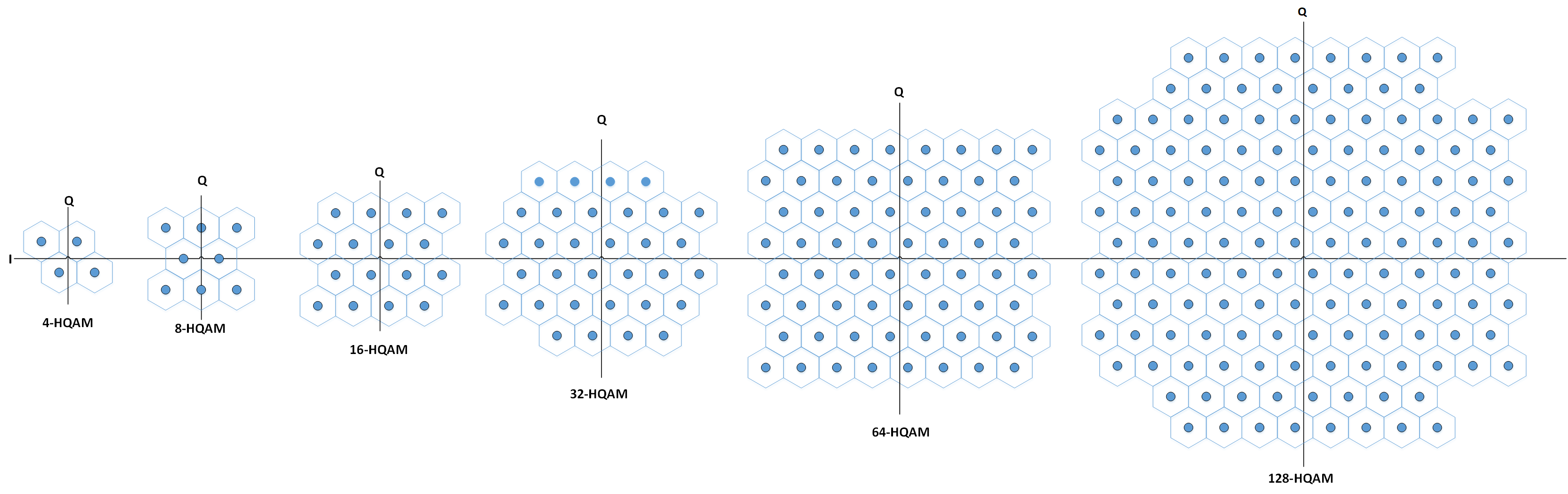}}	
		\hfill
		\subfigure[]{\includegraphics[width=10 cm,height=10.5cm]{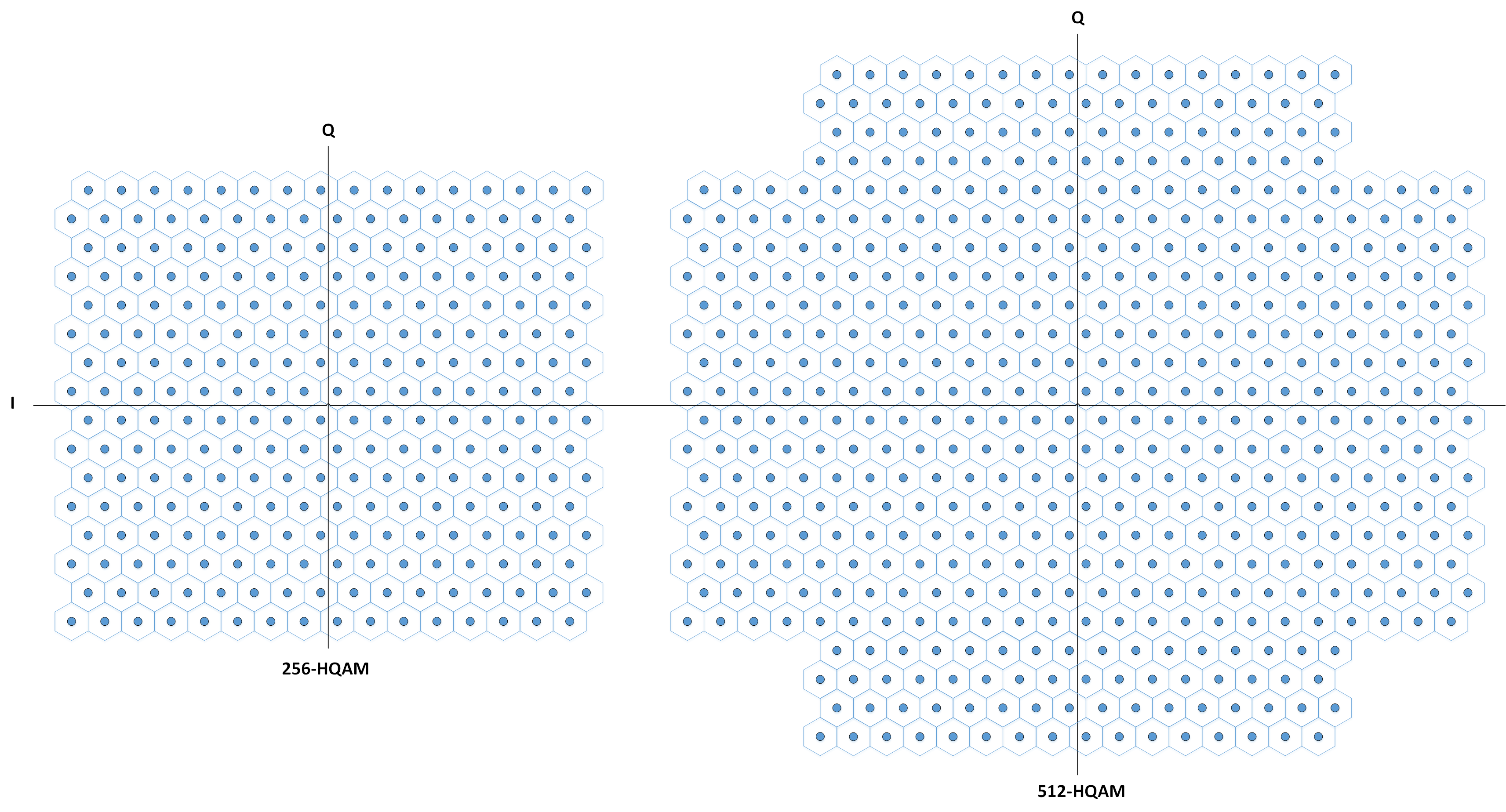}}
		\hfill
		\subfigure[]{\includegraphics[width=8 cm,height=10.5cm]{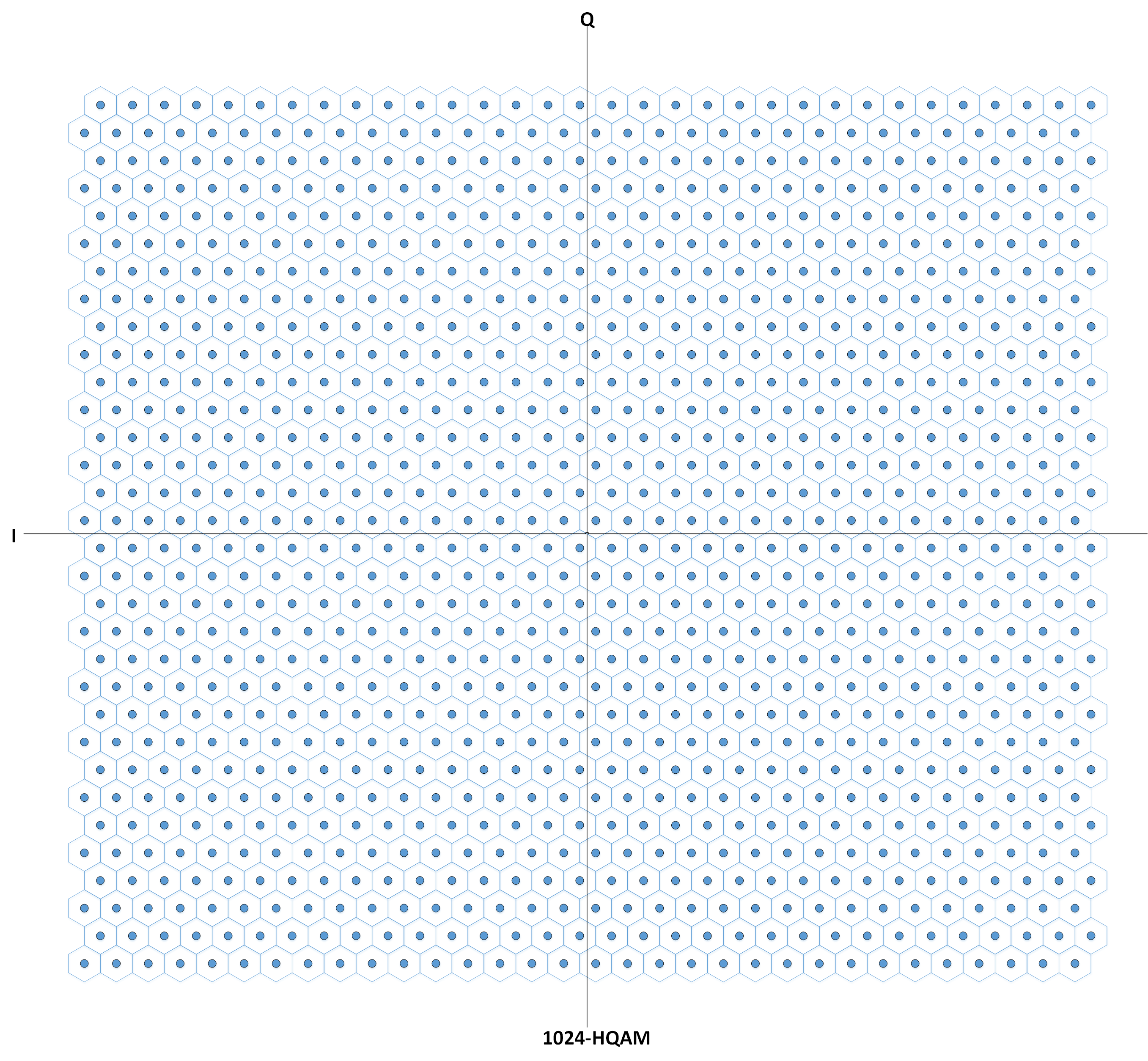}}
		\caption{\small{Various regular HQAM constellations.}}
		\label{HQAMReg}	
		\noindent\makebox[\linewidth]{\rule{18.5cm}{0.4pt}}
	\end{figure*}

\begin{figure*}
		\hfill
		\subfigure[]{\includegraphics[width=18 cm,height=10cm]{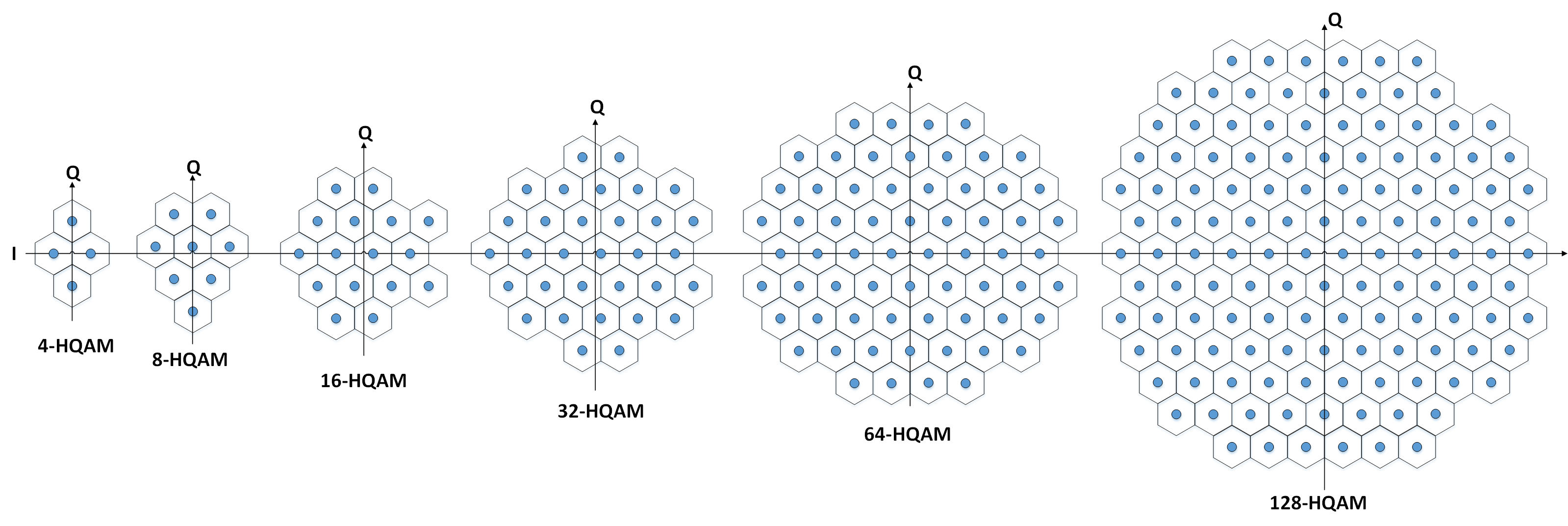}}	
		\hfill
		\subfigure[]{\includegraphics[width=10 cm,height=10cm]{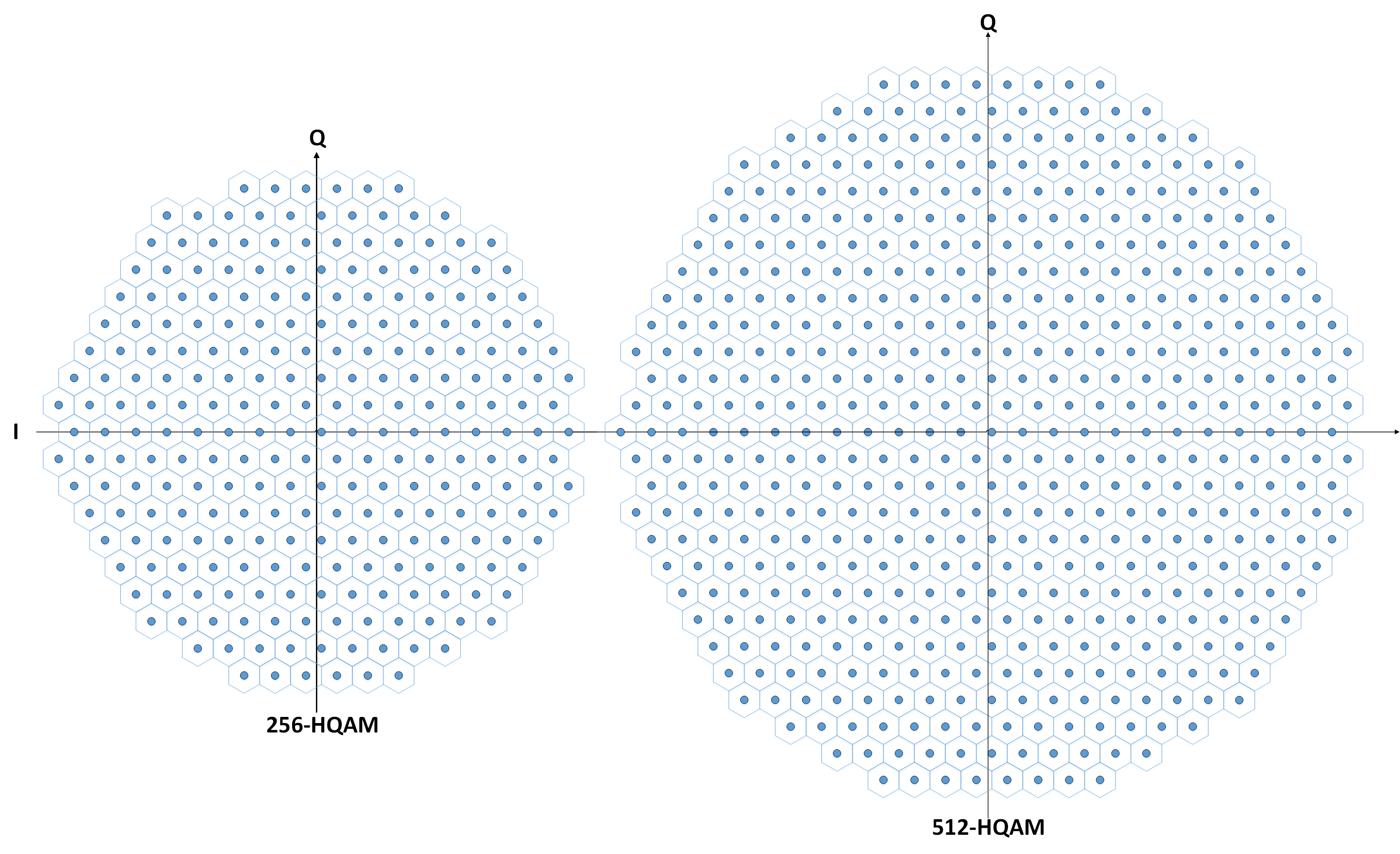}}
		\hfill
		\subfigure[]{\includegraphics[width=8 cm,height=10cm]{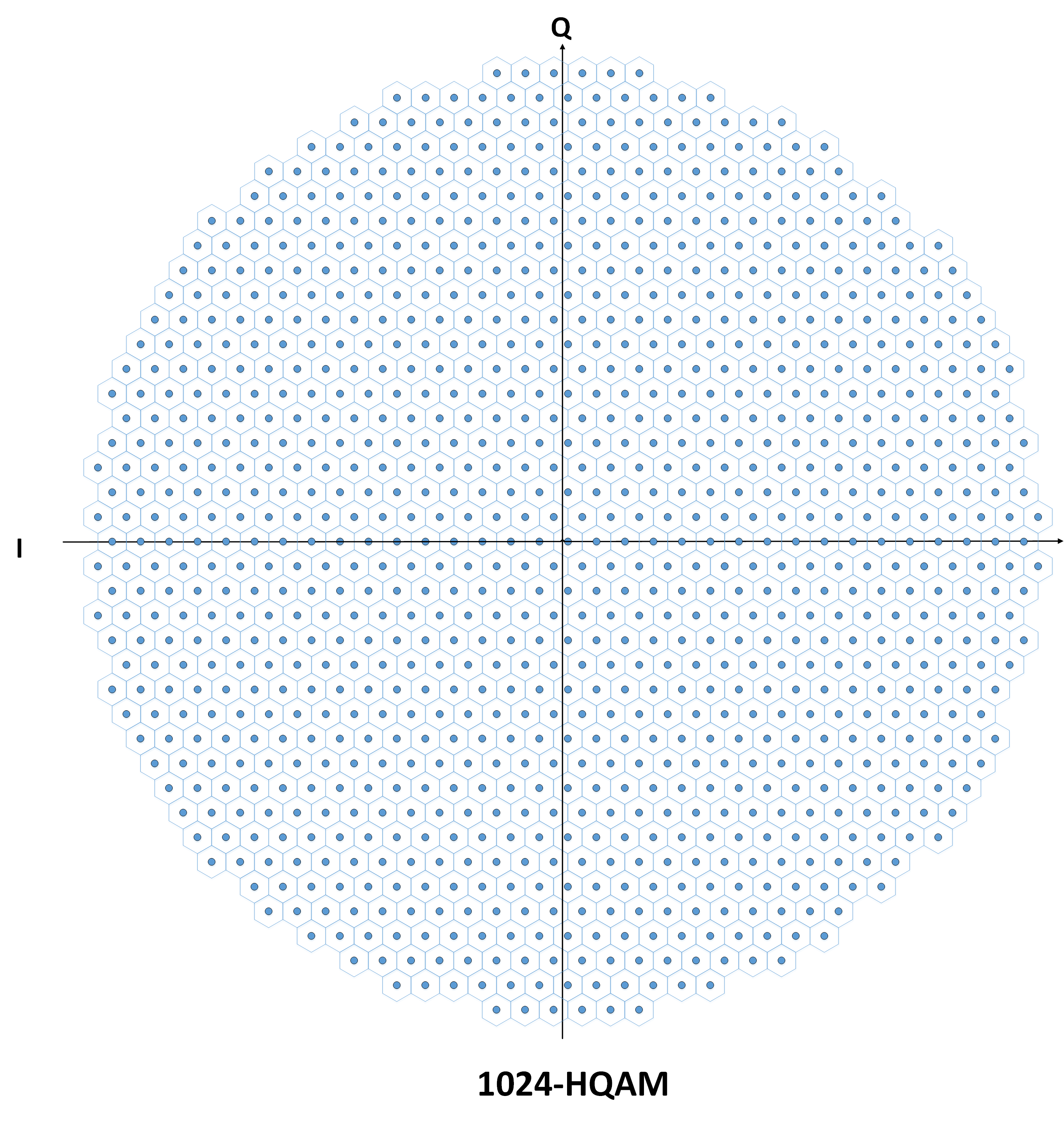}}
		\caption{\small{Various irregular HQAM (optimum) constellations.}}
		\label{HQAMIrr}	
		\noindent\makebox[\linewidth]{\rule{18.5cm}{0.4pt}}
	\end{figure*}

	%
	Energy efficient 2D signal constellation which can save considerable transmission energy is the most basic requirement of the present and future wireless communication systems. For a 2D signal constellation, SER is mainly affected by the minimum separation between the two neighboring constellation points, and average symbol energy which depends upon the mean squared distance of constellation points from the origin.
	Considering these points in mind, an optimum 2D hexagonal lattice based HQAM constellation is proposed where constellation points are situated at the center of the concentric spheres. For a minimum distance separation of $2d$  between the two adjacent constellation points in hexagonal constellation, area of the hexagonal region is $2\sqrt{3}d^2$ which is $0.866$ times the area of the rectangle regions with the same dimensions. This allows hexagonal regions to provide around 0.6 dB gain over the rectangle regions \cite{forney1984efficient}. HQAM is also termed as triangular QAM (TQAM) in the literature. Placement and separation of the constellation points on the 2D plane is same for both the HQAM and TQAM constellations. If the constellation is termed as HQAM, the constellation points are considered at the center of an equilateral hexagon and separated with the neighboring point hexagon with $2d$ distance. If the constellation is termed as TQAM, then the constellation points are considered at the vertices of well connected equilateral triangles of side-length $2d$. However, the decision boundaries are hexagonal shaped as shown in Fig. \ref{16TQAMHQAM}.
	Based on the placement of the constellation points, HQAM constellations are further categorized into regular and irregular HQAM structures. Regular HQAM  has comparatively simpler detection, its power efficiency or BER performance can further be improved for larger values of $M$. The irregular HQAM  provides improved power efficiency and optimum performance, however, at the cost of increased detection complexity \cite{abdelaziz2018triangular}. Since, the distance between the constellation points is equal, in the inner regions, the ML detection boundaries are in equilateral hexagonal shapes. However, at the boundaries, detection regions vary as per the constellation geometry. Constellation for  regular HQAM, are symmetric around the origin \cite{park2007triangular,park2009odd}. For, even power of 2,  regular HQAM constellations are in square shapes \cite{park2007triangular} such as for $M=16, 64, 256, 1024,...$, however, for odd power of 2, regular HQAM constellations are in cross shapes \cite{park2009odd} such as for $M=32, 128, 512,...$. For, irregular HQAM, constellations are in circular shapes as $M$ increases which makes the irregular HQAM constellations more compact than others \cite{park2008irregularly}.  Various regular HQAM and irregular HQAM (optimum) constellations  are shown in Fig. \ref{HQAMReg} and Fig. \ref{HQAMIrr}, respectively.
	Average energy of HQAM constellations is less than the SQAM constellations for any value of $M$ except for $M=4$. This is due to the fact, that 4-HQAM has same average energy of the constellation as 4-SQAM, however is having higher value of NNs.
	Further, irregular HQAM constellations have comparatively less average energy than the regular HQAM constellations, however, are more complex to generate and detect.

	%
	\begin{figure*}[!h]
		\vspace{-0.5em}
		\centering
		\includegraphics[width=7in,height=2.96in]{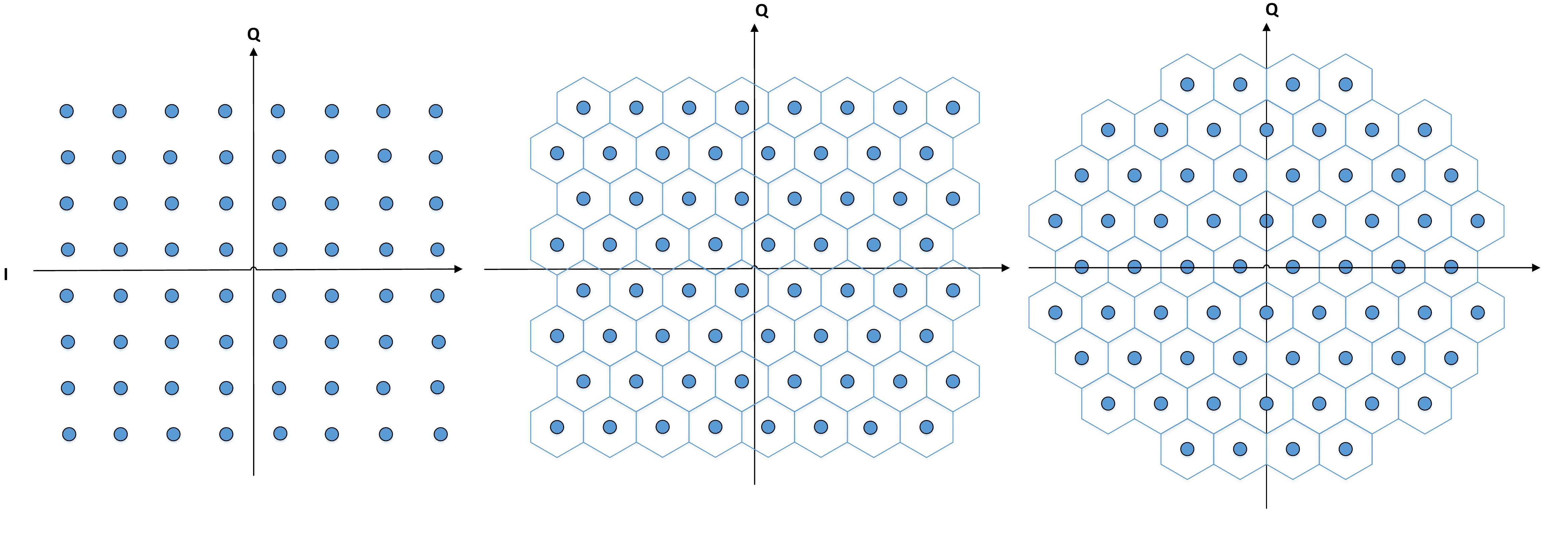}
		\vspace{-1em}
		\caption{\small{Generation of optimum irregular HQAM constellation (M=64).}}
		\label{HQAMgen}
		\vspace{-1em}
	\end{figure*}
In the following subsection, generation of irregular HQAM from the regular HQAM constellation is explained.
\subsubsection{Formation of Irregular HQAM (Optimum) Constellation from Regular HQAM}
Process to generate an irregular HQAM (optimum) constellation (Fig. \ref{HQAMgen}) is shown in Algorithm 1,
\begin{algorithm}[h!]
	\caption{Irregular HQAM generation}
	\begin{algorithmic}[1]
		\Inputs{M=number of constellation points.}
		\Initialize{Generate regular HQAM, set counter $l=1$.}
		\If {$M\neq8$}
		
		Shift one row of the generated regular HQAM to X-axis.
		\Else
		
		Shift one row of the generated regular HQAM to $\sqrt{3}/{4}$ distance above X-axis.
		\EndIf\\
		
		Generate rows on both sides of the x-axis with $d_{vs}=\sqrt{3}$.\\		
		Identify center point of the constellation.\\
		Calculate energy (${E}_{c}$) of all probable constellation points and store it in an array $\mathbf{E}_{c}$.
		\While{\{$l<=M$\}}\\
		Find $\underbrace{\min}_{1\leq j \leq M} \mathbf{E}_{c}(j)$, and place the constellation point at the coordinates of $j^{th}$ index.\\
		$l=l+1.$
		\EndWhile
	\end{algorithmic}
\end{algorithm} 
where, $d_{vs}$ represents the vertical spacing, $E_c$ represents the energy of a constellation point, and $\textbf{E}_c$ represents the array of constellation points energies.
{\renewcommand{\arraystretch}{1.5}
	\begin{table*}[!t]
		\centering
		\caption {Imaginary values of boundary points for region $R_9$.}
		\label{Tabim}
		\begin{tabular}{|c|c|c|c|c|c|c|c|}
			\hline
			\centering
			{$\text{Im}(P_1)$}  & {$\text{Im}(P_2)$}     &  {$\text{Im}(P_3)$}  & {$\text{Im}(P_4)$}  & {$\text{Im}(P_5)$}     &  {$\text{Im}(P_6)$}  & {$\text{Im}(P_7)$}  & {$\text{Im}(P_8)$}  \\
			\hline
			$\frac{11d-x}{\sqrt{3}}$	& $\frac{7d+x}{\sqrt{3}}$   &  $\frac{5d-x}{\sqrt{3}}$  &  $\frac{d+x}{\sqrt{3}}$  & -\Big($\frac{d+x}{\sqrt{3}}$\Big)   &  -\Big($\frac{5d-x}{\sqrt{3}}$\Big)  &  -\Big($\frac{7d+x}{\sqrt{3}}$\Big)   &  -\Big($\frac{11d-x}{\sqrt{3}}$\Big)   \\
			\hline		
		\end{tabular}
	\end{table*}
}
{\renewcommand{\arraystretch}{1.6}
	\begin{table*}[!t]
		\centering
		\caption {Imaginary values of boundary points for various vertical regions.}
		\label{Tabdec}
		\begin{tabular}{|c|c|c|c|c|c|c|c|c|}
			\hline
			\centering
			&{$R_9$, $R_8$}  & {$R_{10}$, $R_7$}     &  {$R_{11}$, $R_6$}  & {$R_{12}$, $R_5$}  & $R_{13}$, $R_4$    & $R_{14}$, $R_3$  & $R_{15}$, $R_2$  & $R_{16}$, $R_1$  \\
			\hline
			{$\text{Im}(P_1)$} &\Big($\frac{11d-x}{\sqrt{3}}$\Big) & \Big($\frac{10d+x}{\sqrt{3}}$\Big) & \Big($\frac{13d-x}{\sqrt{3}}$\Big)&  \Big($\frac{10d+x}{\sqrt{3}}$\Big) & \Big($\frac{11d+x}{\sqrt{3}}$\Big) &-&-& - \\
			\hline		
			{$\text{Im}(P_2)$} & \Big($\frac{7d+x}{\sqrt{3}}$\Big) & \Big($\frac{9d-x}{\sqrt{3}}$\Big) &  \Big($\frac{7d+x}{\sqrt{3}}$\Big) & \Big($\frac{11d-x}{\sqrt{3}}$\Big) & \Big($\frac{7d+x}{\sqrt{3}}$\Big) & \Big($\frac{13d-x}{\sqrt{3}}$\Big) & \Big($\frac{7d+x}{\sqrt{3}}$\Big)& \Big($\frac{x+d}{\sqrt{3}}$\Big) \\
			\hline		
			$\text{Im}(P_3)$ & \Big($\frac{5d-x}{\sqrt{3}}$\Big) & \Big($\frac{4d+x}{\sqrt{3}}$\Big) & \Big($\frac{7d-x}{\sqrt{3}}$\Big)& \Big($\frac{4d+x}{\sqrt{3}}$\Big) & \Big($\frac{9d-x}{\sqrt{3}}$\Big) & \Big($\frac{4d+x}{\sqrt{3}}$\Big) & \Big($\frac{11d-x}{\sqrt{3}}$\Big)& \Big($\frac{x-3d}{\sqrt{3}}$\Big) \\
			\hline	
			{$\text{Im}(P_4)$} & \Big($\frac{d+x}{\sqrt{3}}$\Big) & \Big($\frac{3d-x}{\sqrt{3}}$\Big) & \Big($\frac{x-d}{\sqrt{3}}$\Big) & \Big($\frac{5d-x}{\sqrt{3}}$\Big) & \Big($\frac{x-3d}{\sqrt{3}}$\Big) & \Big($\frac{7d-x}{\sqrt{3}}$\Big) & \Big($\frac{x-5d}{\sqrt{3}}$\Big) & \Big($\frac{9d-x}{\sqrt{3}}$\Big) \\
			\hline	
			{$\text{Im}(P_5)$} & -\Big($\frac{d+x}{\sqrt{3}}$\Big) & -\Big($\frac{3d-x}{\sqrt{3}}$\Big) & -\Big($\frac{x-d}{\sqrt{3}}$\Big) & -\Big($\frac{5d-x}{\sqrt{3}}$\Big) & -\Big($\frac{x-3d}{\sqrt{3}}$\Big) & -\Big($\frac{7d-x}{\sqrt{3}}$\Big) & -\Big($\frac{x-5d}{\sqrt{3}}$\Big) & -\Big($\frac{9d-x}{\sqrt{3}}$\Big) \\
			\hline	
			{$\text{Im}(P_6)$} & -\Big($\frac{5d-x}{\sqrt{3}}$\Big) & -\Big($\frac{4d+x}{\sqrt{3}}$\Big) & -\Big($\frac{7d-x}{\sqrt{3}}$\Big) & -\Big($\frac{4d+x}{\sqrt{3}}$\Big) & -\Big($\frac{9d-x}{\sqrt{3}}$\Big) & -\Big($\frac{4d+x}{\sqrt{3}}$\Big) & -\Big($\frac{11d-x}{\sqrt{3}}$\Big) & -\Big($\frac{x-3d}{\sqrt{3}}$\Big)\\
			\hline	
			{$\text{Im}(P_7)$} & -\Big($\frac{7d+x}{\sqrt{3}}$\Big) & -\Big($\frac{9d-x}{\sqrt{3}}$\Big) & -\Big($\frac{7d+x}{\sqrt{3}}$\Big) & -\Big($\frac{11d-x}{\sqrt{3}}$\Big) & -\Big($\frac{7d+x}{\sqrt{3}}$\Big) & -\Big($\frac{13d-x}{\sqrt{3}}$\Big) & -\Big($\frac{7d+x}{\sqrt{3}}$\Big) & -\Big($\frac{x+d}{\sqrt{3}}$\Big) \\
			\hline		
			{$\text{Im}(P_8)$} & \Big($\frac{11d-x}{\sqrt{3}}$\Big) & -\Big($\frac{10d+x}{\sqrt{3}}$\Big) & -\Big($\frac{13d-x}{\sqrt{3}}$\Big) & -\Big($\frac{10d+x}{\sqrt{3}}$\Big) & -\Big($\frac{11d+x}{\sqrt{3}}$\Big) & -&- &- \\
			\hline						
		\end{tabular}
	\end{table*}
}
\vspace{-1em}
\subsection{Decision Regions}
\begin{figure*}
		\hfill
		\subfigure[]{\includegraphics[width=18 cm,height=10cm]{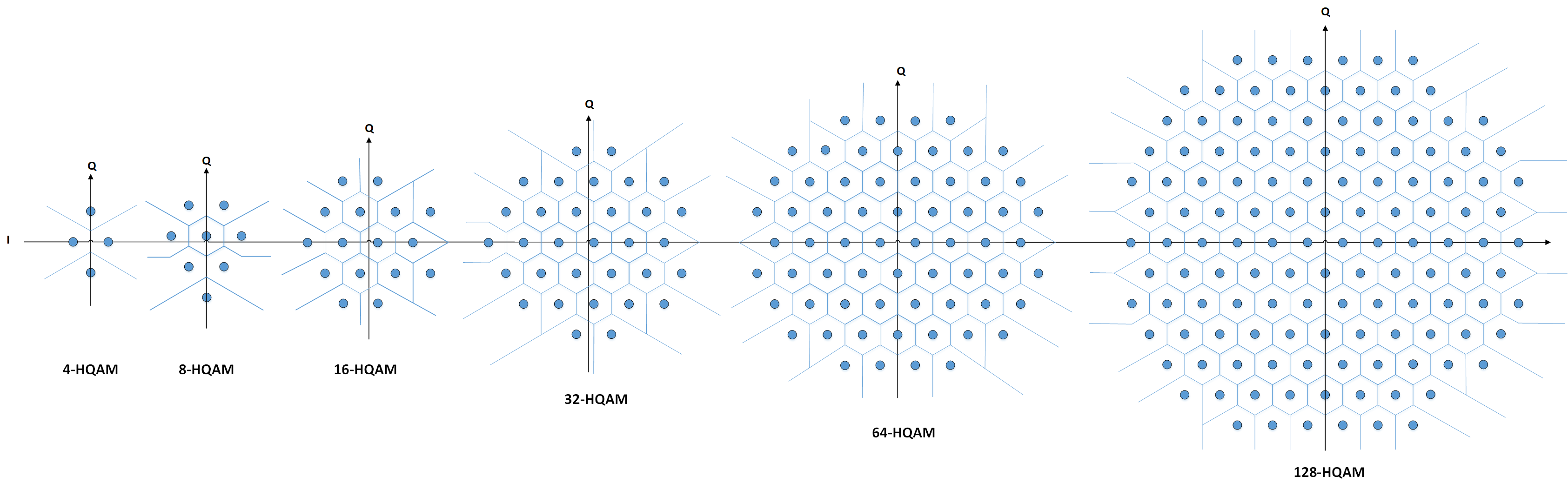}}	
		\hfill
		\subfigure[]{\includegraphics[width=11 cm,height=10cm]{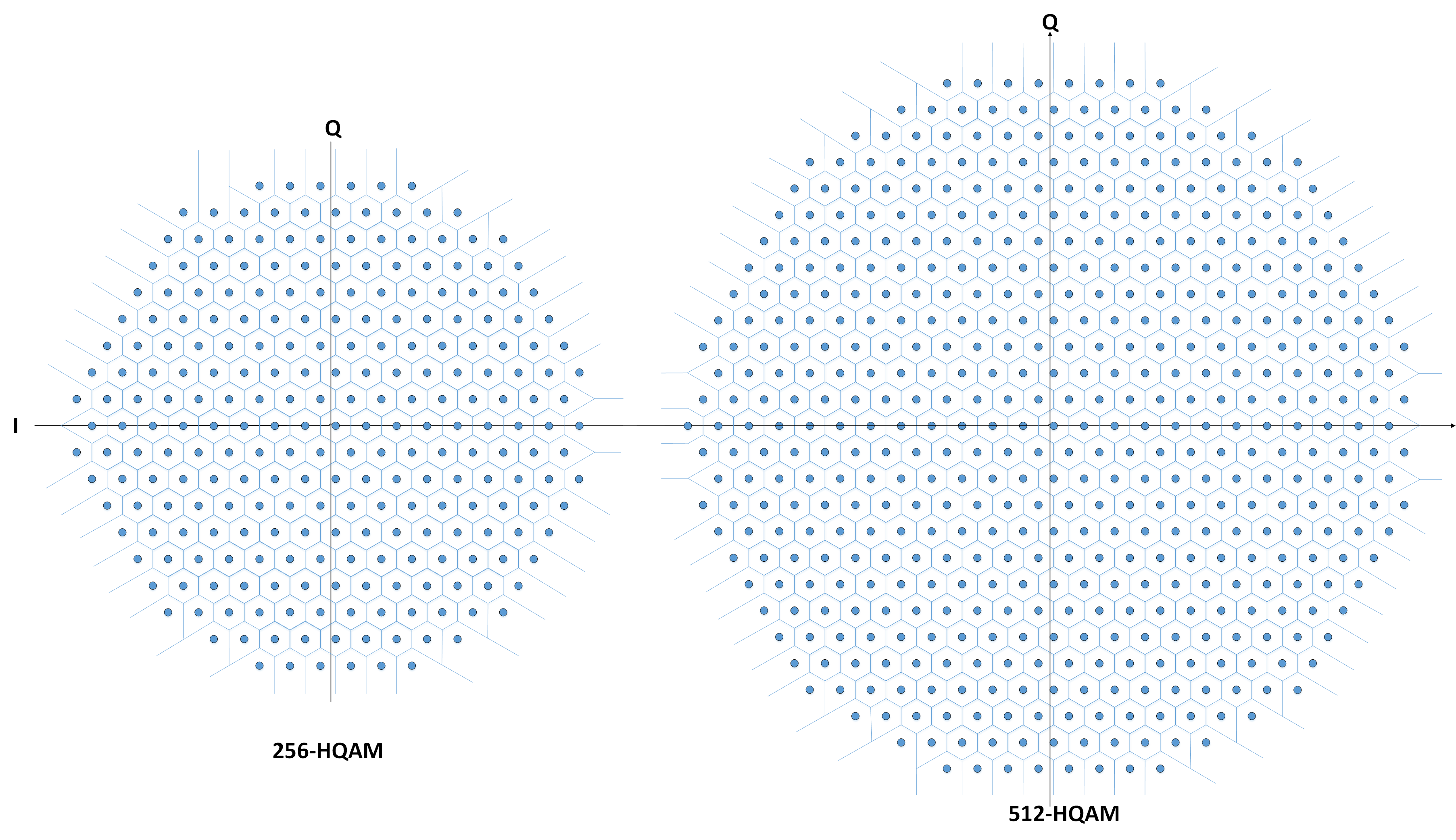}}
		\hfill
		\subfigure[]{\includegraphics[width=8 cm,height=10cm]{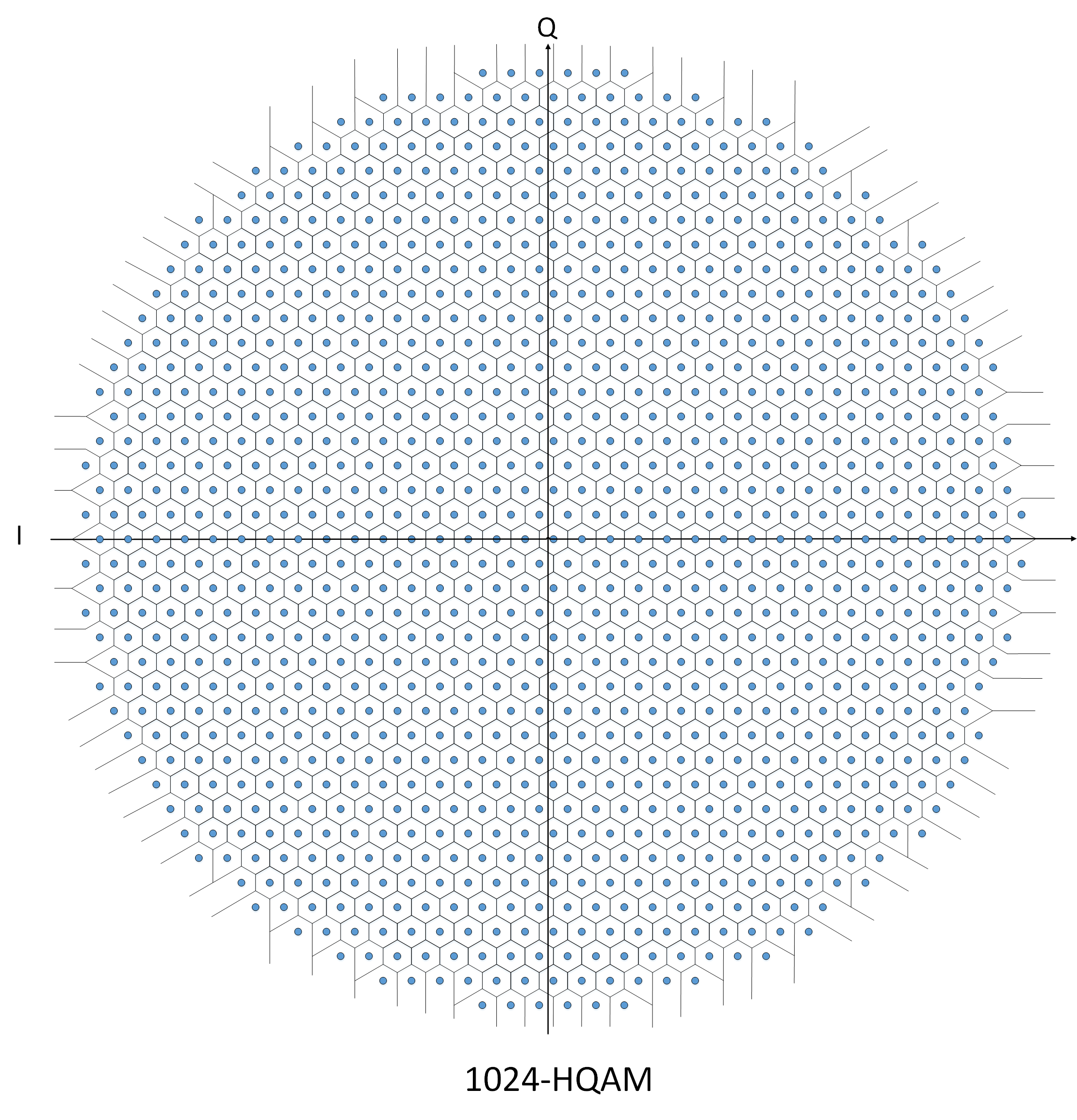}}
		\caption{\small{Various irregular HQAM (optimum) constellations with their decisions boundaries.}}
		\label{HQAMIrrdec}	
		\noindent\makebox[\linewidth]{\rule{18.5cm}{0.4pt}}
	\end{figure*}

%
ML detection is the most popularly used detection scheme where distance between each signal point and the received symbol is calculated, and decision (as an estimate of the transmitted symbol) is made to the nearest symbol. ML detection has less detection complexity however, is not applicable for HQAM detection in a straight-forward manner. Thus, an optimum and simple detection scheme for HQAM constellations is proposed where entire constellation is initially divided into vertical regions. A signal point is first located in a region and then the nearest constellation point in that region is found. This process ultimately reduces the number of distance calculations as compared to the traditional ML detection scheme \cite{park2007triangular}.
\begin{figure}[!h]
	\centering
	\includegraphics[width=3.7in,height=3.2in]{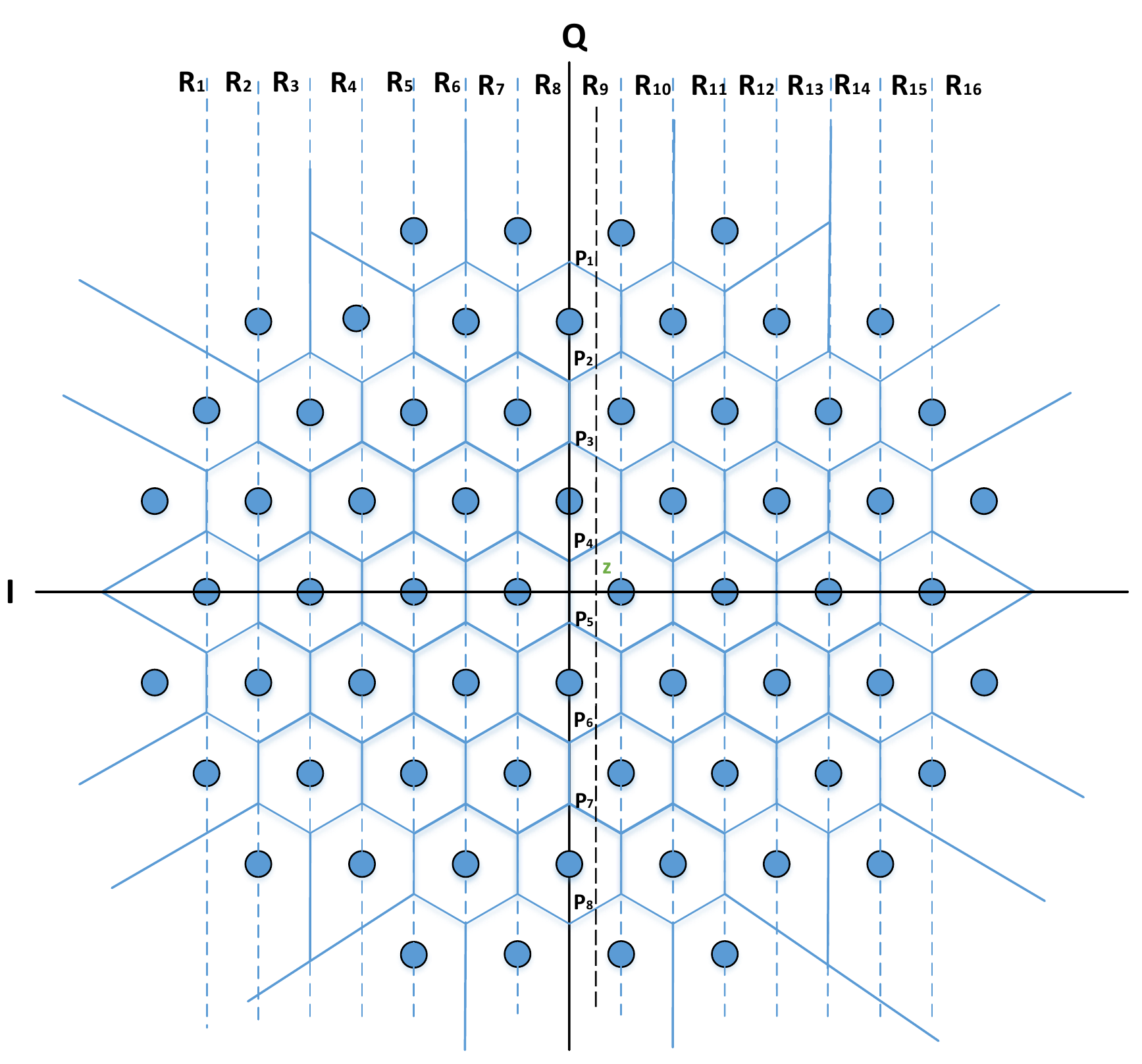}
	\caption{Simplified detection method for irregular 64-HQAM constellation.}
	\label{64decision}
\end{figure}
In Fig. \ref{64decision}, a simplified detection technique for 64-ary irregular HQAM constellation is shown which is explained as follows. Initially, the entire constellation is divided into $R_1-R_{16}$ regions.  When a symbol, lets say $z=x+iy$ is received (where $x$ and $y$ are the constant values and $j=\sqrt{-1}$),  based on the the arbitrary value x, one of the sixteen regions is selected. Further, imaginary values for the decision points are calculated and decision for a symbol is made with a comparison between the imaginary value $y$ of the demodulated symbol and calculated imaginary boundaries. For different regions, calculated imaginary boundaries are shown in Table \ref{Tabdec}. 
For illustration, depending on the arbitrary value of $x$, initially a region ($R_9$) is selected for the demodulated symbol $z$ as shown in Fig. \ref{64decision}. Next, to detect 9 signaling points which lie in $R_9$ region, 8 boundary points ($P_1-P_8$) are marked on the dotted line and their imaginary values are calculated to estimate a signaling point out of the 9 signaling points. The imaginary values of 8 boundary points for $R_9$ region are shown in Table \ref{Tabim}. Following similar procedure for the other regions, imaginary values for the boundaries points are calculated and summarized in Table \ref{Tabdec}.
Similar procedure can be followed for the other HQAM constellations for symbol detection. For various irregular HQAM constellations  ($M$ varies from 4 to 1024), decision regions are shown in Fig. \ref{HQAMIrrdec}.
	
	\subsection{SER Analysis}
	%
	Let us consider a set $S_M\in\{s_0, s_1,...,s_{M-1}\}$ for $M$-ary HQAM constellation. SEP for HQAM constellation is based on the shape of the decision regions. For this, a suitable correction to NN approximation is applied \cite{park2012performance} which is expressed as
	\begin{align}
	\mathcal{P}_{NN}=\tau Q\Big(\frac{d}{2\sigma}\Big).
	\end{align}
	Here $\tau$ represents the average number of NNs and is defined as $\tau=\frac{1}{M}\sum_{k=0}^{M-1}\tau(k)$ wherein $\tau(k)$ is the number of NNs for $k^{th}$ symbol $s_k$, $Q(\cdot)$ represents the Gaussian Q-function, $d$ is the half the distance between constellation points, and $\sigma$ is the AWGN standard deviation. 
	\begin{figure}[!h]
		\centering
		\includegraphics[width=2in,height=1.5in]{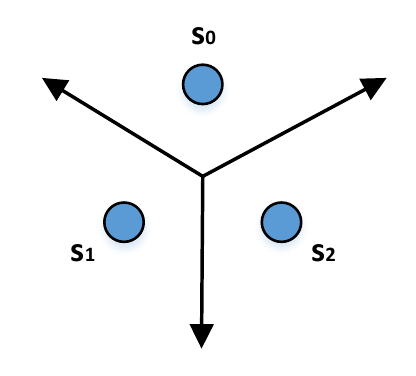}
		\caption{3-PSK constellation.}
		\label{psk}
	\end{figure}
	For illustration, simplest HQAM structure is the  3-PSK structure as shown in Fig. \ref{psk}. The 3-PSK consists of three symbols $s_0$, $s_1$, and $s_2$. Starting with $s_0$ which has two NNs $s_1$ and $s_2$. $s_1$ is also an NN of $s_2$. Thus, $s_1$ and $s_2$ are the couple of NNs of $s_0$. Two half planes with partial overlapping cover the error region in NN approximation which produces an overestimation of SEP. This can be clarified from Fig. \ref{psk}. Thus, exact SEP expression for 3-PSK is given as 
	\begin{align}
	\mathcal{P}^{\text{3-PSK}}=2 Q\Big(\frac{d}{2\sigma}\Big)-C_1,
	\end{align}
	where $C_1$ is a correction term, used to compensate for the double counting of overlapping. For 3-PSK, $C_1$ is given in \cite{proakis2008masoud}. However, for simplification, integral can be avoided and $C_1$ is approximated with $C_2$ as \cite{rugini2015tight}
	\begin{align}\label{R2}
	C_1\approx C_2=2Q\Big(\frac{d}{2\sigma}\Big)Q\Big(\frac{d}{2\sqrt{3}\sigma}\Big)-\frac{2}{3}Q^2\Big(\frac{d}{\sqrt{6}\sigma}\Big).
	\end{align}
	Hence, approximate SEP expression for 3-PSK can be given as
	\begin{align}
	\mathcal{P}^{\text{3-PSK}}=2 Q\big(d/{2\sigma}\big)-C_2.
	\end{align}
	This correction approach is easily applicable for generalized HQAM constellation. In HQAM, multiple overlaps with a $2\pi/3$ angle occur which generate couple of NNs. Therefore, for AWGN channel, approximate SEP expression for HQAM can be given as
	\begin{align}\label{SEP1}
	\mathcal{P}^{HQAM}=\tau Q\Big(\frac{d}{2\sigma}\Big)-\tau_cC_2,
	\end{align}
	where $\tau_c=\frac{1}{M}\sum_{k=0}^{M-1}\tau_c(k)$ is the average of couple of NNs, wherein $\tau_c(k)$ represents couple of NNs for $s_k$ symbol. Further, considering $K=\frac{1}{\gamma}\big(\frac{d}{2\sigma}\big)^2$ with average transmit SNR $\gamma={P_s}/{N_0}$, and substituting $C_2$ from (\ref{R2}) in (\ref{SEP1}), generalized conditional SEP of HQAM is given as
	\begin{align}\label{CHQAM}
	\mathcal{P}^{HQAM}(e|\gamma)&= \tau Q(\sqrt{K\gamma})-2\tau_c Q(\sqrt{K\gamma}) Q\Big(\sqrt{\frac{K\gamma}{3}}\Big)\nonumber\\&
	+\frac{2}{3}\tau_cQ^2\Big(\sqrt{\frac{2K\gamma}{3}}\Big).
	\end{align}

	{\renewcommand{\arraystretch}{1.4}
		\begin{table*}[!h]
			\centering
			\caption {{Different parameters for the various regular and irregular HQAM constellations over the AWGN channel.}}
			\label{tab_HQAM}
			\begin{tabular}{|c|cccccc|cccccc|}
				\hline
				&        \multicolumn{6}{c|}{Regular HQAM} &               \multicolumn{6}{c|}{Irregular HQAM (optimum)}       \\ \hline
				M 	& \multicolumn{1}{c|}{K} & \multicolumn{1}{c|}{$\tau$} & \multicolumn{1}{c|}{$\tau_c$}  & \multicolumn{1}{c|}{$E_s/d^2$}& \multicolumn{1}{c|}{PAPR} &\multicolumn{1}{c|}{$G_p$}	& \multicolumn{1}{c|}{K} & 
				\multicolumn{1}{c|}{$\tau$}   & \multicolumn{1}{c|}{$\tau_c$} &\multicolumn{1}{c|}{$E_s/d^2$} &\multicolumn{1}{c|}{PAPR} &\multicolumn{1}{c|}{$G_p$}\\ \hline
				
				4  &  \multicolumn{1}{c|}{1} & \multicolumn{1}{c|}{$\frac{5}{2}$} &    \multicolumn{1}{c|}{$\frac{3}{2}$} & \multicolumn{1}{c|}{2}& \multicolumn{1}{c|}{1.5} &\multicolumn{1}{c|}{$1.166$} &  \multicolumn{1}{c|}{1} & \multicolumn{1}{c|}{$\frac{5}{2}$} & \multicolumn{1}{c|}{$\frac{3}{2}$} & \multicolumn{1}{c|}{2} & \multicolumn{1}{c|}{1.5} &\multicolumn{1}{c|}{$1.166$}
				\\ \hline
				
				8 	& \multicolumn{1}{c|}{$\frac{2}{6}$} & \multicolumn{1}{c|}{$\frac{7}{2}$} &    \multicolumn{1}{c|}{$\frac{21}{8}$}  & \multicolumn{1}{c|}{6}& \multicolumn{1}{c|}{2.16}  &\multicolumn{1}{c|}{$1.275$}   &   \multicolumn{1}{c|}{$\frac{32}{69}$} & \multicolumn{1}{c|}{$\frac{7}{2}$} & \multicolumn{1}{c|}{$\frac{21}{8}$} & \multicolumn{1}{c|}{4.3125}& \multicolumn{1}{c|}{2.130} &\multicolumn{1}{c|}{$-$}
				\\ \hline
				
				16	& \multicolumn{1}{c|}{$\frac{2}{9}$} & \multicolumn{1}{c|}{$\frac{33}{8}$} &    \multicolumn{1}{c|}{$\frac{27}{8}$}   & \multicolumn{1}{c|}{9}& \multicolumn{1}{c|}{2.11}  &\multicolumn{1}{c|}{$1.237$}  &  \multicolumn{1}{c|}{$\frac{8}{35}$} & \multicolumn{1}{c|}{$\frac{33}{8}$} & \multicolumn{1}{c|}{$\frac{27}{8}$} & \multicolumn{1}{c|}{8.75} & \multicolumn{1}{c|}{1.742} &\multicolumn{1}{c|}{$1.27$}
				\\ \hline
				
				32  & \multicolumn{1}{c|}{$\frac{8}{71}$} & \multicolumn{1}{c|}{$\frac{75}{16}$} &    \multicolumn{1}{c|}{$\frac{33}{8}$}   & \multicolumn{1}{c|}{17.75}& \multicolumn{1}{c|}{2.084}   &\multicolumn{1}{c|}{$1.388$} &  \multicolumn{1}{c|}{$\frac{512}{4503}$} & \multicolumn{1}{c|}{$\frac{75}{16}$} & \multicolumn{1}{c|}{$\frac{33}{8}$} & \multicolumn{1}{c|}{17.59} & \multicolumn{1}{c|}{1.8792} &\multicolumn{1}{c|}{$-$}
				\\ \hline
				
				64  & \multicolumn{1}{c|}{$\frac{2}{37}$} & \multicolumn{1}{c|}{$\frac{161}{32}$} &    \multicolumn{1}{c|}{$\frac{147}{32}$}   & \multicolumn{1}{c|}{37}& \multicolumn{1}{c|}{2.51}   &\multicolumn{1}{c|}{$1.2822$} &  \multicolumn{1}{c|}{$\frac{8}{141}$} & \multicolumn{1}{c|}{$\frac{163}{32}$} & \multicolumn{1}{c|}{$\frac{75}{16}$} &  \multicolumn{1}{c|}{35.25} & \multicolumn{1}{c|}{1.90} &\multicolumn{1}{c|}{$1.351$}
				\\ \hline
				
				128	 & \multicolumn{1}{c|}{$\frac{2}{72}$} & \multicolumn{1}{c|}{$\frac{339}{64}$} &    \multicolumn{1}{c|}{$\frac{159}{32}$}   & \multicolumn{1}{c|}{72}& \multicolumn{1}{c|}{2.347}  &\multicolumn{1}{c|}{$1.363$}  &  \multicolumn{1}{l|}{$\frac{2}{70.56}$} & \multicolumn{1}{l|}{$\frac{343}{64}$} & \multicolumn{1}{c|}{$\frac{81}{16}$} & \multicolumn{1}{c|}{70.562} & \multicolumn{1}{c|}{1.96} &\multicolumn{1}{c|}{$1.48$}
				\\ \hline
				
				256	 & \multicolumn{1}{c|}{$\frac{2}{149}$} & \multicolumn{1}{c|}{$\frac{705}{128}$} &    \multicolumn{1}{c|}{$\frac{675}{128}$}    & \multicolumn{1}{c|}{149}& \multicolumn{1}{c|}{2.74}  &\multicolumn{1}{c|}{$-$} &    \multicolumn{1}{c|}{$\frac{2}{141}$} & \multicolumn{1}{c|}{$\frac{711}{128}$} & \multicolumn{1}{c|}{$\frac{171}{32}$} & \multicolumn{1}{c|}{141} & \multicolumn{1}{c|}{2.03}  &\multicolumn{1}{c|}{$-$}
				\\ \hline
				
				512	 & \multicolumn{1}{c|}{$\frac{2}{289.06}$} & \multicolumn{1}{c|}{$\frac{2895}{512}$} &    \multicolumn{1}{c|}{$\frac{5619}{1024}$}   & \multicolumn{1}{c|}{289.06} & \multicolumn{1}{c|}{2.535}   &\multicolumn{1}{c|}{$-$} &    \multicolumn{1}{c|}{$\frac{200}{28217}$} & \multicolumn{1}{c|}{$\frac{2911}{512}$} & \multicolumn{1}{c|}{$\frac{5667}{1024}$} & \multicolumn{1}{c|}{282.17} & \multicolumn{1}{c|}{2.01} &\multicolumn{1}{c|}{$-$}
				\\ \hline
				
				1024 & \multicolumn{1}{c|}{$\frac{2}{597}$} & \multicolumn{1}{c|}{$\frac{2945}{512}$} &    \multicolumn{1}{c|}{$\frac{2883}{512}$}    & \multicolumn{1}{c|}{597}& \multicolumn{1}{c|}{2.86}  &\multicolumn{1}{c|}{$-$} &    \multicolumn{1}{c|}{$\frac{100}{28227}$} & \multicolumn{1}{c|}{$\frac{2955}{512}$} & \multicolumn{1}{c|}{$\frac{1449}{256}$} & \multicolumn{1}{c|}{564.54} & \multicolumn{1}{c|}{1.99} &\multicolumn{1}{c|}{$-$}
				\\ \hline
			\end{tabular}
		\end{table*}
	}
	Here, $K$ is related to the average energy of the constellation which varies with $M$. Further,  $\tau$ and $\tau_c$ are the average number of NNs and couple of NNs, respectively, which also vary according to $M$. Various regular HQAM constellations and irregular HQAM constellations (optimum) are shown in Fig. \ref{HQAMReg} and Fig. \ref{HQAMIrr}, respectively. Based on the constellations (regular or irregular HQAM) and  modulation order $M$; different values of $K$, $\tau$, and $\tau_c$ are given in Table \ref{tab_HQAM} which are used to select various HQAM constellations with different constellation orders. 

	For a well connected HQAM, $\tau_c=3T/M$, where T represents the number of equilateral triangles with sides of length $2d$, and $\tau=2\Big[\frac{\tau_c}{3}+1-\frac{1}{M}\Big]$ \cite{rugini2016symbol}. For regular HQAM with even power of 2 constellations, $K=\frac{24}{7M-4}$, $\tau=2\big(3-\frac{4}{\sqrt{M}}+\frac{1}{M}\big)$, and $\tau_c=6\Big(1-\frac{1}{\sqrt{M}}\Big)^2$. However, for an arbitrary  power of 2 constellation, no such pattern can be found in the literature for different values of $K$, $\tau$, and $\tau_c$ for both regular and irregular HQAM constellations. A detailed study on various HQAM or TQAM constellations can be seen in \cite{rugini2016symbol,abdelaziz2018triangular,forney1984efficient,park2012performance,park2009odd,park2008irregularly}. 
	Solving the equations given for $K$, $\tau$, and $\tau_c$, generalized approximate values of $K$, $\tau$, and $\tau_c$, for a well connected $M$-ary irregular HQAM can be given as
	\begin{align}\label{HQAMapp}
	K&=\frac{7}{2M-1},\nonumber\\
	\tau&=6.07-6.733 M^{-0.456},\nonumber\\
	\tau_c&={3}\Big[\frac{\tau}{2}-1+\frac{1}{M}\Big].
	\end{align}

		\begin{figure*}[!t]
		\centering
		\includegraphics[width=7.3in,height=3.2in]{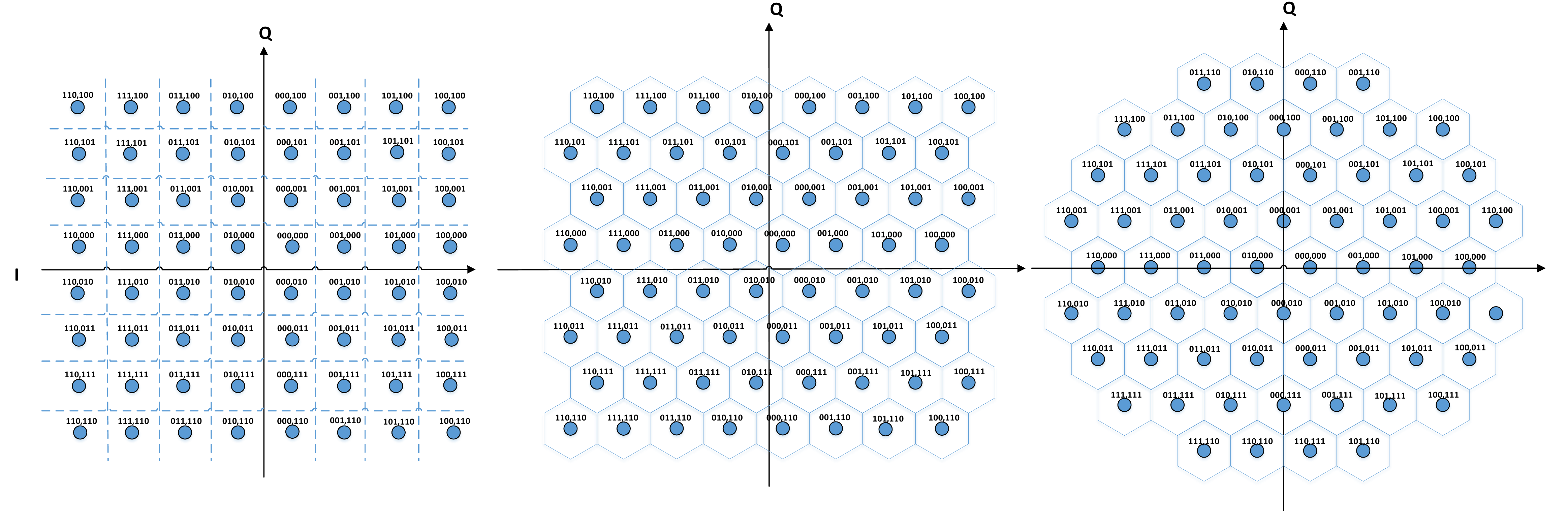}
		\caption{Bit mapping for 64-HQAM.}
		\label{bitmap}
	\end{figure*}
	\begin{table*}[!t]
		\centering
		\caption {Comparison of various QAM constellations.}
		\label{tab_comp}
		\begin{tabular}{|c|cc|cc|cc|cc|cc|}
			\hline
			&      \multicolumn{2}{c|}{SQAM} & \multicolumn{2}{c|}{RQAM} & \multicolumn{2}{c|}{XQAM} &  \multicolumn{2}{c|}{Regular HQAM} &               \multicolumn{2}{c|}{Irregular HQAM (optimum)}       \\ \hline
			M  & \multicolumn{1}{c}{$E_s/d^2$}& \multicolumn{1}{c|}{PAPR} &
			\multicolumn{1}{c}{$E_s/d^2$}& \multicolumn{1}{c|}{PAPR} &
			\multicolumn{1}{c}{$E_s/d^2$}& \multicolumn{1}{c|}{PAPR} & \multicolumn{1}{c}{$E_s/d^2$}& \multicolumn{1}{c|}{PAPR} 	& \multicolumn{1}{c}{$E_s/d^2$} &\multicolumn{1}{c|}{PAPR} 
			\\ \hline
			4 &\multicolumn{1}{c}{2}& \multicolumn{1}{c|}{1} &
			-& - &
			-& - &
			\multicolumn{1}{c}{2}& \multicolumn{1}{c|}{1.5} 	& \multicolumn{1}{c}{2} &\multicolumn{1}{c|}{1.5} 
			\\ \hline
			8 & -& - &
			\multicolumn{1}{c}{6} & \multicolumn{1}{c|}{1.666} &
			\multicolumn{1}{c}{-}   & \multicolumn{1}{c|}{-} &
			\multicolumn{1}{c}{4.5}& \multicolumn{1}{c|}{1.55} &
			\multicolumn{1}{c}{4.312}& \multicolumn{1}{c|}{2.13} 
			\\ \hline
			16 &\multicolumn{1}{c}{10}& \multicolumn{1}{c|}{1.80} &
			-& - &
			-& - &
			\multicolumn{1}{c}{9}& \multicolumn{1}{c|}{2.11} 	& \multicolumn{1}{c}{8.75} &\multicolumn{1}{c|}{1.742} 
			\\ \hline
			32 & -& - &
			\multicolumn{1}{c}{26}& \multicolumn{1}{c|}{2.23} &
			\multicolumn{1}{c}{20}& \multicolumn{1}{c|}{1.70} &
			\multicolumn{1}{c}{17.75}& \multicolumn{1}{c|}{2.084} 	& \multicolumn{1}{c}{17.59} &\multicolumn{1}{c|}{1.879} 
			\\ 
			\hline
			64 &\multicolumn{1}{c}{42}& \multicolumn{1}{c|}{2.333} &
			-& - &
			-& - &
			\multicolumn{1}{c}{37}& \multicolumn{1}{c|}{2.51} 	& \multicolumn{1}{c}{35.25} &\multicolumn{1}{c|}{1.90} 
			\\ 
			\hline
			128 & -& - &
			\multicolumn{1}{c}{106}& \multicolumn{1}{c|}{2.584} &
			\multicolumn{1}{c}{82}& \multicolumn{1}{c|}{2.073} &
			\multicolumn{1}{c}{72}& \multicolumn{1}{c|}{2.347} 	& \multicolumn{1}{c}{70.56} &\multicolumn{1}{c|}{1.96} 
			\\ 
			\hline
			256 &\multicolumn{1}{c}{170}& \multicolumn{1}{c|}{2.647} &
			-& - &
			-& - &
			\multicolumn{1}{c}{149}& \multicolumn{1}{c|}{2.74} 	& \multicolumn{1}{c}{141.023} &\multicolumn{1}{c|}{2.03} 
			\\ 
			\hline
			512 & -& - &
			\multicolumn{1}{c}{426}& \multicolumn{1}{c|}{2.784} &
			\multicolumn{1}{c}{330}& \multicolumn{1}{c|}{2.28} &
			\multicolumn{1}{c}{290}& \multicolumn{1}{c|}{2.494} 	& \multicolumn{1}{c}{282.17} &\multicolumn{1}{c|}{2.01} 
			\\ 			\hline
			1024 &\multicolumn{1}{c}{682}& \multicolumn{1}{c|}{2.81} &
			-& - &
			-& - &
			\multicolumn{1}{c}{597}& \multicolumn{1}{c|}{2.86} 	& \multicolumn{1}{c}{564.54} &\multicolumn{1}{c|}{1.99} 
			\\ \hline
		\end{tabular}
	\end{table*}
	%
	\subsection{Bit Mapping, Gray Code Penalty, and BER}
	BER performance is significantly affected by the bit mapping of the constellation points. For an $M$- ary constellation, bit mapping should be done  such 
	that the bit difference between the neighboring points is minimum. The optimum bit mapping is achieved with the least average number of bit difference between neighboring symbols.
		In SQAM, each constellation point shares its boundaries with maximum 4 constellation points and differ with its neighbors by only a single bit. Thus, SQAM has the perfect Gray coding and its  $G_P$ is always 1 \cite{gray1953pulse}.
 However, for HQAM, perfect Gray mapping is not possible \cite{park2007bit}, since each symbol shares its boundaries with maximum of six neighboring symbols.  This generates $G_P$ which  is expressed as 
	\begin{align}
	G_P=&\frac{1}{M}\sum_{k=1}^{M}{G_P^{s_k}},\nonumber\\
	=&\frac{1}{M}\sum_{k=1}^{M}\frac{\sum_{j=1}^{N(s_k)}b_d(s_k,s_j)}{N(s_k)},
	\end{align}
	where $s_k$ denotes the $k^{th}$ data symbol, $G_P^{s_k}$ is the Gray code penalty of $k^{th}$ data symbol,
	$b_d(s_k, s_j)$ represents the bit difference of $k^{th}$ data symbol with its neighboring symbol $s_j$, and $N(s_k)$ is the NN count of $k^{th}$ data symbol $s_k$. 
	
	In any constellation, bit mapping should be performed very precisely and carefully to minimize the BEP for a given SEP for the constellation. For a given SEP, approximated BEP expression is given in (\ref{BER})
	which is directly proportional to  $G_P$. Hence, optimum  bit mapping is required to minimize $G_P$ to get minimum BER. Optimum bit mapping can only be obtained with exhaustive search over all the possible bit mappings for the constellation. Further, it should be noticed that there is not an unique optimum bit mapping, others may exist which have the same $G_P$.
	Practically, for an $M$-ary signal constellation, there are $M!$ bit mappings and finding an optimum bit mapping for optimum $G_P$ for a large signal constellation  is extremely tedious and time consuming. Thus,  a sub-optimum bit mapping is preferred for a signal constellation with $M>16$.
	For an $M$-ary signal constellation, each symbol is represented with $n$ bits, and $2^n$ bit streams appear to represent all the symbols in the constellation.
	In practice, a sub-optimum bit mapping is preferred which is performed in three steps.
	Initially, regular HQAM constellations are generated with the same bit mapping as preferred for SQAM (or RQAM) constellations. Further, irregular HQAM constellations are generated by shifting the regular HQAM constellations such that the maximum signal points get matched on I/Q coordinates.	
	Next, for the matched signaling points of irregular HQAM, the same bit mapping is preferred as used for regular HQAM constellation. Finally,  
	for the remaining symbols, a partially optimized bit mapping is considered where each symbol picks its favorable NNs symbols from the available ones such that it has minimum bit difference with its NNs.	
	As an example, for $M=64$, bit mapping for irregular HQAM is shown in Fig. \ref{bitmap}. For $M=64=2^6$, 6 bits are required for mapping. Initially, 8 rows and 8 columns are selected, and 3 bits are assigned to represent the rows and rest of the 3 bits represent columns. Lower three bits represent the rows and upper three bits represent the columns. Bit mapping for irregular HQAM is completed in three steps. In the first step, regular HQAM is formed similar to SQAM formation. The same bit mapping is adopted for regular HQAM constellation as preferred by SQAM constellation. Further, irregular HQAM constellation is formed with a little shift in constellation points of regular HQAM on I/Q plane. For the matched points in irregular HQAM constellation, the same bit mapping is preferred as in regular HQAM constellation. Finally, for the remaining points, a sub-optimum bit mapping is performed by considering that  $s_k$ consists of the NNs with minimum $G_P$ from the available signaling points. Thus, the $G_P$ for irregular 64-HQAM is 1.35.
	Finally, substituting $G_P$ and SER in (\ref{BER}), approximate BER expression for HQAM can be obtained.
	As the $G_P$ of regular and irregular HQAM constellations are always greater than 1, the bit error performance of HQAM constellations is always poor as compared to the symbol error performance because $G_P$ of SQAM constellations is always 1.
	%

	\subsection{Comparison of Irregular HQAM Constellation with Other QAM Constellations}

	A comparative analysis between the irregular HQAM and other QAM constellations for different constellation orders is illustrated through Table \ref{tab_comp}. Here, a fixed Euclidean distance of $2d$ is maintained between the neighborhood constellation points for all the constellations and their average constellation energy or power and PAPR are calculated. From the  Table \ref{tab_comp}, it can be concluded that for all the constellation orders (even as well as odd bits constellations), irregular HQAM has the reduced peak and average energies or powers. Hence, irregular HQAM can be concluded as the optimum constellation which provides better ASER performance than the other constellations.  For various cooperative relay networks, this can be verified in \cite{singya2018impact,singya2019performance,singya2018performance1,parvez2019impact}.
	
	\section{Probabilistic Shaping}
	%
	 The development of optical transport networks is pushed forward due to the growing demands of Internet traffic in past decades. Advanced modulation schemes (specially the family of QAMs) are adopted for better spectral efficiency; and various multiplexing techniques are implemented for improved capacity. It is observed, for transmission and detection in the modern day optical networks, usually uniform QAM constellations are used which results in a loss of around 1.53 dB gain towards the Shannon limit \cite{qu2019probabilistic}. However, the Shannon limit can be achieved by implementing shaping and coding techniques.
	 Shaping gain can be achieved through probabilistic and geometric amplitude shaping. In uniform QAM constellation, all the constellation points are equiprobable and equidistant on the 2D plain. However, in geometrically shaped QAM, constellation points are non-equidistant on 2D plain, whereas in probabilistically shaped QAM, constellation points are non-equiprobable on 2D plain. In this work, all the constellations discussed in the previous sections are geometrically shaped. However, we can further improve their performance by probabilistic shaping. 	 
	Considerable works on the probabilistically shaped QAM constellations are observed in the literature. Works on the comparative study of both the probabilistically  and geometrically shaped QAM constellations are presented below.
	  
In \cite{yankov2014constellation,fehenberger2016sensitivity}, probabilistically shaped QAM constellations are compared with the uniform QAM constellations for the optical communication system. 
	In \cite{bocherer2015bandwidth}, for equidistant constellation points, probabilistic amplitude shaping is applied to bipolar ASK which is applied to SQAM. In \cite{fehenberger2016probabilistic}, for  non-linear fiber channel, probabilistic shaping of QAM is performed.
	In \cite{steiner2017comparison}, for AWGN channel, both the geometric and probabilistic shaping are studied.
	In \cite{zhang2018constellation},  generalized pairwise optimization scheme is reviewed for geometrically and probabilistically shaped constellations. 
	In \cite{qu2018hybrid,qu2019two,qu2019probabilistic}, comparison to geometrically shaped QAM; the probabilistically shaped QAM is presented for optical communication system.	
	Most of the works focus on the probabilistic and geometric shaping in optical communication system, however, this can also be implemented in wireless systems for improved power efficiency along with the spectral efficiency.
		
\section{Application of QAM Constellations in Wireless Communication Systems}
	%
	The QAM  was first proposed by C. R. Cahn \cite{cahn1960combined}  in 1960 and later extended by Hancock and Lucy in \cite{hancock1960performance}, who termed Type I to the Cahn's constellation and Type II to their proposed constellation. 
	Later in 1962, Campopiano and Glazer  proposed  properly organized square QAM constellation and denoted it as Type III constellation \cite{campopiano1962coherent}. 
	There were some initial works \cite{salz1971data,simon1973hexagonal,foschini1973selection,foschini1974optimization,thomas1974digital,smith1975odd}, where various QAM constellations were proposed. 
Since then, such constellations have received tremendous attention for high data-rates in various fields of communication with different applications. These areas of communication with applications are described below in details.

Further, except for some initial works on HQAM or TQAM constellations \cite{simon1973hexagonal,foschini1974optimization,thomas1974digital,forney1984efficient,dong1999signaling}, significant research attention in BER/SER analysis of HQAM or TQAM constellations is reported recently in \cite{park2007bit, park2007triangular, park2008irregularly, park2009odd, park2012performance, tanahashi2009multilevel, lee2012error, rugini2016symbol, markiewicz2017construction, abdelaziz2018triangular, kumar2018aser, singya2018impact, singya2018performance1, sadhwani2018, sadhwani2018IET, singya2019performance, parvez2019impact, shaik2019performance, garg2019performance, Parvez_D2D_Access_2019,  Shaik_SPIE2020_UVC_R452}.
	In \cite{park2007bit,park2007triangular,park2008irregularly,park2009odd,park2012performance},  Park has presented an extensive research for various regular and irregular TQAM (or HQAM) constellations with different constellation orders. In these works, various TQAM constellations with their peak and average powers, bit mapping, detection regions, etc. are presented separately.
	In \cite{tanahashi2009multilevel}, for non-power of 2 HQAM constellations, multilevel coded modulation scheme is presented.
	In \cite{pappi2010theta}, performance of $\theta$-QAM family (including SQAM and TQAM) is analyzed in both the AWGN and Nakagami-m fading channels.
	In \cite{rugini2016symbol}, a non-relay system is considered and  SEP of HQAM over Rayleigh fading channel is derived. 
	For a multi-relay network over independent and non-identically distributed (i.n.i.d.) Nakagami-m fading links, ASER performance of  various signal constellations such as HQAM, RQAM, XQAM, De-QPSK,  and $\pi/4$-QPSK are analyzed in \cite{kumar2018aser}.
	In \cite{singya2018impact}, impact of imperfect CSI is observed on the performance of amplify-and-forward (AF) relay system over i.n.i.d. Nakagami-m fading links and closed-form expressions of outage probability and ASER of HQAM and RQAM constellations are derived.
	In \cite {abdelaziz2018triangular}, Abdelaziz and Gulliver have examined both power of 2 and non-power of 2 constellations. Further, BER performance with detection complexities of regular, irregular, and sub-optimum TQAM constellations are analyzed.
	Work done in \cite{kumar2018aser, singya2018impact} is extended in \cite{singya2019performance}, where a multi-relay AF system over i.n.i.d. Nakagami-m fading channels with both integer and non-integer fading parameters is considered in presence of imperfect CSI and NLPA at the relay, and analytical expressions of 	outage probability, asymptotic outage probability, and ASER of HQAM, RQAM, and XQAM constellations are derived.

The major areas of applications for various QAM constellations are:

\begin{itemize}
	\item Mobile communications
	\item Broadcasting-cable, satellite, microwave
	\item Spatial diversity-multicarrier-(OFDM MIMO), cooperative relaying
	\item Optical wireless communications
	\item Miscellaneous
\end{itemize}	
\subsection{Mobile Communications}
%

Due to increase in the number of wireless devices, the requirements of voice, data, and video applications are increasing day-by-day. However, to access the limited spectrum, efficient spectrum utilization techniques are required. At the same time, Internet and other data related wireless applications requires very high data-rates. Spectrally efficient high data-rate can be achieved through adaptive digital modulation schemes at an affordable cost. Hence, research is directed towards various spectrally and power efficient compact 2D constellations. In this subsection, we have explored a detailed survey on various signal constellations, adopted in mobile communications from the early days till now.

In \cite{salz1971data}, authors presented performance of a digital data modem transmission, employing combined amplitude and phase modulation.
In \cite{dupuis197916}, authors experimentally implemented the BER of  4-PSK, 8-PSK, and 16-QAM, for a bit-rate of 140 Mbps by considering high capacity digital radio-relay system.
In \cite{komaki1979characteristics}, authors investigated the impact of multipath fading on a 200 Mbps 16-QAM digital radio system experimentally. 
In \cite{oetting1979comparison}, Oetting compared various constellations up-to 16 constellation orders over AWGN channel which were applicable to digital radio systems.
In \cite{prabhu1981cochannel}, Prabhu determined the co-channel interference immunity of high capacity QAM constellations  for digital radio applications.
In\cite{giger1981effects}, the authors  analyzed the impact of multipath fading over a high-speed digital radio systems operating in $4-11$ GHz band by considering offset 4-PSK, conventional 8-PSK and 16-QAM constellations.
In \cite{Joindot1981BasebandAE}, authors concluded that the decision feedback equalizer with few taps is efficient in the presence of multipath fading and constructed a 140 Mb/s 16-QAM modem. 
In \cite{hill1983performance},  Hill et al. presented the  performance of nonlinearly amplified (NLA) 64-QAM and  the sensitivity over group delays and amplitude distortions was discussed for digital radio systems. 
In \cite{feherdigital, noguti19836ghz, saito1984feasibility, mcnicol1984design}, authors presented their works on QAM constellations in digital radio systems.
In \cite{benveniste1984blind}, authors proposed a blind equalization technique and verified its performance of one and two carrier systems with QAM and V29 constellations.
In \cite{forney1984efficient}, Forney et al. presented a detailed survey on various power efficient RQAM and HQAM constellations for band limited channels like telephone channels. 	
In \cite{steele1985towards}, authors proposed a futuristic mobile communication system based on the integration of a high-capacity digital cellular mobile radio system with a packet-switched routing network and discussed various modulation formats including QAM.
In \cite{borgne1985comparison}, authors presents a comparative study of 16, 32, 64, and 128-QAM  in a digital radio system environment by considering the effects of filtering, interference, amplifier non-linearities, and selective fading.
In \cite{wu1985256}, authors analyzed the performance of 256-QAM over distorted channels by considering  amplitude distortions such as linear, parabolic, and  sinusoidal characteristics along with group delay.
In \cite{mathiopoulos1986performance}, authors analyzed the performance of 512-QAM by considering the effects of linear channel distortions.
In \cite{shafi1986further}, performance of transversal and decision feedback equalizers for 16-QAM and 64-QAM digital radio systems with tap spacing  was analyzed.
In 1987, Sundberg et al. \cite{sundberg1987logarithmic,steele1987transmission} considered SQAM constellation for voice communication over a Rayleigh fading channel. This was the first major consideration of QAM in mobile radio applications.
In \cite{nakamura1987256}, authors described the  performance of a 400 Mbps transmission capacity 256-QAM modem  and  introduced some novel approaches such as forward-error-correction (FEC) to reduce residual errors, good phase jitter performance and no false lock phenomenon to attain good system performance. 
In \cite{rustako1987using}, authors investigated the performance of $M$-ary QAM by using times-four method of carrier recovery in a digital radio link. 
In \cite{pupolin1987performance}, authors evaluated the effects of transmit amplifier non-linearities on digital radio link performance by considering 256-level constellations. 
In \cite{andrisano1987adaptive}, authors presented the performance of multilevel QAM radio system over selective fading by considering  adaptive baseband equalization to reduce the signatures.
In \cite{vogel1987orthogonal}, authors proposed a progressive filter concept for co-channel transmission with negligible interference employing 16-QAM with a 0.19 roll-off  at 140 Mbps rate.
In \cite{chamberlin1987design}, technical and design requirements for 256-QAM data in voice modems such as signal processing for implementation, error correction codec etc. were described.
In\cite{collins1987application}, authors reviewed the importance of coded modulation for a 1.544 Mbps data-rate in voice modems employing 256-QAM.
In \cite{duvoisin1987operation}, authors presented the performance of a dual polarized $M$-ary QAM up-to 64 levels by considering coherence between cross talk and the desired signal.
In \cite{sebald1987advanced}, authors proposed an equalization technique for multilevel QAM systems which eliminates linear distortions with easy implementation at high data-rates. 
In \cite{andrisano1987adaptive}, authors presented the performance of multilevel QAM radio system in selective fading by considering  adaptive baseband equalization to reduce the signatures.
In \cite{borgne1987new}, performance of dual 16-QAM system over dispersive radio channels was analyzed by considering the cross-polarization interference cancelers. 
In \cite{baccetti1987full}, analysis of 140 Mbps 64-QAM is performed by considering the digital adaptive equalization technique.
In \cite{agusti1987performance}, comparative study of 16, 32, 64, 128-QAM constellations at 140 Mbps by considering  both linear and non-linear equalizers was presented.
In \cite{matsue1987digitalized}, highly precise cross-polarization interference canceler was developed by using digital transversal filter and performance of 256-QAM at 12.5 Mbps was examined.
In \cite{VUCETIC198869}, authors performed a simulation case study of $M$-ary QAM constellations for a high capacity digital radio link, subjected to frequency selective fading with different types of channel encoding.
In \cite{carpine1988comparative}, authors presented the performance of 16-QAM, 64-QAM, 49-QPR, and 81-QPR over a 45 Mbps line-of-sight digital radio systems. 
In \cite{chuang1989effects}, authors investigated the impact of delay spread over non-linearly filtered QAM. 

In the initial studies, the decision regions for most of the signal constellations were straight boundaries (including some special cases of QAM constellations), and exact and simple BER expressions were obtained. However, more compact 2D signal constellations such as family of QAM do not have boundaries with perpendicular angles. For such constellations exact BER expressions are not simple and consists of special functions during the integral calculation. Considering this point in mind, Craig \cite{craig1991new} presented simple and exact BER expressions with polygonal decision boundaries for such more practical signal constellations. This approach was also considered in \cite{simon1995digital}.
In \cite{hanzo1990subband}, authors examined the performance of 16-QAM by using sub-band coded speech and BCH error correcting codes. 
In \cite{webb199016}, authors presented the differential encoding technique and circular star QAM constellation mapping to improve the BER performance.
In \cite{webb1992qam}, author discussed about the application of variable rate square and star QAM constellations for microcellular network based mobile radio applications and the hardware development of a QAM modem.
In \cite{adachi1992performance}, authors examined the BER performance of various 16-ary QAM constellations and observed that 16-SQAM outperforms 16-star QAM.
In \cite{webb1992modulation}, author presented the details of best suitable QAM constellations for micro-cellular and macro-cellular personal communication systems. 
In \cite{sunaga1993performance}, authors proposed a post-detection maximum ratio combining (MRC) for space diversity in digital land mobile communications by considering $M$-ary QAM constellations over Rayleigh fading channels.
In \cite{sampei1993rayleigh}, authors examined the performance of $M$-ary QAM constellations by considering pilot symbol aided scheme to compensate Rayleigh fading for land mobile communications.
In \cite{du1993new}, authors proposed a trellis code based 32-QAM constellation to improve the BER performance of the system for fading channels. 
In \cite{jalloul1994performance}, authors analyzed the performance of DS/CDMA system over multipath fading channel with RAKE receiver by employing non-coherent $M$-ary orthogonal modulation.
In \cite{aghamohammadi1990error, shayesteh1995error}, considering the amplitude and phase errors, performance of various coherently modulated non-constant envelope like $M$-ary RQAM constellations were studied over the AWGN and various fading channels.
In \cite{otsuki1995square}, authors proposed an SQAM constellation for land mobile communications to achieve high data-rate with high QoS for flat and selective fading channels.
In \cite{webb1995variable}, authors investigated the performance of  adaptive QAM constellation in the presence of co-channel interference.
In \cite{ue1995symbol}, authors proposed a system to control both modulation level and SER according to channel conditions and monitor the channel condition based on the delay profile measurement.
In \cite{torrance1996optimisation},  authors optimized the switching levels for BPSK, QPSK, 16-QAM, and 64-QAM based on received signal strength.
In \cite{goldsmith1998adaptive}, a method of superimposing the closet codes  with adaptive $M$-ary QAM over fading channels was proposed.
In \cite{lu1998error}, authors investigated the performance of L-branch diversity reception system and the SER expressions of $M$-ary QAM over Rayleigh fading and AWGN channels were derived.
In \cite{alouini1999unified}, analytical exact ASER expressions for different constellations over various fading channels were derived.
In \cite{dong1999signaling}, Dong et al. extended the work done by Craig, and  performance of various coherent 8 and 16-ary signal constellations was analyzed for both the AWGN and fading channels. 
In \cite{lau1999variable}, authors employed non-coherent orthogonal $M$-ary modulation with RAKE receiver for CDMA over fast fading and combined fast fading, shadowing and power control channels. Further, authors determine the BER of the adaptive signaling constellations from $2^2$ to $2^9$ symbols.
In \cite{wong1999upper}, authors mitigated the effects of wideband multipath Rayleigh fading channel with a decision feedback equalizer for adaptive modulation and obtained the BER and bits per symbol performance of BPSK, QPSK, 16-QAM, and 64-QAM constellations. 
In  \cite{alouini1999adaptive}, authors proposed adaptive modulation schemes for simultaneous transmission of voice and data transmission over Nakagami-m fading channels. 
In \cite{qiu1999performance}, authors illustrated the impact of power control over adaptive modulation schemes in presence of interference in a multi-user environment. 
In \cite{torrance1999interference}, authors investigated the performance of adaptive modulation over slow fading channels by considering co-channel interference.
In \cite{tang1999effect}, authors analyzed the performance of adaptive $M$-ary QAM techniques by considering channel estimation errors over Rayleigh fading channel.
In \cite{yang2000recursive}, authors proposed the recursive algorithm for $M$-ary QAM constellations to obtain the accurate BER over AWGN channels. 
In \cite{hole2000adaptive}, authors introduced the adaptive modulation over Nakagami-m multipath fading channels and determined the average spectral efficiency for 2L-dimensional trellis codes. 
In \cite{ormeci2001adaptive}, authors introduced adaptive bit-interleaved coded  modulation to combat the estimation errors.
In \cite{blogh2001dynamic}, authors analyzed cellular network scenario by considering adaptive antennas based on fixed and channel allocation schemes.
In \cite{chung2001degrees}, authors determined the adaptive modulation schemes under constraints of data-rate, transmit power, and instantaneous BER to maximize the spectral efficiency.
In \cite{cho2002general}, authors presented the generalized expression for 1D and 2D amplitude modulations schemes encoded with gray mapping.
In \cite{wang2002soft}, authors derived the generalized log likelihood ratio for soft decision channel decoding for QAM constellations.
In \cite{jafar2003adaptive}, authors analyzed the performance of multi-rate CDMA system and determined the optimal adaptive rate and power allocation strategies  to maximize the throughput of the system.
In \cite{choi2003optimum}, authors proposed the optimal modulation-mode switching levels for the various PSK and QAM constellations and their BER performance were analyzed.
In \cite{vishwanath2003adaptive}, authors analyzed the performance of a turbo coded system and optimize the adaptive turbo codes and transmitter power. Further, BER performance under average power constraint was obtained.
In \cite{baum2003performance}, authors analyzed the performance of a cellular system with different link adaptation strategies by considering practical constraints and proposed a hybrid link to overcome losses due to the practical constraints.
%
In \cite{falahati2004adaptive}, authors optimized the adaptive modulation schemes assisted with channel prediction of Rayleigh fading channels for uncoded $M$-ary QAM.
In \cite{1268365}, authors proposed  a soft decision equalization techniques with low complexity and better error performance for MIMO frequency selective channels and BER of QAM constellations was evaluated. 
In \cite{oien2004impact}, authors analyzed the impact of CSI prediction employing the linear fading predictor over trellis coded Rayleigh fading channel with adaptive modulation and  closed-form expressions of BER and spectral efficiency were obtained.
In \cite{cao2004exact}, authors derived the BER expressions of $M$-ary QAM with pilot symbol assisted channel estimation of static and Rayleigh fading channels by considering the impact of noise and the estimator de-correlation.
In \cite{cai2005adaptive}, authors analyzed the effect of linear MMSE channel estimation and channel prediction errors on BER and developed the adaptive pilot symbol assisted modulation schemes to maximize the spectral efficiency. 
In \cite{yoo2006throughput}, authors developed a framework for single user throughput maximization by optimizing the symbol rate, packet length, and M-QAM constellation size over AWGN and Rayleigh fading channels.
In \cite{lozano2006optimum}, authors proposed an optimum power allocation policy for the discrete signaling constellations limited in PAPR to maximize the mutual information with arbitrary input distributions.
In \cite{song2006adaptive}, authors analyzed the bit-interleaved coded OFDM packet transmission with adaptive modulation and determined the pairwise error probability over slow fading channels.
In \cite{conti2007slow}, authors investigated the performance of a slow adaptive modulation scheme employing $M$-ary QAM constellations with antenna diversity.
In \cite{meshkati2007game}, authors analyzed the impact of constellation size of an $M$-ary QAM constellations, and the impact of trellis coded modulation on energy efficiency of wireless network through game theoretic approach was analyzed.
Further, a number of research works on the BER/SER performance of various QAM constellations in wireless communication systems, operating under different fading environments have been addressed in \cite{vitthaladevuni2001ber,dixit2012symbol, dixit2012performance,dixit2014performance,pena2015performance,dixit2017performance}.

	%
	 \subsection{Cable, Satellite, and Microwave Communications}
	 %
	 In a limited frequency band, the high data transmission through satellite leads towards the use of highly spectral efficient modulation schemes in satellite communication. Further, digital microwave radio system is an efficient way of transmission when digitizing a communication system.  In designing a microwave radio system, RF spectrum must be used in an efficient way, which can be done through adaptive modulation scheme. Hence, in this subsection, a survey on various modulation schemes used in satellite and microwave radio communications is presented.
	
	 For the very first time, Thomas et al. \cite{thomas1974digital} mentioned QAM application in satellite communication and considered the non-linear distortion effect due to the use of traveling wave tube amplifier in transponder.
	 In \cite{miyauchi1976new}, authors implemented multi-level QAM constellations using PSK modems for realizing high baud-rate multi-level transmission for high speed digital-microwave, millimeter-wave, and satellite communication system. 
	 In \cite{horikawa1979design}, authors proposed a long-haul 16-QAM digital radio system with 200 Mbps in 40 MHz bandwidth as an application for satellite and  microwave links.
	 In \cite{prabhu1980detection}, Prabhu evaluated the performance of 16-QAM in presence of co-channel interference over 8-PSK and 16-PSK by developing a theory in evaluating the error probabilities as an application for terrestrial and satellite communications. 
	 In \cite{yamamoto1981advanced}, from spectral efficiency point of view, 16-SQAM constellation was adopted for digital microwave radio system operating in 4-5 GHz band.
	 For satellite communication, a turning point came in 1982, when K. Feher proposed a new method of SQAM generation with the use of nonlinear amplifiers which was termed as NLA-QAM \cite{morais1982nla}, where 16 and 64-ary NLA-QAM constellations were proposed.  
	 In \cite{feher1983digital}, authors presented the work on digital communications in satellite communications.
	 In \cite{foschini1985equalizing}, author proposed the equalization techniques for digital terrestrial radio systems employing QAM techniques by tap adaptation.
	 In \cite{feher19871024} author addressed FEC coded 1024-QAM modems, staggered 1024-QAM, and 256-QAM modems for microwave and cable system application  for the first time, where practical constraints of phase noise, amplitude distortions, and group delays were considered.
	 In \cite{sinnreich1987system}, authors considered 64-256 QAM constellations to calculate the system parameters for high speed data transmission to deploy 1.54 Mbps modems in microwave radio.
	 In \cite{choi2002predicting}, authors analyzed the satellite channels by considering the weather induced impairments such as scintillation and rain attenuation, and developed predictors using autoregressive models to improve the efficiency along with the adaptive modulation.
	 In \cite{diaz2007non}, authors investigated the performance of non-linear precoding to mitigate the interference in the broadband MIMO satellite communications and performed BER of uncoded BPSK and QAM constellations.
	 Further, in \cite{etsi1998,  Dirk, zhang2010exact, reimers2013dvb, ANSI/SCTE072013, yang2014star, DVB-C}, QAM constellations were standardized for digital signal broadcasting over cable and satellite links. 
	 %
	\subsection{Spatial Diversity:} Multipath fading is one of the prominent constraint which severely degrades the reliability, robustness, and coverage of a wireless communication system. To combat this, spatial diversity is one of the solutions which improves the performance of a wireless communication system. Spatial diversity can be achieved through multi-carrier system by
	deploying multiple	antennas at the communication nodes. A communication system with multiple antennas at the transmitter and receiver is referred to as MIMO system. In MIMO systems, multiple paths between the source and destination are established with the help of multiple antennas to improve  diversity of the communication system.
	However, multiple antennas at each
	node are not always possible due to hardware limitations and device size (such as in mobile handsets and sensor networks). In
	such cases, cooperative communication is preferred as an alternative to improve the spatial diversity of the communication system. 
	Next, we will study the performance of such constellations.

		\subsubsection{Multi-carrier System-(OFDM, MIMO)}
	Multicarrier systems are explored by using OFDM and MIMO systems. A detailed study of OFDM systems is given in \cite{hwang2008ofdm,molisch2012wireless,singya2017mitigating}. Various applications employing OFDM techniques are listed as below:
		
	\paragraph{OFDM}
	In \cite{hirosaki1981orthogonally}, author proposed DFT based orthogonally multiplexed QAM system for multichannel system. 
	In \cite{saito1984feasibility}, authors investigate the multi-carrier system and the application of higher-order QAM constellations. 
	In \cite{robing2001construction}, authors realized the 16-QAM from scaled 4-PSK signals to reduce the peak-to-average  envelope power ratio employing Golay sequences for OFDM systems.
	In \cite{song2006adaptive}, authors analyzed the bit-interleaved coded OFDM packet transmission with adaptive modulation and determined the pairwise error probability of the system over slow fading channels.
	In \cite{marques2006optimizing}, authors developed loading algorithms to minimize the transmit power of OFDM system with rate and error probability constraints under three different CSI scenarios for QAM constellations. 
	In \cite{taha2007low}, authors presented the Golay code based 16-star QAM constellation constructed from sum of two scaled QPSK to reduce the peak-to-mean envelope power for OFDM systems.
	In \cite{miao2010energy}, authors investigated OFDM system with adaptive transmission for maximizing energy efficiency with transmit power and its allocation constraint.

\paragraph{MIMO}	
	In \cite{annamalai1999exact}, authors investigated the performance of $M$-ary QAM techniques by considering the L-fold antenna diversity over Nakagami-m fading channels with MRC receiver for independent and correlated channels and equal gain combining receiver for independent channels. 
	In \cite{catreux2001attainable}, authors examined the performance of MIMO system with adaptive modulation and adaptive array processing at the receiver in interference limited cellular systems. 
	In \cite{catreux2002data}, authors developed a framework for mean spectral efficiency by considering mean SNR at the cell boundary, propagation exponent, coding technique and the number of antennas. Further, authors highlighted the potential benefits of adaptive modulation with multiple transmissions in terms of spectral efficiency.
	In \cite{xia2004adaptive}, authors illustrated the impact of partial CSI over adaptive MIMO OFDM QAM mapped symbols with 2D space-time coder beamformer.
	In \cite{zhou2004adaptive}, authors investigated the impact of imperfect CSI and outdated CSI over a multi-antenna system with adaptive modulation.
	In \cite{falahati2004adaptive}, authors optimized the adaptive modulation schemes assisted with channel prediction of Rayleigh fading channels for uncoded $M$-ary QAM.
	In \cite{zhou2004accurate}, authors presented a pilot symbol assisted modulation channel predictor for MIMO Rayleigh fading channels, the impact of prediction errors on BER performance was analyzed. 
	In \cite{zhou2004adaptive}, authors investigated the impact of imperfect and outdated CSI on a multi-antenna system based on adaptive modulation.
	In \cite{windpassinger2004precoding}, authors examined the multi-user multi-antenna scenario and emphasized the advantages of non-linear pre-equalization, and SER analysis was performed for QAM constellations.
	In \cite{zhou2005mimo}, authors investigated the impact of both the perfect and imperfect CSI at both the transmitter and receiver under average transmit power and instantaneous constraint for MIMO systems. 
	In \cite{zhang2005efficient}, authors proposed adaptive resource allocation based on subcarrier allocation, power, and bit distribution for multi-user MIMO/OFDM systems based on channel conditions.
	In \cite{xia2005multiantenna}, authors optimized the performance of a multi-antenna system with bandwidth limited feedback link based on adaptive modulation and transmit beamforming.
	In \cite{palomar2005designing}, authors analyzed a point-to-point MIMO system with perfect CSI at transmitter and receiver, and optimized the constellation selection and linear transceiver based on BER. 
	In \cite{xia2005analytical}, authors investigated the multicode CDMA systems with interference cancellers to support high data-rate communications and BER of $M$-ary QAM was obtained.
	In \cite{ko2006orthogonal}, authors analyzed the performance of rate-adaptive $M$-ary QAM constellations with orthogonal space time block codes (STBC) and derive the closed-form BER expressions.
	In \cite{vicario2006cross}, authors investigated cross-layer approach for  transmit antenna selection to maximize the throughput of the system and also adaptive modulation to improve systems performance. 
	In \cite{yang2007optimal}, authors investigated half-duplex non-orthogonal AF MIMO multi-relay network with STBC and performed diversity multiplexing trade-off and frame-error-rate  analysis employing QAM constellations. 
	In \cite{haleem2007opportunistic}, authors developed a framework for opportunistic encryption to maximize the throughput with desired security constraints based on channel characteristics. Further, authors optimized the encryption, modulation, and employed FEC codes to combat bit errors. 
	In \cite{ding2007amplify}, authors investigated the performance of a half-duplex cooperative MIMO system and asymptotic closed-form pairwise error probability expression was derived. 
	In \cite{yu2008performance}, authors investigated the performance of space time codes based MIMO system over Rayleigh fading channels with adaptive modulation by considering imperfections in channel estimation. 
	In \cite{ho2008optimal}, authors investigated QAM mapping based multi-user MIMO OFDM system and optimized transmit power with user rate constraint.
	In \cite{chae2008mimo}, authors investigated multi-user MIMO relay cellular system with adaptive modulation and derived the upper and lower bounds for achievable sum rate. 
	In \cite{jeganathan2009space}, authors introduced space shift keying (SSK)  employing QAM mapping for MIMO systems, and analyzed the system performance under spatial correlation and estimation errors. 
	In \cite{lee2010joint}, authors investigated the MIMO half-duplex AF one and two-way relay systems and proposed algorithms to optimize linear transceivers at both source and relay nodes with sum-rate and mean-square error constraints, and determined the BER of QAM constellations. 
	%
	In \cite{shen2011best}, authors presented a VLSI architecture for MIMO decoder which supports $ 4\times 4$ antenna configuration with 64-QAM based on  depth-first and breadth-first approach and proposed an algorithm which reduces complexity for both the soft and hard decoding. 
	In \cite{datta2011hybrid}, authors proposed a hybrid algorithm combining technique using reactive tabu search algorithm and belief propagation algorithm to improve the performance of large MIMO detection with higher order QAM constellations.
	In \cite{sugiura2011reduced}, authors proposed a near optimum detection algorithm for coherently detected space time shift keying techniques with L-point PSK and QAM constellations. 
	In \cite{zhou2004adaptive, sadek2006multinode, 4099545, kim2008performance,ma2009error, lee2009performance, kim2009bit, romero2009performance, chen2009error, chen2009performance,kim2010new, suraweera2010amplify,kim2010performance, wang2011power, song2011performance, liu2011exact,  suraweera2011amplify, 5706356, 5682209,  6129378, bansal2012low,eraslan2013performance, arti2014maximal,6679187, 6987541, song2017designs, al2017performance}, authors presented the works considering multiple antennas in various system models including multi-node communication and performed SER analysis of QAM constellations.
	In \cite{yeoh2012unified}, authors presented a unified asymptotic framework for transmit antenna selection (TAS)-MIMO multi-relay network with various fading channels and derived the closed-form expressions of the outage probability and SER of $M$-ary PSK and $M$-ary QAM constellations.  
	In \cite{yang2016transmit}, authors proposed the TAS algorithms and obtained the BER performance of the QAM constellation.
	In \cite{siohan2002analysis}, authors presented an analysis of OFDM with offset QAM.
	In \cite{xia2004adaptive}, authors presented the impact of  partial CSI over adaptive MIMO OFDM QAM mapped symbols with 2D space-time coder-beamformer.
	In \cite{zhang2005efficient}, authors proposed the adaptive resource allocation based on subcarrier allocation, power and bit distribution for multi-user MIMO/OFDM systems based on channel conditions.
	In \cite{ho2008optimal}, authors investigated QAM mapping based multi-user MIMO OFDM system and optimized transmit power with user rate constraint.
	In \cite{mesleh2008spatial}, authors developed a framework for OFDM based spatial modulation over Rayleigh fading channels and derived the SER expressions for QAM constellations.
	In \cite{hong2014frequency}, authors analyzed an OFDMA based cellular downlink by considering non-Gaussian intercell interference  to  improve the transmission rates of cell edge users, and proposed new modulation scheme frequency and QAM by combining FSK and QAM constellations.
	In \cite{ha2018machine}, authors analyzed the performance of a MIMO OFDM system with adaptive modulation based on machine learning as an application to 5G new radio systems.
	%
	%
	In \cite{tarokh1999space}, authors introduced STBC using multiple transmit and receiver antennas. Further, the authors determined the maximum diversity order attained and also maximum achievable code rate for complex constellations such as QAM and PAM.
		In \cite{liu2006error}, authors investigated the STBC diversity systems over Rayleigh fading channels and SEP of RQAM constellations was obtained.
	In \cite{han2009performance}, authors analyzed the performance of half-duplex two-way AF relaying system by considering the Alamouti coding and derived the lower and upper bounds for average sum-rate and pairwise error probability by considering QAM constellations.
	In \cite{yang2012semidefinite}, authors investigated a MIMO system and proposed a virtual antipodal detection scheme for gray coded higher order RQAM based on semi-definite programming relaxation.
		In \cite{wang2004orthogonal, yang2006performance, bansal2012decoding, kulkarni2014performance}, authors analyzed STBC systems by considering QAM constellations.
	%
	In \cite{wiesel2005semidefinite, sidiropoulos2006semidefinite, yang2007mimo}, authors proposed semi-definite relaxation (SDR) approach for MIMO detection with higher order QAM constellations.
	In \cite{mao2007semidefinite}, for $M$-ary QAM constellations based systems,  a semi-definite programming  relaxation (SDPR) approach is proposed to combat the multiuser detection problems.
	In \cite{mobasher2007near}, authors proposed SDR based quasi-maximum likelihood algorithm for MIMO systems and introduced several relaxation models with increasing complexity and performed the SER analysis for QAM and PSK constellations. 
	In \cite{ma2009equivalence}, authors investigated the performance of three different SDR such as polynomial-inspired SDR, bound-constrained SDR, and virtually antipodal SDR over MIMO channels for BPSK, QPSK, and higher order QAM constellations. 
	%
	\subsubsection{Cooperative Relaying}
	%
	Cooperative relaying is one of the prominent techniques to achieve spatial diversity in wireless communication. Cooperative relaying is extensively explored because of its capability to provide improved capacity, coverage, and power and spectral efficiency due to the diversity gain without using multiple antennas at different communication nodes. Hence, it is also considered as a virtual MIMO system. A detailed study of cooperative relaying can be seen in \cite{laneman2004cooperative,liu2009cooperative}. In this subsection, we have explored various works on the SER or BER performance of different cooperative relaying networks in various fading environments.
	
	In \cite{annamalai2001error}, authors investigated single and multi-channel communication, and presented the BER analysis for $M$-ary constellations over Nakagami-m and Rayleigh fading channels.
	In \cite{win2001virtual}, authors investigated the SER of $M$-ary PSK and QAM constellations over Nakagami-m fading channels for hybrid selection/MRC diversity.
		In \cite{cao2004exact}, authors investigated the effect of estimator de-correlation on the BER performance of $M$-ary QAM constellations over Rayleigh fading channels.
	In \cite{nabar2004fading}, authors investigated the performance of cooperative relay system for three different time division multiple access transmission protocols and determined the SER of 4-QAM over fading channels.
	In \cite{annamalai2005general}, authors derived the average error rates of various $M$-ary constellations and outage probability using characteristic function based approach over various fading channels.
	In \cite{ribeiro2005symbol}, authors analyzed cooperative diversity by considering arbitrary cooperative branches with arbitrary hops per branch over various fading channels and derived the SEP for various constellations including QAM.
	In \cite{su2005ser}, authors derived the closed-form expressions of QPSK and QAM for a cooperative systems and optimized the power allocation.
	In \cite{zhao2006symbol}, author analyzed the AF cooperative multi-relay system employing selection combining and obtained the SER analysis of QAM constellations.
	In \cite{sadek2006multinode}, authors investigated multi-node cooperative relay system and derived the SER expressions for M-PSK and M-QAM constellations and optimized the power allocation.
	In \cite{le2007multihop}, authors investigated the performance of multi-hop cellular network and proposed the routing and power allocation algorithms to maximize the throughput and BER is evaluated for QAM constellations.
	In \cite{ma2007effect}, authors analyzed the impact of channel estimation error on the BER performance of RQAM and SQAM constellations.
	In \cite{ikki2007performance}, authors analyzed the performance of cooperative multi-relay system over i.n.i.d. Nakagami-m fading channels and derived the outage probability and error rate employing QAM constellations.
	In \cite{su2008cooperative}, authors investigated the performance of cooperative relay system with DF and AF relaying protocols and analyzed the SER for $M$-ary QAM and PSK constellations with optimum power allocation.  
	In \cite{ma2008bit}, authors investigated the selection detection and forward  and AF half-duplex OFDM relay systems and proposed margin adaptive bit and power loading techniques to minimize the transmit-power consumption, and determined the SER of QAM constellations.
	In \cite{zhou2008energy}, authors investigated the performance of clustered relay network based on STBC and evaluated the packet error rate employing BPSK and 16-QAM constellations. 
	In \cite{maaref2009exact}, authors investigated the single and multi-channel diversity reception over i.n.i.d. Nakagami-m fading channels and average SEP expression of RQAM was derived.
	In \cite{lee2009performance}, authors investigated the SER performance of multi-relay DF system over Nakagami-m fading channels for QAM constellations and optimized the power allocation.
	In \cite{bhatnagar2011decode}, author analyzed  DF based single and multi-relay systems and proposed maximum likelihood  and piece-wise linear decoders for complex-valued unitary and non-unitary transmissions, and performed the SER analysis of $M$-ary QAM and PSK constellations.
	In \cite{amin2011optimal}, authors analyzed the performance of OFDM based cooperative relay system and optimized the BER of QAM techniques with respect data rate based on optimal power, bit, and joint power and bit loading.
	In \cite{bhatnagar2011ml}, authors investigated the decode and forward cooperative system and derived the maximum likelihood decoder for $M$-ary PSK, PAM, and QAM constellations.
	Apart from the works presented above and reference therein, a lot of work is presented in the literature on the BER/SER performance of various constellations for various relay networks over different fading channels till now.
	
\subsection{Optical Wireless Communications}
%

Optical wireless communications (OWC) operates in $350-1550$ nm band, through visible light (VL), infra-red, and ultraviolet communication (UVC) bands; has  gained significant research attention for the future wireless broadband access. OWC is affordable with high data-rates ($30$ Gbps) and virtually infinite bandwidth \cite{ghassemlooy2015emerging, mitra2017precoded}. Various OWC technologies are VL communications (VLC),  free space optical (FSO) communication, optical camera communication, light fidelity, 
and light detection and ranging \cite{chowdhury2018comparative}. OWC operating in VL band  ($390-750$ nm) is known as VLC. FSO is also known as  terrestrial point-to-point OWC and operates near infra-red frequencies ($750 – 1600$ nm) \cite{uysal2014optical}. UVC is an non-line-of-sight (NLoS) communications operates in ultraviolet band  ($200 – 280$ nm). OWC is employed for both indoor and outdoor communications. OWC also finds its potential application in underwater communications \cite{uysal2014optical, zeng2016survey}.

In VLC, achievable high data-rates are limited by the low modulation bandwidth of the GaN-based white light emitting diodes (LEDs) due to the slow response time. Bandwidth can be increased by using blue LEDs at the cost of intensity loss of 10 dB \cite{khalighi2017pam}. Spectral efficiency can be increased by employing MIMO, massive MIMO, and higher order modulation schemes like optical-OFDM (O-OFDM) \cite{mitra2018minimum}. However, MIMO channel are limited by the receiver position, making channel matrix either ill conditioned or rank-deficit which makes  data recovery almost impossible. O-OFDM limited by  Hermitian symmetry constraints,  cyclic prefix, PAPR, and dynamic range of LEDs. As an alternative to O-OFDM, carrier-less amplitude and phase modulation has gained significant interest in optical communications for high data-rate transmissions \cite{carruthers1996multiple}. Bandpass CAP modulation is a QAM modulation where two orthogonal finite impulse response digital filters are employed to achieve it. Various seminal works using CAP modulation schemes to improve the spectral efficiency of the band-limited LEDs are presented in   \cite{vuvcic2010513, ntogari2011combining, khalid20121, hanzo2012wireless, dimitrov2012information, azhar2012gigabit, wang2013demonstration, wu2013performance,   sung2014dimming, wang2014enhanced, sung2014blue, pathak2015visible, huang20151,  karunatilaka2015led,  wang20154,  chun2016led,  jain2018adaptive, zhang2018design}.

In underwater communications, OWC provides very high data-rates and bandwidth when compared with acoustic, RF, and optical wave communications. In underwater OWC, spectral efficiency is increased by employing coherent modulation, where information is transmitted through amplitude, phase, or polarization of the optical field \cite{khalighi2014survey}. Coherent modulation includes $M$-ary PSK, $M$-ary QAM, and multilevel polarization shift keying. In underwater OWC, coherent modulation schemes are very attractive due to high receiver sensitivity, spectral efficiency, and robust to background noise and interfering signals with increased cost and implementation complexity \cite{khalighi2014survey, zeng2016survey}. Some of the notable works which presented the advantages of QAM schemes over intensity-modulation direct-detection systems are \cite{ochi2005basic, cochenour2007phase,  kuschnerov2009dsp, song2013efficient,wan2014adaptive, mizukoshi2014underwater,  oubei20154,  xu2016underwater, zeng2016survey, sun201771, al2017real, saeed2019underwater}.
		
In UV NLoS communications, several notable works were performed employing QAM constellations to improve the spectral efficiency \cite{peppas2010average,  hassan2012subcarrier, hassan2013performance, trung2013performance, trung2013performance1, hassan2014exact,  alheadary2016performance, ardakani2016relay, ai2016pointing, trung2017performance, garg2019aser, garg2019performance, Shaik_SPIE2020_UVC_R452}.
In \cite{ djordjevic2006multilevel,  djordjevic2009communication,djordjevic2010adaptive, niu2011error, tang2012coherent, vu2012performance,  trung2013performance, vu2013bit, trung2014performance,  khalighi2014survey, ai2016af}, authors have presented the application of QAM and adaptive modulation in FSO communications.
%

\subsection{Miscellaneous-Emerging Areas for High Data-rate Communications}
\subsubsection{Machine Learning-5G, Security}
In \cite{yang2005statistical}, authors addressed the application of adaptive modulation schemes for 3G wireless communication systems and developed method to select the modulation and coding based on the channel conditions with minimum frame error rate constraint.
In \cite{haleem2007opportunistic}, authors developed a framework for the opportunistic encryption to maximize the throughput with the desired security constraints based on channel characteristics. Further, authors optimized the encryption, modulation and employ FEC codes to combat bit errors.
In \cite{ermolova2009useful, asghari2009symbol, yu2011error, shi2012some, dixit2013performance, dixit2017performance}, authors investigated the SER performance of different QAM constellations over $\eta-\mu$, $\kappa$-$\mu$, and two wave with diffuse power fading channels.
In \cite{aggarwal2019survey}, author presented a survey on approximation of $\text{Q}$ function and its significance in error probability evaluation. Further, author presented the closed-form SEP of various constellations such as QPSK, QAM, HQAM, and XQAM with approximated Q-functions.
In \cite{li20145g, agiwal2016next, cai2017modulation}, authors presented a survey on 5G new radio and also discussed the application of QAM constellations.
	In \cite{kim2015pre, han2016enhanced, kim2018new, singh2018ser, singh2019semi, singh2019probability, singh2019receivers}, authors presented the work based on filter bank multi-carrier systems using QAM constellations as an application for future wireless communications.
%
%
In \cite{mesleh2008spatial}, authors developed a framework for MIMO systems with spatial modulation schemes over generalized fading channels and computed the average BEP for PSK and QAM constellations.
In \cite{di2011bit}, authors analyzed the performance of MIMO system with SSK and generalized SSK modulation by considering multiple-access interference for PSK and QAM constellations.
In \cite{di2013spatial}, authors presented a state-of-the-art survey on spatial modulation based MIMO system with QAM for the emerging and future wireless communications and discussed the advantages, research challenges, and experimental activities. 
 	In \cite{yang2014star}, authors investigated the performance of star constellation based spatial modulation.
In \cite{narayanan2018wireless}, authors investigated the energy harvesting based distributed spatial modulation and evaluated the error probability of energy recycling DSM with QAM constellations.
In \cite{wen2019survey}, authors presented a survey on spatial modulation schemes and the  application of spatially modulated QAM is presented.
In \cite{6644231, 6423761, 6692187}, authors investigated spatial modulation techniques employing QAM.
\subsubsection{Edge Caching, Energy Harvesting, Full-Duplex Communications, Game Theory}
%
In \cite{cui2005energy}, authors developed a framework to optimize the modulation scheme to be employed by reducing the energy consumption over energy constraint nodes for both coded and uncoded systems. 
In \cite{cui2005joint}, authors investigated sensor network and proposed variable length TDMA schemes to minimize energy consumption for transmission of QAM constellations by optimizing routing, scheduling, and link adaptation strategies.
In \cite{meshkati2007game}, authors analyzed the impact of constellation size of $M$-ary QAM constellations and also the impact of trellis coded modulation on energy efficiency of wireless network through game theoretic approach.
In \cite{zhou2008energy}, authors analyzed the STBC based clustered wireless networks using cooperative communications and performed packet error rate analysis for QAM constellations to minimize the energy consumption with transmit power allocation, number of sensors, and distance between clusters as constraints.
In \cite{cui2007cross, li2009distributed, miao2011low, miao2013energy}, authors investigated the wireless sensor networks and presented various techniques for minimizing the energy consumption by considering QAM transmissions.
In \cite{rodriguez2014performance}, authors investigated the performance of full-duplex AF relay system under residual self interference and derived the pairwise error probability for uncoded systems and BER for coded systems. 
In \cite{hoque2012energy}, authors presented a survey on solutions provided for improving energy efficiency of multimedia based battery operated hand-held devices by considering various modulation schemes for transmission including QAM constellations.
In \cite{zhang2016performance}, authors investigated the performance of spatial modulation based full-duplex two-way relay systems and derived the tight upper bound for average BEP and asymptotic BEP for QAM constellations. 
In \cite{kumbhani2015mgf, zhang2016performance, rajashekar2017generalized, narayanan2017performance, jin2017full}, authors analyzed full-duplex cooperative relay system by considering spatial modulation based QAM.
In \cite{naresh2018performance}, authors investigated the performance of full-duplex relay system and compared the performance of media based modulation schemes with QAM.

\section{Conclusion}

			 As bandwidth and power efficiency are prominent  constraints for 5G and beyond wireless communication systems,  a detailed study of various QAM constellations has been presented in this work. This study started with an introduction to QAM constellations used since early 1960s to more complex higher order power and bandwidth efficient QAM constellations with their applications in various wireless communication systems and IEEE standards. A detailed study of star QAM is performend. Further, constellation modeling, bit mapping, Gray code penalty, decision regions, peak and average powers, and PAPR of the advance XQAM and HQAM constellations have also been discussed in details. 
			 Study of HQAM includes regular and irregular HQAM constellations with various constellation orders.  Finally, a comparison of various QAM constellations has been presented which concludes the irregular HQAM as the optimum constellation which is highly power efficient than the other constellations employed in the existing wireless communication systems and standards. From this, it can be claimed that the HQAM can be adopted in various wireless applications targeted for 5G and beyond wireless communication systems where high data-rates with good QoS is required within the limited power and bandwidth. Performance can be  improved further by probabilistically shaping the HQAM constellations in the future.

	\ifCLASSOPTIONcaptionsoff

\newpage
\fi

\bibliographystyle{IEEEtran}
\tiny
\bibliography{ref_Tuto}

\end{document}